\documentclass[prd, aps, reprint, amsfonts, amssymb, amsmath, preprintnumbers, showpacs, nofootinbib]{revtex4-1}
\usepackage{graphicx, amsthm, multirow,soul}
\usepackage[citecolor=blue]{hyperref}
\usepackage[all]{hypcap}
\usepackage{url}
\usepackage{color}
\usepackage{subfigure}

\newcommand{\equaref}[1]{Eq.~(\ref{#1})}
\newcommand{\equasref}[2]{Eqs.~(\ref{#1})~and~(\ref{#2})}
\newcommand{\equassref}[3]{Eqs.~(\ref{#1}), (\ref{#2})~and~(\ref{#3})}
\newcommand{\figref}[1]{Fig.~\ref{#1}}

\newcommand{\secref}[1]{Section~\ref{#1}}

\newcommand{\tabref}[1]{Table~\ref{#1}}
\newcommand{\refref}[1]{Ref.~\cite{#1}}
\newcommand{\refsref}[2]{Refs.~\cite{#1}~and~\cite{#2}}
\newcommand{\deltasns}[1]{\ensuremath{\Delta(#1)}}

\def\phig{\ensuremath{\varphi_\text{g}}}
\def\abelian{abelian}
\def\nonabelian{non-abelian}
\def\lagrangian{lagrangian}
\def\eg{\emph{e.g.}}

\def\aka{\emph{a.k.a.}}

\begin{document}

\title{Mixing angle and phase correlations from A$_5$ with generalised CP and their prospects for discovery}

\author{Peter Ballett}
\email{peter.ballett@durham.ac.uk}

\author{Silvia Pascoli}
\email{silvia.pascoli@durham.ac.uk}

\author{Jessica Turner}
\email{jessica.turner@durham.ac.uk}

\affiliation{Institute for Particle Physics Phenomenology, Department of
Physics, Durham University, South Road, Durham DH1 3LE, United Kingdom}

\date{\today}

\begin{abstract}

The observed leptonic mixing pattern could be explained by the presence of a
discrete flavour symmetry broken into residual subgroups at low energies. In
this scenario, a residual generalised CP symmetry allows the parameters of the
PMNS matrix, including Majorana phases, to be predicted in terms of a small set
of input parameters. In this article, we study the mixing parameter
correlations arising from the symmetry group A$_5$ including generalised CP
subsequently broken into all of its possible residual symmetries.
Focusing on those patterns which satisfy present experimental bounds, we then
provide a detailed analysis of the measurable signatures accessible to the
planned reactor, superbeam and neutrinoless double beta decay experiments. We
also discuss the role which could be played by high-precision measurements from
longer term projects such as the Neutrino Factory. 
This work provides a concrete example of how the synergies of the upcoming
experimental programme allow flavour symmetric models to be thoroughly
investigated. Indeed, thanks to the rich tapestry of observable correlations,
we find that each step of the experimental programme can make important
contributions to the assessment of such flavour-symmetric patterns, and
ultimately all patterns that we have identified can be excluded, or strong
evidence found for their continued relevance.
 
\end{abstract}

\preprint{IPPP/15/10, DCPT/15/20} 

\maketitle

%%%%%%%%%%%%%%%%%%%%%%%%%%%%%%%%%%%%%%%%%%%%%%%%%%%%%%%%%%%%%%%%%%%%%%%%%%%%%%%
%%%%%%                          Introduction                             %%%%%%
%%%%%%%%%%%%%%%%%%%%%%%%%%%%%%%%%%%%%%%%%%%%%%%%%%%%%%%%%%%%%%%%%%%%%%%%%%%%%%%

\section{Introduction}

The existence of three families of fermions in the Standard Model (SM),
identical in all properties apart from their masses, is as yet unexplained by
any physical principle or mechanism. Moreover, the discovery that both quarks
and leptons permit complementary but distinct descriptions in terms of the
flavour states which diagonalise the weak interactions and the states which
diagonalise their mass terms has shown that the connection between families
betrays a precise structure which is an essential component in our description
of the physical world.
Explaining the origins of this flavour structure has been a recurring theme in
proposed extensions of the SM. One such programme is the application of
discrete flavour symmetries, predominately in the lepton sector, where the
flavour quantum numbers are associated with a new symmetry and particles are
assigned to its irreducible representations. This can provide a way to unify
the three families into a single mathematical object. However, as the lepton
masses are known to be distinct, any \nonabelian\ symmetry can only be exact
above the scale of mass generation. Nevertheless, its existence at high energy
shapes the theory, and the residual symmetries which survive the breaking
procedure at low energies can play an important role in the structure of
flavour observables. 

The paradigm of a \nonabelian\ flavour symmetry breaking into residual
symmetries has been used by many authors to make predictions about the six
mixing angles and phases which constitute the Pontecorvo-Maki-Nakagawa-Sakata
(PMNS) matrix and parameterize neutrino mixing: $\theta_{12}$, $\theta_{13}$,
$\theta_{23}$, the Dirac phase $\delta$ and two Majorana phases $\alpha_{21}$
and $\alpha_{31}$.
For a recent review of such models see \eg\ \refref{King:2013eh}.
Although many of the earliest models were designed to predict a very small
value of $\theta_{13}$, a prediction now firmly ruled out
\cite{An:2012eh,*An:2013uza,Ahn:2012nd,Abe:2011sj,*Abe:2011fz}, a number of
models remain consistent with the current data, many of which are based on
groups taken from the $\deltasns{6n^2}$ family: $\deltasns{96}$
\cite{Toorop:2011jn,Ding:2012xx}, $\deltasns{150}$ \cite{Lam:2012ga},
$\deltasns{600}$ \cite{Lam:2013ng} and
$\deltasns{1536}$\cite{Holthausen:2012wt}. This connection was strengthened in
\refref{Fonseca:2014koa} which showed that, based on only a few generic model
building assumptions, if the full PMNS matrix is to be specified by the
symmetry alone (so-called ``direct'' models \cite{King:2013eh}), the only
possible predictions which would agree with current data are those arising
(minimally) from $\deltasns{6n^2}$. 
In general, however, the existence of residual symmetries amongst the leptonic
mass terms may not fully specify the mixing pattern. In these ``semi-direct''
models \cite{King:2013eh} the symmetries reduce the degrees of freedom
necessary to describe the mixing parameters by defining a correlation between
previously independent parameters (sometimes known as mixing sum rules
\cite{King:2007pr,King:2005bj,Masina:2005hf,Antusch:2005kw}). Often these
correlations between mixing angles and phases can be derived from quite generic
analyses of the residual symmetries present in a system without needing to
specify a full UV-complete theory \cite{Hernandez:2012ra, Hernandez:2012sk,
Ballett:2013wya, Meloni:2013qda, Hanlon:2013ska, Petcov:2014laa,
Ballett:2014dua}. As such, focusing on these relations can be an effective way
to compare a wide class of models to data \cite{Antusch:2007rk,
Ballett:2013wya, Meloni:2013qda, Hanlon:2013ska, Ballett:2014uia,
Petcov:2014laa, Girardi:2014faa, Ballett:2014dua}.

To date, the fine structure of the PMNS matrix has been inaccessible to
experiment, preventing the study of subtle parameter correlations. 
Current measurements of the mixing angles have $3\sigma$ uncertainties of
around $6.9\%$ on $\theta_{12}$, $7.3\%$ on $\theta_{13}$ and $16.7\%$ on
$\theta_{23}$ \cite{Gonzalez-Garcia:2014bfa}\footnote{For alternative global
analyses of oscillation data, see \refref{Capozzi:2013csa,*Forero:2014bxa}.}.
All three of the CP phases are unconstrained at this significance level;
although, some low-significance hints for a maximally CP-violating value of the
Dirac phase $\delta \approx 3\pi/2$ have been observed
\cite{Gonzalez-Garcia:2014bfa,Capozzi:2013csa,*Forero:2014bxa}.
Therefore the upcoming experimental work will focus on two key topics: the
precision determination of the mixing angles and the first measurements of the
CP phase $\delta$. 
This will open the door for studies of the correlations between mixing angles,
and between mixing angles and phases, that are predicted by models of flavour
symmetries.

In the most popular formulation of models with discrete flavour symmetries, the
constraints on the mass matrices used to derive the PMNS matrix cannot remove a
number of complex phase degrees of freedom. This results in an inability to
predict the Majorana phases and, in general, lessens the predictivity of the
model.
However, by imposing a \emph{generalised CP symmetry} (GCP), phase information
may be accessible and dictated by the flavour structure itself.  This can lead
to very predictive scenarios, where all $6$ mixing parameters are related to a
small number of input parameters \cite{Feruglio:2012cw}. GCPs were first
explored in the context of discrete and continuous groups in
\refref{Ecker:1981wv, *Bernabeu:1986fc,*Ecker:1989ay}; however, they have
recently been revived due to the question of consistency between a CP symmetry
and discrete flavour
group~\cite{Holthausen:2012dk,Feruglio:2012cw,Chen:2014tpa}. This has lead to
interesting work studying the predictions of models with imposed flavour and
GCP symmetries for a number of groups such as A$_4$
\cite{Ding:2013bpa,Holthausen:2012dk}, S$_4$
\cite{Feruglio:2012cw,Feruglio:2013hia,Li:2013jya,Ding:2013hpa,Li:2014eia},
$\deltasns{48}$\cite{Ding:2013nsa,Ding:2014hva} and $\deltasns{96}$
\cite{Ding:2014ssa} along with more comprehensive analyses of the families
$\deltasns{3n^2}$ and $\deltasns{6n^2}$ \cite{Hagedorn:2014wha,
Ding:2014ora,King:2014rwa}. (See also \refref{Everett:2015oka} and
\refref{Chen:2014wxa} for further applications of GCP symmetries.)
 
In this article, we present a detailed analysis of a single group: the
alternating group on $5$ elements, A$_5$. 
This was first introduced to leptonic flavour physics in
\refref{Everett:2008et} via the study of Golden Ratio mixing
\cite{Kajiyama:2007gx,Datta:2003qg}: a possible pattern of the PMNS matrix with
$\theta_{13}=0$, $\theta_{23}=\pi/4$ and a value of $\theta_{12}$ related to
the golden ratio $\varphi = \frac{1+\sqrt{5}}{2}$, $\tan\theta_{12}=1/\varphi$.
This pattern has been shown to be a prediction of a number of different models
based on A$_5$ \cite{Everett:2008et, Everett:2010rd, Datta:2003qg, Ding:2011cm,
Feruglio:2011qq}; however, it is not the only fully-specified mixing pattern
associated with direct models based on this group. If a $\mathbb{Z}_3$ subgroup
is preserved among the charged leptons and a $\mathbb{Z}_2\times\mathbb{Z}_2$
is preserved among the neutrinos, a pattern with $\theta_{13}=0$,
$\theta_{23}=\pi/4$ and $\cos\theta_{12} = \varphi/\sqrt{3}$ can be found
\cite{Lam:2011ag,deAdelhartToorop:2011re,Ballett:2014dua}. Further patterns are
also possible when a $\mathbb{Z}_2\times\mathbb{Z}_2$ symmetry is preserved in
the charged-leptons whilst a different $\mathbb{Z}_2\times\mathbb{Z}_2$ remains
in the neutrino sector, which predict a large value of $\theta_{13}$,
$\theta_{13}\approx 17.9^\circ$ \cite{deAdelhartToorop:2011re}.  

Needless to say, the patterns above are in severe tension with the current
global data by dint of their $\theta_{13}$ predictions alone; however, the
possibility remains that symmetries such as these do not completely survive at
low energies and that a semi-direct approach may remain viable.
In this work, we consider the group A$_5$ with a GCP symmetry, deriving the
most general GCP transformation which could be implemented for this group. We
assume that the flavour group with GCP is broken into a set of residual
symmetries at low energies insufficient to fix all of the oscillation
parameters. We compute all possible predictions for the induced correlations
amongst the mixing parameters. These are compared to the current data, and we
identify those which are compatible with the current bounds. The viable
patterns that we identify are highly predictive, expressing all six parameters
of the PMNS matrix in terms of a single unphysical angle.
We take particular care in assessing the phenomenology of the correlations
between mixing angles and phases for these viable models: discussing their
accessibility to reactor, long-baseline and neutrinoless double beta decay
experiments, and highlighting particularly interesting signatures to be tested.
Although we have restricted our attention to the group A$_5$, this can be seen
as an illustrative choice and we would like to stress the rich but moreover
\emph{readily testable} phenomenology which exists in the residual symmetry
framework, much of which arises from the predictions taken as a whole instead
of resting on single generic types of parameter correlation.

The work presented in this paper is divided into two main parts. In
\secref{sec:intro} and \secref{sec:A5}, we discuss the assumptions behind our
framework, and explain the technical steps in our derivation of the
correlations. In \secref{sec:pheno}, we focus on the predictions
themselves, presenting some simplified formulae for the correlations between
observable quantities and identifying their most interesting phenomenological
signatures. We also study a number of ways that the correlations
can be tested by upcoming reactor, superbeam and neutrinoless double beta decay
experiments, as well as possible longer-term experiments such as neutrino
factories.  We make our concluding remarks in \secref{sec:conclusions}.

\section{\label{sec:intro}Residual flavour and generalised CP symmetries}

We assume the presence of a discrete flavour symmetry, $G$, at high energies.
To unify the three flavours, we assume that the fields are assigned to a
$3$-dimensional irreducible representation of this group, and the general
multiplets of leptons $\Psi$ transform as
\[ \Psi_\alpha \to  \rho(g)_{\alpha\beta}\Psi_\beta, \]
where $\rho\,:\,G\to\text{GL}(3,\mathbb{C})$ represents a unitary representation
of $G$\footnote{Throughout this paper we will assume that all representations 
are unitary and therefore all group elements are represented by unitary 
matrices.}. 
Ensuring the existence of the $3$-dimensional irreducible representation $\rho$
restricts us to \nonabelian\ groups. 

As neutrinos are known to oscillate, they cannot have degenerate masses and
therefore the \nonabelian\ flavour group, $G$, cannot be a symmetry of our
low-energy effective \lagrangian. Therefore, we assume that the full flavour
symmetry must be broken at low energies into two \abelian\ residual symmetry
groups, $G_e$ and $G_\nu$, which are unbroken in the charged-lepton and
neutrino sectors, respectively.
We denote the leptonic mass terms in the low-energy effective theory by
\begin{equation} \begin{aligned} \!\!\!-\mathcal{L}_{\lambda\nu} = 
\overline{(e_\text{L})_\alpha} (m_\lambda)_{\alpha\beta}
(e_\text{R})_\beta~&+~\overline{(e_\text{R})_\alpha}
(m^\dagger_\lambda)_{\alpha\beta} (e_\text{L})_\beta\\
+~\frac{1}{2}\overline{(\nu^c_\text{L})_\alpha} (m_\nu)_{\alpha\beta}
(\nu_\text{L})_\beta~&+~\frac{1}{2}\overline{(\nu_\text{L})_\alpha}
(m^\dagger_\nu)_{\alpha\beta} (\nu^c_\text{L})_\beta.
\label{eq:lagrangian}\end{aligned} \end{equation} 
where the Greek indices are flavour indices  and the mass matrices are
$3\times3$ and complex-valued. 
Due to the anti-commutation of the fermionic fields, one can show that the
Majorana mass matrix $m_\nu$ is restricted to being complex symmetric.  

We assume that there exist residual symmetries acting on the left-handed
charged- and neutral-leptons. If we denote a general element of these subgroups
by $g_e \in G_e$ and $g_\nu \in G_\nu$, the fields transform according to the
following rules
\begin{align*} (e_\text{L})_\alpha \to \rho\left(g_e\right)_{\alpha\beta} (e_\text{L})_\beta,\quad\text{and}\quad(\nu_\text{L})_\alpha \to \rho\left(g_\nu\right)_{\alpha\beta} (\nu_\text{L})_\beta.  \end{align*}
Combining these relations with the \lagrangian\ in \equaref{eq:lagrangian} leads
us to matrix relations which the mass terms must satisfy if the residual
symmetries are to be preserved at low energies 
\begin{align} m_\lambda m_\lambda^\dagger &= \rho\left(g_e\right)^\dagger (m_\lambda
m_\lambda^\dagger)\rho\left(g_e\right), \label{eq:T_mlambda}\\ m_\nu &= \rho\left(g_\nu\right)^\text{T}m_\nu
\rho\left(g_\nu\right).\label{eq:S_mnu} \end{align}
These relations constrain the forms of the mass matrices and, as we will show,
knowledge of their existence can be used to derive a form of the PMNS matrix,
$U_\text{PMNS}$.

Working from a bottom-up perspective, we would like to deduce the
phenomenological consequences of a given choice of residual symmetries $G_e$
and $G_\nu$. The choice of residual flavour groups is constrained in two ways.
Firstly, $G_e$ and $G_\nu$ must be subgroups of the unbroken group $G$.
Secondly, the possible residual flavour symmetries must be subgroups of the
largest symmetry allowed by the mass terms in \equaref{eq:lagrangian}. 
These maximal symmetries are best identified in the basis where both mass terms
are diagonal. In this basis, the most general symmetry of the charged lepton
mass matrices is $U(1)^3$, which has discrete subgroups of the form
$G_e=\mathbb{Z}_m$ for any $m$ or a direct product of such groups. For the
neutrino residual symmetry, the argument changes due the assumed Majorana
nature of the mass term. In this case the largest possible symmetry is smaller,
$\mathbb{Z}_2\times\mathbb{Z}_2$, which leaves us with only two choices,
$G_\nu=\mathbb{Z}_2$ or $G_\nu=\mathbb{Z}_2\times\mathbb{Z}_2$. 

In addition to a flavour symmetry $G$, we also assume the presence of a GCP
symmetry. A CP symmetry is understood as a combination of charge-conjugation
and a parity transformation; however, a generalised CP symmetry is one which
also acts on the flavour indices whilst making this transposition. In
\refref{Holthausen:2012dk}, it was shown that ensuring the consistency of a
discrete flavour symmetry and a CP symmetry often requires the introduction of
a non-trivial generalised CP symmetry. 

We define our generalised CP symmetry to act on a set of fields $\Psi_\alpha$
as 
\[ \Psi_\alpha \to X_{\alpha\beta}\Psi^c_\beta,   \]
where $X_{\alpha\beta}$ is assumed to be a unitary matrix so as to preserve the
kinetic terms in the \lagrangian\ and $\Psi^c$ denotes the conventional CP
conjugate appropriate for the Lorentz representation of the field $\Psi$. If a
discrete flavour symmetry is present, $G$, then a generalised CP symmetry must
satisfy a consistency equation \cite{ Holthausen:2012dk,Feruglio:2012cw} 
\begin{equation}\label{consistency} X {\rho(g)}^{*}{X}^{*}=\rho(g^\prime),
\end{equation}
where $g$ and $g^\prime$ are elements of $G$. For a faithful representation
$\rho$, this relation can be seen as establishing a mapping from $g$ to
$g^\prime$ which preserves the structure of the group and therefore defines a
group automorphism. In \refref{Chen:2014tpa}, it was pointed out that a
physical GCP transformation must be restricted to a single irreducible
representation, and is related to a class-inverting automorphism of $G$,
meaning that $g^\prime$ is mapped to an element in the conjugacy class of
$g^{-1}$. 
We restrict our consideration to involutory GCP transformations, requiring that
the application of the transformation twice is equivalent to the identity, and
therefore the $X$ matrix satisfies an additional constraint
\begin{equation}  XX^* = 1. \label{eq:XXstar} \end{equation} 

As with the flavour symmetry $G$, for our GCP symmetry to leave the
\lagrangian\ invariant the mass matrices must satisfy further constraints. The
GCP symmetry exchanges the hermitian conjugate terms in the lagrangian of
\equaref{eq:lagrangian}, which remains invariant if the mass matrices obey the
relations 
\begin{align}  X^\text{T} m_\nu X &= m^*_\nu, \label{eq:Xconst} \\ X^\dagger
(m_\lambda  m^\dagger_\lambda) X &= (m_\lambda
m^\dagger_\lambda)^*.\label{eq:XCLconst} \end{align}

If \equasref{eq:Xconst}{eq:XCLconst} are unbroken relations at low energies, it
can be shown that all CP violating effects of the PMNS matrix vanish
\cite{Holthausen:2012dk,Feruglio:2012cw}.  However, if only one of these
relations is preserved, the consistency of the flavour and CP symmetries leads
to novel constraints on the PMNS matrix. In this paper, we assume that the GCP
symmetry is broken in the charged-lepton sector but is preserved in the
neutrino sector. For this to be consistent, the $X$ matrix must map the 
elements of the neutrino residual symmetry to themselves, 
\begin{equation} X \rho(g_\nu)^* X^* = \rho(g_\nu). \label{eq:SX_commute}\end{equation}

In summary, we assume a discrete flavour symmetry $G$ and a GCP symmetry
implemented by $X$ at high energy scales which is assumed to break into a
subgroup $G_e$ acting on the charged-lepton mass terms and another subgroup
$G_\nu$ which along with the GCP symmetry acts on the neutrino mass terms. This
leads to a system of constraints which the mass matrices must satisfy:
\equaref{eq:T_mlambda}, \equaref{eq:S_mnu},  \equaref{eq:Xconst} and
\equaref{eq:SX_commute}. In the next section we will show how knowledge of
these constraints alone can be used to predict the PMNS matrix, including its
Majorana phases.

\subsection{Constructing the PMNS matrix using symmetry constraints}
Constraints on the mass matrices, as derived above, lead to restrictions on
their allowed form, and subsequently to the matrices required to diagonalise
them.
We focus first on the charged leptons. A constraint on the charged-lepton mass
matrix of the form in \equaref{eq:T_mlambda} can be rephrased as a statement of
commutation
\[ [\rho(g_e),(m_\lambda m^\dagger_\lambda)] = 0.  \]
As the matrix $\rho(g_e)$ is unitary and the matrix $H=m_\lambda
m^\dagger_\lambda$ is hermitian, there exists a basis such that they are
simultaneously diagonalised,
\[ \exists U_e~\text{s.t.}~U_e^\dagger U_e = 1,~~~ \rho(g_e)_\text{d} =
U_e^\dagger \rho(g_e) U_e,~~~ H_\text{d} = U_e^\dagger H U_e, \]
where $\rho(g_e)_\text{d}$ and $H_\text{d}$ denote diagonal forms of the
matrices $\rho(g_e)$ and $H$. 
As the charged leptons have distinct masses, $H$ is full rank.  This implies
that $U_e$ is unique (up to re-phasing and re-ordering of columns). If
$\rho(g_e)$ is also known to be full rank, $U_e$ can also be found by
diagonalising this operator.  In this way, by insisting on the relation in
\equaref{eq:T_mlambda}, we can compute $U_e$ solely from the group element
$\rho(g_e)$, and the symmetry alone specifies the mixing matrix.
However, a complication arises if $\rho(g_e)$ is not full rank. In this case,
$\rho(g_e)$ does not have a unique diagonalising matrix, as in any basis in
which it takes diagonal form, further SU(2) transformations can be performed
freely in its degenerate eigenspace. Without knowledge of the mass matrix, our
knowledge of $\rho(g_e)$ will only allow the identification of the family of
diagonalising matrices of $\rho(g_e)$, and $U_e$ must take a more general form
\[ U_e = U_0 R_e(\phi,\gamma)\Phi, \]
where $U_0$ is any matrix which diagonalises $\rho(g_e)$, $R_e(\phi,\gamma)$ is
a complex rotation in the degenerate subspace of $\rho(g_e)$ by an angle $\phi$
with a phase $\gamma$ and $\Phi$ is a diagonal matrix of phases.

In the neutrino sector, we have three constraints to consider on the mass
terms: one from the flavour symmetry, one from the GCP symmetry and one
ensuring their consistency.  Under a change of flavour basis, the matrix $X$ is
mapped to 
\[ X \to U^\dagger XU^*.  \]
As $X$ is unitary and symmetric, Takagi factorization allows us to express it
as $X=\Omega\Omega^\text{T}$ for some unitary matrix $\Omega$ which implies
that we can choose a basis where $X$ becomes trivial  \cite{Feruglio:2012cw}.
In fact, this basis is not unique and the remaining freedom can be used to
further diagonalise $\rho(g_\nu)$. In this basis the constraint in
\equaref{eq:Xconst} implies that the mass matrix is real-valued
\[ \left(\Omega^\text{T}m_\nu\Omega\right)_{\alpha\beta} \in \mathbb{R}. \]
As $\rho(g_\nu)$ is diagonal and commutes with this matrix, we know that the
mass matrix in this basis must be diagonal up to a basis change in the
degenerate subspace of $\rho(g_\nu)$. As it is purely real, the most general
additional basis transformation required to bring it into diagonal form is a
rotation in 2-dimensions  
\[   U_\nu = \Omega R_\nu(\theta), \]
where $\theta$ is the angle describing the real rotation. There remains the
possibility that the diagonal mass matrix is not positive definite, in which
case a diagonal re-phasing must occur.  
Without further knowledge of the mass matrix this cannot be predicted, and in
consequence, the Majorana phases can be predicted only up $\pm\pi$ or multiples
thereof\footnote{We will always work with the Particle Data Group
parameterization of the PMNS matrix \cite{Agashe:2014kda}, in which the
Majorana phases are defined by the diagonal matrix
$\text{diag}\left(1,e^{\mathrm{i}\frac{\alpha_{21}}{2}},
e^{\mathrm{i}\frac{\alpha_{31}}{2}}\right)$ which take physical values on the
intervals $\alpha_{ij}\in[0,2\pi)$.}. 

We see that the GCP symmetry in the neutrino sector has specified a special
basis in which the residual flavour symmetry elements are diagonal and the mass
matrix is real. In this way, GCP symmetries help to fix some of the re-phasing
degrees of freedom associated with diagonalising matrices and allow for the
prediction of Majorana phases.
Combining the results for the charged-lepton sector and the neutrino sector, we
find the full PMNS matrix is given by 
\begin{equation} U_\text{PMNS} = \Phi R_e(\phi, \gamma)U^\dagger_0 \Omega
R_\nu(\theta), \label{eq:PMNS_decomp} \end{equation}
where $R_e$ and $R_\nu$ denote two unspecified rotations ($R_e=1$ if
$\text{ord}\left(g_e\right)>2$). We make two further simplifications: $\Phi$ is
removed by re-phasing the charged leptons, and we note that the angles $\theta$
and $\phi$ need only be defined over the interval $\theta,\phi\in[0,\pi)$, as
shifts by $\pi$ can be absorbed by unphysical redefinitions of the complex
phases. 

\section{\label{sec:A5}Mixing patterns from A$_5$}

The preceding section showed how the assumed residual flavour and GCP
symmetries can lead to expressions for the PMNS matrix. In this section, we
will derive the possible mixing matrices which arise by this method for the
group A$_5$. First we will discuss the structure of A$_5$ and the possible
subgroups eligible to be taken as residual symmetries. Then we will derive the
form of the most general GCP transformation. In the subsequent subsections, we
consider all viable combinations of CP and residual subgroups and present those
patterns which are consistent with the current global data \cite{Gonzalez-Garcia:2014bfa}.

\subsection{Subgroups and GCP symmetries}

The group A$_5$ can be defined as the group of even permutations on $5$
elements. It has the abstract presentation 
\[ \langle S, T\,|\,S^2 = T^5 = (ST)^3 \rangle,  \]
where $S$ and $T$ are the \emph{generators} of the group and all group elements
can be expressed by a word made from these distinguished elements.
The structure of this group and its representation theory have been discussed
in the physics literature before (see \eg\
\refsref{Everett:2008et}{Ding:2011cm}) and we abstain from deriving the
explicit representations, deferring the reader to these references instead.
However, we will briefly recap those features of the group and its
representations most pertinent to our subsequent analysis.

We assume that the lepton doublets are assigned to a
$3$-dimensional representation. A$_5$ has two distinct $3$-dimensional
irreducible representations, and in the following we will always work with the
representation $3$ of \refref{Ding:2011cm}. We have checked that the final
results do not change if we choose the alternative $3$-dimensional
representation instead. For our chosen representation, the
generators $S$ and $T$ can be expressed by  
\begin{align*} 
S = \left ( \begin{matrix} -1 & 0 & 0 \\ 0 & -1 & 0 \\ 0 & 0 & 1
\end{matrix}\right )\quad\text{and}\quad T = \frac{1}{2}\left ( \begin{matrix}
1 & -\varphi & -\phig \\ \varphi & -\phig & -1 \\ -\phig & 1 & \varphi
\end{matrix}\right ),
\end{align*}
where $\phig=\frac{1-\sqrt{5}}{2}$ represents the Galois conjugate\footnote{In
this case Galois conjugation exchanges the two solutions of the minimal
polynomial over the rationals $x^2 - x - 1 =0$.} of $\varphi$.  We note that
this is a real representation and in our chosen basis all group elements are
real; it also forms a subgroup of SU($3$).

The non-identity elements of A$_5$ are either of order $2$, $3$ or $5$ and can
be partitioned into $4$ conjugacy classes: one of order $2$ elements ($15$
members), one of order $3$ elements ($20$ elements) and two of order $5$
elements ($12$ members each). The centre of A$_5$ is trivial, and the identity
alone forms one additional conjugacy class. 

As described in \secref{sec:intro}, to compute the PMNS matrix from residual
symmetries, we must first find $U_e$, the matrix which diagonalises the
generator of the residual symmetry of the charged-lepton mass term. To do so,
we must identify the eligible residual symmetry groups for $G_e$. This symmetry must be
an \abelian\  subgroup of A$_5$, of which there are 4 kinds (up to
isomorphism). Three of these subgroups are cyclic groups: $\mathbb{Z}_2$,
$\mathbb{Z}_3$ and $\mathbb{Z}_5$. These are the groups generated by a single
element $g$, and its members are those powers of $g$ less than or equal to its
order 
\[\langle g\rangle= \{g^n\,|\,\text{s.t.}~n\le\text{ord}(g)\}.\] 
As any element can be taken as the generator of a cyclic group, there are $15$
distinct subgroups of $\mathbb{Z}_2$ in A$_5$, $10$ of $\mathbb{Z}_3$ and $6$
of $\mathbb{Z}_5$. The diagonalizing basis for these groups is simply the basis
which diagonalises the generator $\rho(g)$.

In addition to the cyclic subgroups, there are also non-cyclic \abelian\
subgroups in A$_5$. These are isomorphic to the Klein four group
$\mathbb{Z}_2\times\mathbb{Z}_2$ which is generated by distinguished pairs of
order-2 elements, 
\[ \langle g_1, g_2\,|\,g_1^2=g_2^2=(g_1g_2)^2 \rangle.  \]
In fact, the $15$ order-2 elements in A$_5$ can be divided into $5$ triplets
which (with the identity) define distinct 
four-element groups. For these non-cyclic groups, the diagonalizing basis is
defined as that which diagonalizes the two generators simultaneously.

Therefore, the different choices for the residual symmetry of the
charged-lepton mass term can be divided into four categories depending on the
preserved subgroup: $G_e\in\{\mathbb{Z}_2, \mathbb{Z}_3,
\mathbb{Z}_5,\mathbb{Z}_2\times\mathbb{Z}_2\}$. For the residual symmetry of
the neutrino Majorana mass term, we are restricted to taking subgroups of
$\mathbb{Z}_2\times\mathbb{Z}_2$, leaving us with two options: a single
$\mathbb{Z}_2$ or the full Klein group $\mathbb{Z}_2\times\mathbb{Z}_2$.

\subsection{Deriving $X$}

The matrix $X$ which implements the generalised CP symmetry, as discussed in
\secref{sec:intro}, must satisfy $XX^*=1$ and be related to a class-inverting
automorphism of the group,
\[ \forall g \in \mathrm{A}_5,~\exists h_g\in \mathrm{A}_5~~\text{s.t.}~~
(X^*\rho(g)X)^* = \rho(h_g^{-1})\rho(g^{-1})\rho(h_g), \]
where $\rho$ is our chosen irreducible representation, generated by the
matrices $S$ and $T$.
We shall derive the most general form of $X$ for the group A$_5$ by exploiting
our knowledge of the automorphism structure of the group. The automorphism
group of A$_5$ is S$_5$ (see \eg\ \cite{simple}), and we identify two important
subgroups: \emph{inner} and \emph{outer} automorphisms. The inner automorphism
group, $\text{Inn}(\mathrm{A}_5)$, comprises those automorphisms which can be
represented by conjugation by a group element,
\[ \phi_h \in \text{Inn}(\mathrm{A}_5)~~~\iff~~~
\forall\,g\in\mathrm{A}_5,~~\phi_h(g) = h^{-1}gh. \]
This group can be found by considering the map from element ($h\in
\mathrm{A}_5$) to inner automorphism ($\phi_h(g) = h^{-1}gh$), and applying the
first isomorphism theorem,
\[ \mathrm{Inn}(\mathrm{A}_5) \cong \mathrm{A}_5/\mathcal{Z}(\mathrm{A}_5)
\cong \mathrm{A}_5, \]
where the final step uses the fact that A$_5$ has trivial centre,
$\mathcal{Z}(\mathrm{A}_5)=1$. Therefore, the inner automorphisms of A$_5$ are
given by A$_5$ itself.
The outer automorphism group is defined as the quotient of the full
automorphism group by the inner automorphism group. For A$_5$ it follows from
our discussion above that this is the unique group of two elements
$\mathbb{Z}_2$. Our derivation of $X$ is greatly simplified by A$_5$ being an
\emph{ambivalent} group, where each element is conjugate to its inverse. For
such groups, the class-inverting automorphisms are also class-preserving. All
inner automorphisms of a group are class-preserving, but the two properties are
not equivalent as there do exist class-preserving outer automorphisms for some
groups \cite{Burnside:1913aa,Brooksbank:2014aa}.  However, for the case of
A$_5$ we have a single non-trivial outer automorphism to check, and this
automorphism maps elements of order 5 from one conjugacy class to the other.
Therefore, in the present case, we conclude that the class-preserving
automorphisms are precisely the inner automorphisms.

We can therefore simplify our defining constraint on $X$, 
\[ \exists h\in\mathrm{A}_5,~\forall g \in \mathrm{A}_5 \qquad (X^*\rho(g)X)^*
= \rho(h^{-1})\rho(g)\rho(h), \]
where the element $h$ is the same for all elements $g$. As we are working with
a real representation, we can always change basis so that all group elements
are given by real matrices, and we use this fact with \equaref{eq:XXstar} to
make further simplifications
\begin{align*} \forall g \in \mathrm{A}_5 \qquad X\rho(g)X^* &=
\rho(h^{-1})\rho(g)\rho(h),\end{align*}
which is equivalent to a commutation relation,
\begin{align*} [\rho(h)X, \rho(g)] &=  0. \end{align*}
We can then invoke Schur's lemma to infer that as $\rho(h)X$ commutes with all
the elements of an irreducible representation, it must be a scalar matrix:
$\rho(h)X = \lambda 1$, for some complex constant $\lambda$. 
Requiring that
$XX^*=1$ constrains $\rho(h^2)=1/|\lambda|^2$. However, by closure the element
on the left must be a member of A$_5$ and, as our representation is unitary, we
conclude that $\lambda$ is just a complex phase, $\lambda=e^{\mathrm{i}\theta}$
for $\theta\in\mathbb{R}$.  Therefore, $h$ must be an order 2 element, and the
most general form of $X$ which implements an involutory class-inverting
automorphism for A$_5$ is given by 
\[ X = e^{\mathrm{i}\theta} \rho(h) \qquad\text{s.t.}\qquad \text{ord}(h)=2. \]

In our basis, the consistency relation in \equaref{eq:SX_commute} implies that
the $X$ matrix must commute with the generator $S$ of the residual
$\mathbb{Z}_2$ symmetry in the neutrino sector. Therefore not all choices of
$h$ can be consistently implemented, and there will be only $3$ non-trivial $X$
matrices (up to global phases) for any given $S$. These are the three elements
of the Klein four group associated with $S$. If we work in the basis where this
group is diagonal, we find that 
\[ X_1=e^{\mathrm{i}\theta}\left(\begin{matrix} 1 & 0 & 0 \\  0 & -1 & 0 \\ 0 &
0 & -1 \end{matrix}\right),\]
and $X_2$ and $X_3$ can be defined as permutations of this matrix, where the
row of the positive entry is denoted by the subscript. It is necessary for us
to find the basis in which $X$ is trivial. The necessary change of basis is
given for $X_i$ by $\Omega_i$, where
\[\Omega_1 = e^{\mathrm{i}\theta/2}\left(\begin{matrix} 1 & 0 & 0 \\ 0 &
\mathrm{i} & 0 \\ 0 & 0 & \mathrm{i} \end{matrix}\right), \] 
and similar definitions hold for $i=2$ and $i=3$. In this basis, the generator
of the residual symmetry in the neutrino sector is diagonal and the GCP action
is trivial. However, due to the degenerate subspace in the $\mathbb{Z}_2$
generator, it remains possible that the neutrino mass matrix is only block
diagonal and requires an orthogonal transformation to fully diagonalise it.
This rotation must be in the plane of the degenerate subspace for the matrix
$S$. Given these two elements, the most general form of the matrix which maps
between neutrino flavour and mass bases is given by
\[U_\nu = \Omega R(\theta), \]
where $R(\theta)$ is an orthogonal matrix effecting a rotation in either the
12-, 13- or 23-plane. We also note at this point that the overall phase
included in our definition of $\Omega$ can be seen to have no physical effect,
and will be set to zero in what follows.

In the following subsections, we consider all possible residual symmetry groups
in the charged-lepton sector. In the first three sections we consider the
charged-lepton residual symmetry to be given by each of the $10$ $\mathbb{Z}_3$
subgroups, $6$ $\mathbb{Z}_5$ subgroups and $5$
$\mathbb{Z}_2\times\mathbb{Z}_2$ subgroups.  For each of these subgroups, the
basis is found which diagonalises its elements, $U_e$. Due to the order of
these subgroups, this diagonalizing matrix is predicted exactly with no
remaining degrees of freedom (\eg\ $R_e = 1$ in \equaref{eq:PMNS_decomp}). The
PMNS matrix is then constructed combining $U_e$ with one of the forms of
$U_\nu$ found above by consideration of the residual CP symmetry, 
\[ U_\text{PMNS} = U^\dagger_e\Omega R(\theta).  \]
Finally in \secref{sec:2and3degrees}, we consider less restrictive symmetries
when $U_e$ is not fully specified by symmetry alone ($R_e \neq 1$).

For each configuration considered in this section, the arbitrariness in
eigenvector ordering and phasing is accounted for by considering all
permutations of rows and columns of the PMNS matrix. From these permuted
matrices, we compute the mixing angles and phases, and these are compared to
global data. We report all patterns of mixing angles found by this process
which are consistent with the current $3\sigma$ regions as reported in
\refref{Gonzalez-Garcia:2014bfa}. 

\subsection{\label{sec:Z3}Predictions from $G_e=\mathbb{Z}_3$ and $G_\nu=\mathbb{Z}_2$}

\begin{figure}[t]
\includegraphics[width=1.0\linewidth, clip, trim=55 60 50 65]{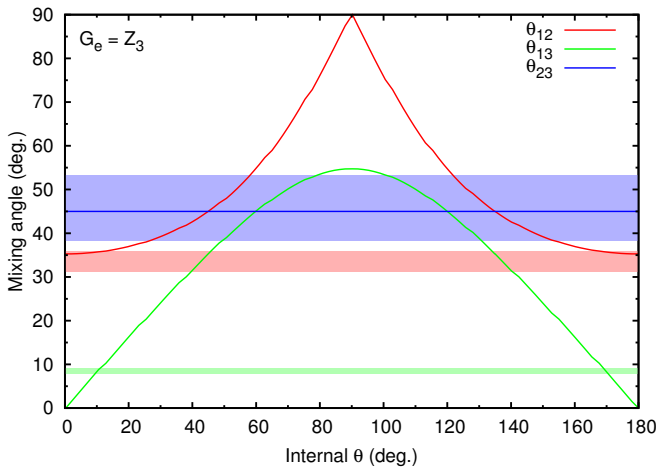}
\caption{\label{fig:Z3}Mixing angles for $\mathbb{Z}_3$ as a function of the internal parameter $\theta$. This pattern predicts $|\sin\delta|=1$ and $\sin\alpha_{21}=\sin\alpha_{31}=0$. The shaded regions show the $3\sigma$ allowed region for the corresponding mixing angle according to current global data~\cite{Gonzalez-Garcia:2014bfa}.}
\end{figure}

\begin{figure*}[t]
\centering
\includegraphics[width=0.45\textwidth, clip, trim=55 60 50 65]{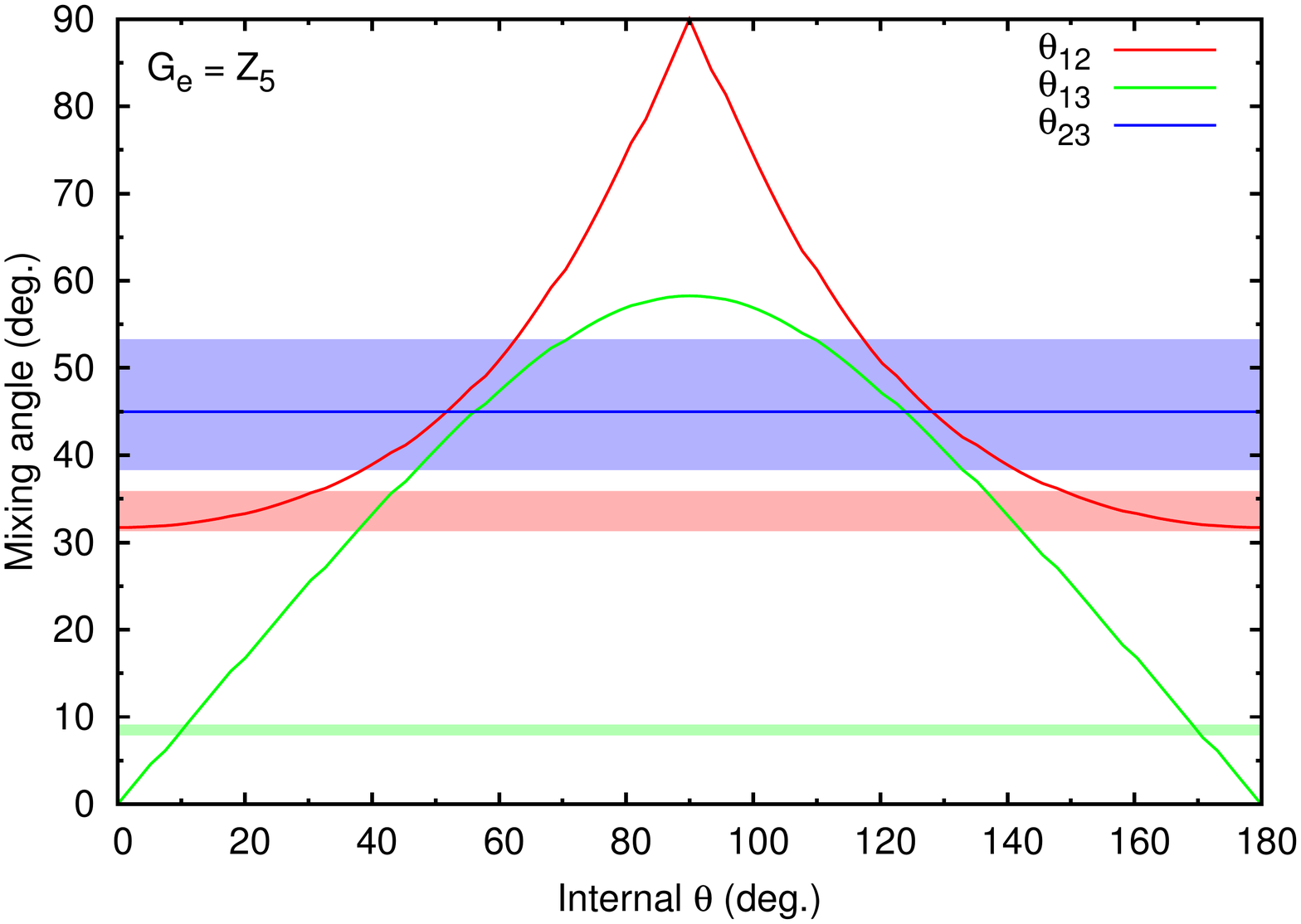}\hspace{0.6cm}% 
\includegraphics[width=0.45\textwidth, clip, trim=55 60 50 65]{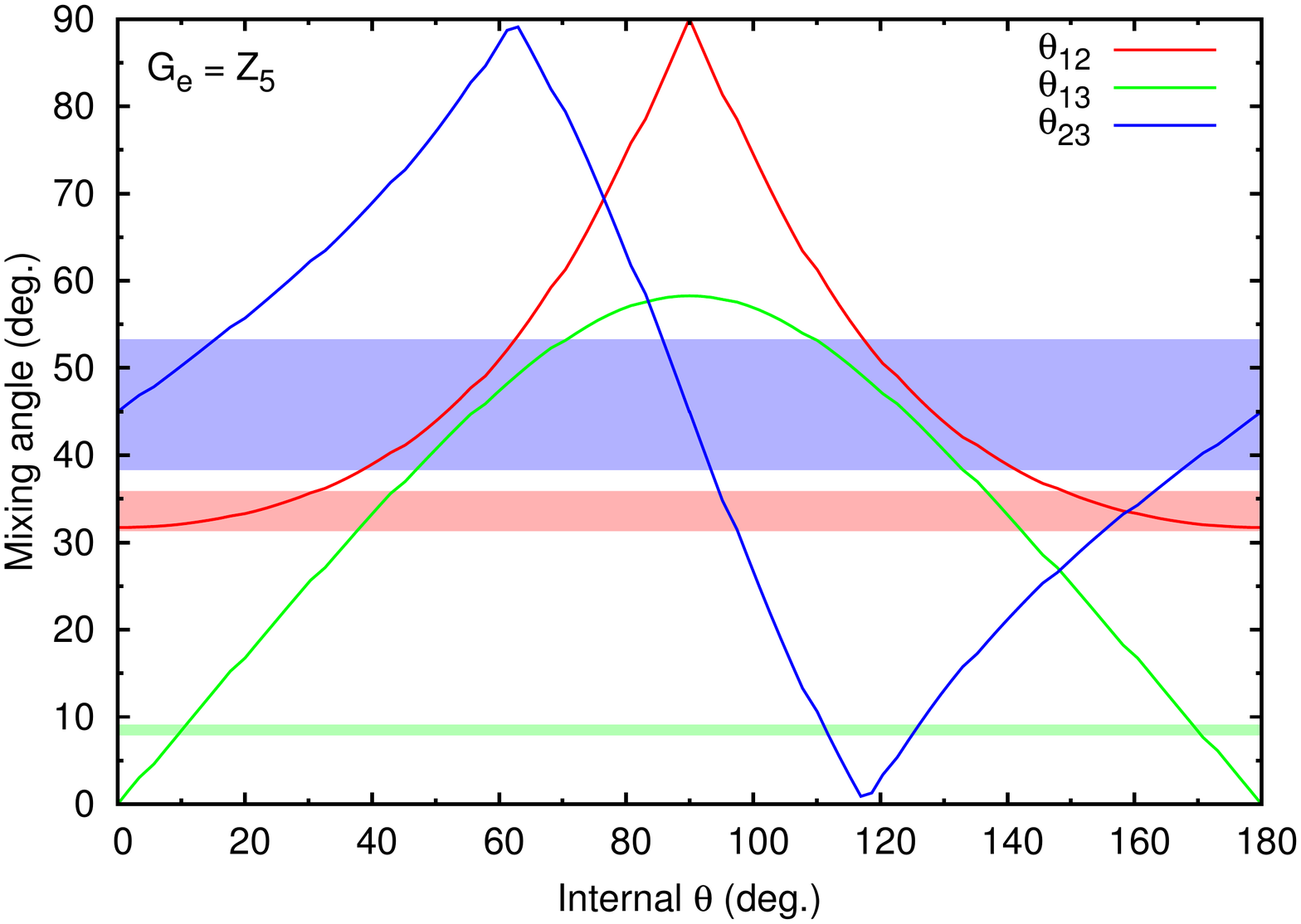}

\caption{\label{fig:Z5}The two patterns of mixing angles for $\mathbb{Z}_5$ as
a function of the internal parameter $\theta$.  Both patterns predict
$\sin\alpha_{21}=\sin\alpha_{31}=0$.  The right pattern predicts
$|\sin\delta|=0$ whilst the left pattern predicts $|\sin\delta|=1$.  The shaded regions show the $3\sigma$ allowed region for the corresponding mixing angle according to current global data \cite{Gonzalez-Garcia:2014bfa}.
 }
\end{figure*}

When the residual symmetry in the charged leptons is taken as $\mathbb{Z}_3$,
the diagonalising matrix of the residual symmetry generator $T$ is uniquely
specified (up to diagonal rephasings and permutations). We have considered the
$10$ $\mathbb{Z}_3$ subgroups of A$_5$ which could act as the residual symmetry
of the charged leptons. 
Although many different group elements lead to viable mixing patterns, all
viable solutions can be described by a single matrix after a suitable
permutation and redefinition of the unphysical parameters. This leads to a
single viable set of correlations between the angles. The angles can be derived
from the PMNS matrix, 
\begin{equation}  \label{eq:Z3_matrix} U_\text{PMNS} = \left
(\begin{matrix}\sqrt{\frac{2}{3}} & -\frac{\mathrm{i}}{\sqrt{3}} & 0 \\
-\frac{\mathrm{i}}{\sqrt{6}} & \frac{1}{\sqrt{3}} & \frac{1}{\sqrt{2}} \\
\frac{\mathrm{i}}{\sqrt{6}} & -\frac{1}{\sqrt{3}} & \frac{1}{\sqrt{2}}
\end{matrix} \right ) R_{13}\left(\theta\right),\end{equation}
where $R_{13}\left(\theta\right)$ denotes a rotation in the 13-plane by an
angle $\theta$. This leads to the following expressions for the mixing angles
\begin{align*} \sin^2\theta_{12} = \frac{1}{3-2\sin^2\theta}, \qquad
\sin^2\theta_{13} = \frac{2}{3}\sin^2\theta, \end{align*}
\vspace{-0.6cm}
\begin{align*} \sin^2\theta_{23} = \frac{1}{2}.  \end{align*}
This pattern is continuously connected to the tribimaximal mixing pattern
\cite{Harrison:2002er} which is recovered at $\theta=0$, and is an explicit
example of a trimaximal pattern \cite{Haba:2006dz,*He:2006qd,*Grimus:2008tt}
where $\left|U_{\alpha 2}\right| =
1/\sqrt{3}~~\forall\,\alpha\in\{e,\mu,\tau\}$.
We have plotted these mixing angle predictions for the full range of the
unobservable parameter $\theta$ in \figref{fig:Z3} in which the coloured
regions show the current $3\sigma$ global intervals from
\refref{Gonzalez-Garcia:2014bfa}.

The Dirac phase for this pattern depends discretely on the value of $\theta$.
It can be shown that 
\[ \delta = \left\{ \begin{matrix} \frac{3\pi}{2} &\quad
\theta\in(0,\frac{\pi}{2}),  \\
\frac{\pi}{2} &\quad \theta\in(\frac{\pi}{2},\pi), \end{matrix}\right. \]
whilst the Majorana phases can be shown to take CP conserving values
$\{\alpha_{21},\alpha_{31}\}\subseteq\{0,\pi\}$ for all values of $\theta$.
From \figref{fig:Z3}, we see that there are two intervals of the unphysical
parameter which lead to mixing angles which satisfy the current global
$3\sigma$ bounds. Due to the symmetry of the expressions, these two solutions
offer identical predictions for values of $\theta_{12}$, $\theta_{13}$ and
$\theta_{23}$; however, one of these values lies in a region with $\delta =
\frac{\pi}{2}$ whilst the other predicts $\delta=\frac{3\pi}{2}$. Therefore,
there are two sets of predictions from the order-3 elements, differing only in
their prediction for the Dirac CP phase.

\subsection{\label{sec:Z5}Predictions from $G_e=\mathbb{Z}_5$ and $G_\nu=\mathbb{Z}_2$}

If the residual charged-lepton symmetry is assumed to be $G_e=\mathbb{Z}_5$,
there are 6 subgroups which we need to consider which could lead to a viable
set of mixing parameters. We find two distinct sets of correlations which are
viable for some range of the unphysical parameter, leading to three distinct
sets of mixing angle predictions. 

The first set of predictions can be derived from the following matrix, 
\begin{equation} U_\text{PMNS} = \left
(\begin{matrix}\frac{\varphi}{\sqrt{2+\varphi}} &
-\frac{\mathrm{i}}{\sqrt{2+\varphi}} & 0 \\
-\frac{\mathrm{i}}{\sqrt{4+2\varphi}} & \frac{\varphi}{\sqrt{4+2\varphi}} &
\frac{1}{\sqrt{2}} \\ \frac{\mathrm{i}}{\sqrt{4+2\varphi}} &
-\frac{\varphi}{\sqrt{4+2\varphi}} & \frac{1}{\sqrt{2}} \end{matrix}
\right)R_{13}(\theta), \label{eq:Z5_matrix_1} \end{equation}
which leads to mixing angles expressed by
\begin{align*} \sin^2\theta_{12} = \frac{1}{1+\varphi^2\cos^2\theta}, \qquad
\sin^2\theta_{13} = \frac{\sin^2\theta}{1+\phig^2}, \end{align*}
\vspace{-0.6cm}
\begin{align*} \sin^2\theta_{23} = \frac{1}{2}.  \end{align*}
The mixing angle predictions from this pattern are shown on the left of
\figref{fig:Z5} as a function of the unphysical parameter $\theta$. In this
case, the Dirac phase is maximally CP violating with $\cos\delta =0$. However,
as with the order-3 elements, the sign of $\sin\delta$ depends on the parameter
$\theta$, 
\[ \delta = \left\{ \begin{matrix} \frac{3\pi}{2} &\quad
\theta\in(0,\frac{\pi}{2}),  \\
\frac{\pi}{2} &\quad \theta\in(\frac{\pi}{2},\pi), \end{matrix}\right. \]
and the Majorana phases are again given by CP conserving values,
$\{\alpha_{21},\alpha_{31}\}\subseteq\{0,\pi\}$, although the precise values
cannot be determined in this framework.

The second viable pattern arising from $\mathbb{Z}_5$ is shown on the right
panel of \figref{fig:Z5}.  This can be derived from a matrix similar to
\equaref{eq:Z5_matrix_1} but distinct in the relative phasing between the
columns,
\begin{equation} U_\text{PMNS} = \left
(\begin{matrix}\frac{\varphi}{\sqrt{2+\varphi}} & \frac{1}{\sqrt{2+\varphi}} &
0 \\ -\frac{1}{\sqrt{4+2\varphi}} & \frac{\varphi}{\sqrt{4+2\varphi}} &
\frac{1}{\sqrt{2}} \\ \frac{1}{\sqrt{4+2\varphi}} &
-\frac{\varphi}{\sqrt{4+2\varphi}} & \frac{1}{\sqrt{2}} \end{matrix}
\right)R_{13}(\theta).
\label{eq:Z5_matrix_2} \end{equation}
This relative phase difference, which arises from the choice of alignment
between the matrix implementing the GCP symmetry $X$ and the generator $S$ of
the residual $\mathbb{Z}_2$ symmetry of the neutrino mass term, crucially
affects the mixing angle $\theta_{23}$ and leads to mixing angles which can be
expressed by
\begin{align*} \sin^2\theta_{12} = \frac{1}{1+\varphi^2\cos^2\theta}, \qquad
\sin^2\theta_{13} = \frac{\sin^2\theta}{1+\phig^2},\end{align*}
\vspace{-0.6cm}
\begin{align*} \sin^2\theta_{23} = \frac{1}{2}\frac{\left(\sin\theta +
\sqrt{1+\varphi^2}\cos\theta\right)^2}{1+\varphi^2\cos^2\theta}.  \end{align*}
All CP phases take CP conserving values for this pattern of mixing parameters.
The precise value of the Dirac phase again depends on $\theta$,
\[ \delta = \left\{ \begin{matrix} 0 &\quad \theta\in(0,\frac{\pi}{2}),  \\ 
\pi &\quad \theta\in(\frac{\pi}{2},\pi). \end{matrix}\right. \]
The appearance of CP conservation can be explained as, although it was not
imposed explicitly, the generalised CP symmetry remains accidentally unbroken
in the charged-lepton sector.
This second pattern leads to two distinct allowed intervals in $\theta$ once we
restrict the mixing angles to lie in the current $3\sigma$ intervals. Due to
the symmetry of the curves, these two viable sets of mixing angles are
distinguished only by their predictions for $\theta_{23}$.

Both of the above mixing patterns are continuous extensions of the well known
GR mixing pattern (\aka\ GR1 or GRA) \cite{Kajiyama:2007gx,Datta:2003qg}, which
is found at $\theta=0$. 
If we expand in the small parameter $r\equiv\sqrt{2}\sin\theta_{13}$
\cite{King:2007pr}, we find that both patterns arising from $\mathbb{Z}_5$ lead
to the prediction for $s\equiv\sqrt{3}\sin\theta_{12}-1$ \cite{King:2007pr}
given by, 
\begin{align} s&= \sqrt{\frac{3}{2+\varphi}}-1 +
\sqrt{\frac{3}{2+\varphi}}\frac{r^2}{4} +\mathcal{O}\left(r^4\right).
\label{eq:Z5_th12}\end{align}
The two patterns are distinguished by their predictions for $\theta_{23}$;
however, both can be seen as subcases of a more general model predicting the
$\theta_{12}$ correlation in \equaref{eq:Z5_th12} and further obeying a
linearised atmospheric sum rule first derived in \refref{Ballett:2013wya}, 
\[a = \frac{1-\varphi}{\sqrt{2}}r\cos\delta + \mathcal{O}(r^2,a^2). \]
From this relation, it is clear how the maximal angle of the first pattern
$a=0$ is associated with the vanishing of $\cos\delta$, while the contrasting
non-trivial predictions of $\theta_{23}$ in the second pattern is due to the
CP-conserving value of $\delta$, $\sin\delta=0$. However, the correlation of
maximal atmospheric mixing and CP conserving values of $\delta$ is not an
artefact of linearisation and holds exactly, as is shown in the right panel of
\figref{fig:Z5}. We shall consider these correlated maximal predictions as a
measurable signature in \secref{sec:th23_delta}.

\subsection{Predictions from $G_e=\mathbb{Z}_2\times\mathbb{Z}_2$ and $G_\nu=\mathbb{Z}_2$}

\begin{figure}[t]
\includegraphics[width=1.0\linewidth, clip, trim=55 60 50 65]{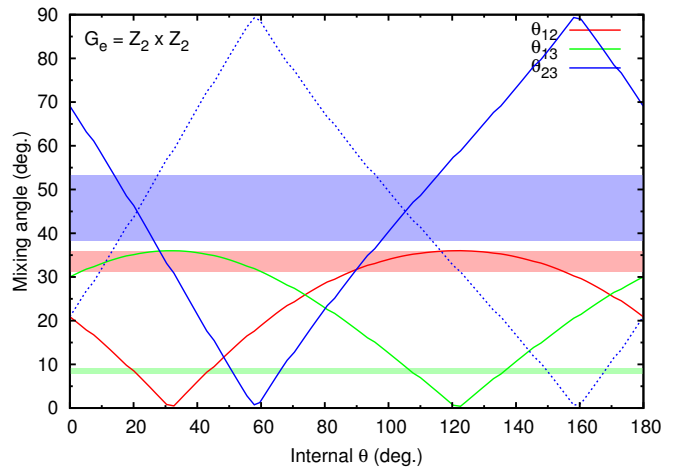}
\caption{\label{fig:K4}Allowed mixing angles for
$G_e=\mathbb{Z}_2\times\mathbb{Z}_2$ as a function of the unphysical parameter
$\theta$. There are two possible sets of predictions of the mixing angles which
have the same $\theta_{12}$ and $\theta_{13}$ predictions but distinct
$\theta_{23}$ predictions (solid and dotted lines) related by the mapping
$\theta_{23}\to \frac{\pi}{2}-\theta_{23}$. All complex phases are CP
conserving for these patterns: $\sin\delta=\sin\alpha_{21}=\sin\alpha_{31}=0$.
The shaded regions show the $3\sigma$ allowed region for the corresponding
mixing angle according to current global data \cite{Gonzalez-Garcia:2014bfa}.}
\end{figure}

The only non-cyclic Abelian subgroup in A$_5$ is the Klein four group
$\mathbb{Z}_2\times\mathbb{Z}_2$. If we take this as the residual symmetry in
the charged lepton sector, we find two patterns of mixing angles which differ
only by their predictions for $\theta_{23}$. Consistent predictions exist for
all of the $3\sigma$ range of $\theta_{13}$, and the predictions for the other
mixing angles can be seen in \figref{fig:K4}.

The first pattern can be derived from the following mixing matrix, 
\begin{equation} U_\text{PMNS} = \frac{1}{2}\left (\begin{matrix}\varphi &
\phig & -1 \\ \phig & 1 & - \varphi \\ -1 & -\varphi & \phig \end{matrix}
\right )\left(\begin{matrix} 1 & 0 & 0 \\ 0 & \cos\theta & \sin\theta \\ 0 &
-\sin\theta & \cos\theta \end{matrix}\right).\label{eq:Klein_matrix_1}
\end{equation}
From this matrix, we find the mixing angles can be expressed by, 
\begin{align*} \sin^2\theta_{12} &= \frac{1+\phig\left[\cos^2\theta +
\sin\left(2\theta\right)\right]}{3-\phig\left[\sin^2\theta -
\sin\left(2\theta\right)\right] },\\ \sin^2\theta_{13} &=
\frac{1+\phig\left[\sin^2\theta - \sin\left(2\theta\right)\right]}{4},\\
\sin^2\theta_{23} &=
\frac{1+\varphi\left[\cos^2\theta-\sin\left(2\theta\right)\right]}{3-\phig\left[\sin^2\theta
- \sin\left(2\theta\right)\right] }.  \end{align*}
For this pattern the complex phases are given by CP conserving values:
$\sin\delta =0$ and $\{\alpha_{21},\alpha_{31}\}\subseteq\{0,\pi\}$. The true
value of $\delta$ can be shown to depend on $\theta$
\[ \delta = \left\{ \begin{matrix} 0 &\quad
31.7^\circ<\theta<58.3^\circ~\text{or}~121.7^\circ<\theta<159.1^\circ, \\ 
\pi &\quad \text{else}. \end{matrix}\right. \]
This dependence on $\theta$ looks complex, but the boundaries of the $\delta=0$
regions can be seen in \figref{fig:K4} to be those values of $\theta$ for which
one mixing angle is either $0^\circ$ or $90^\circ$, and closed form expressions can
be derived for these values from the mixing angle formulae above. For the
matrix shown here, this means that only the $\delta = \pi$ solution is agrees
with the global data. However, when considering all permutations the alternative
CP conserving solution can also be found.  

The prediction for $\theta_{23}$ can be expressed as an atmospheric sum rule to
first order in $r$, 
\[a = \sqrt{\frac{2}{1+\varphi^2}}-1 + \frac{\varphi}{1+\phig^2}r +
\mathcal{O}(r^2). \]
This relation, being derived from a non-cyclic symmetry in the charged-lepton
sector, has to the best of our knowledge not been presented before in general
analyses of atmospheric sum rules \cite{Ballett:2013wya}.

A permutation in the $\theta_{23}$ plane acting from the left of the PMNS
matrix effects a mapping of $\theta_{23} \to \frac{\pi}{2} - \theta_{23}$. For
this reason, both of the expressions defined above have a complementary pattern
with an inverted $\theta_{23}$. These alternative patterns are shown by dashed
lines in \figref{fig:K4}.

\subsection{\label{sec:2and3degrees}Predictions with two and three degrees of freedom}
 
So far we have analysed the cases when $G_e\in\{\mathbb{Z}_3, \mathbb{Z}_5,
\mathbb{Z}_2\times\mathbb{Z}_2\}$ and $G_\nu=\mathbb{Z}_2$. The patterns
resulting from these groups have a single degree of freedom, the unphysical
angle $\theta$, controlling their mixing parameter predictions. There are
however, more general scenarios where the $3$ angles and $3$ phases of the PMNS
matrix are specified by $2$ or $3$ input parameters.

The cases with $2$ degrees of freedom arise from the choice $G_e=\mathbb{Z}_2$
while the neutrino symmetry is enlarged to the full Klein group,
$G_\nu=\mathbb{Z}_2\times\mathbb{Z}_2$. The symmetry of the charged-lepton mass
terms is insufficient to uniquely specify the diagonalising matrix of the mass
matrix, and requires in general a further $2$-dimensional complex rotation,  
\[ R_e(\theta,\gamma) = \left(\begin{matrix} 1 & 0 & 0 \\ 0 & \cos\theta &
\sin\theta e^{\mathrm{i}\gamma} \\  0 & -\sin\theta e^{-\mathrm{i}\gamma} &
\cos\theta \\ \end{matrix}   \right).\]
However, the neutrino symmetry has been enlarged, and so $U_\nu$ is uniquely
specified by the symmetry generators. 
We have scanned over all such combinations and found that no viable patterns
arise from this scenario.

There is one more combination of residual symmetries possible in our
construction: $G_e=\mathbb{Z}_2$ and $G_\nu=\mathbb{Z}_2$. This is an extension
of the previous case, where the charged-lepton symmetry introduces two
parameters but now the neutrino residual symmetry also requires a single real
parameter to diagonalise the most general mass matrix. 
In principle, there is no reason to discount these patterns. They are
consistent with the idea that the full flavour group has broken into residual
subgroups implying correlations on the flavour observables. However, the
increased number of parameters reduces the predictivity of the theory (3
inputs, 6 outputs). We have scanned over such groups and verified that there
are eligible patterns which match all the global data. Some of these patterns
feature non-constant Majorana and Dirac phase predictions, and many are not
simply related to the patterns that we have found in more restrictive schemes.
However, due to the larger parameter space and reduced predictivity, we will
not attempt to present any results of this type. 

\section{\label{sec:pheno}Phenomenological prospects}

In the preceding sections we have derived all patterns of mixing angles and
phases which are possible with an A$_5$ symmetry with generalised CP broken
into residual symmetries.
They depend upon a single real angle, $\theta$, and can all be brought into
agreement with current global data \cite{Gonzalez-Garcia:2014bfa} for a
suitable restriction of its range.
Eliminating the unphysical parameter $\theta$ leads to a set of correlated
predictions between observables which are testable by oscillation experiments
and searches for neutrinoless double-beta decay. In this section, we shall
discuss the prospects for present and future experiments to constrain these
patterns and derive simple versions of the predicted parameter correlations
which may be useful experimentally. We would like to stress that the
correlations identified in this paper will be tested at almost every stage in
the experimental programme of the next few decades. Near term results from T2K
\cite{Abe:2014tzr} and NO$\nu$A \cite{Patterson:2012talk,*NOVA_web} can be
expected on the maximality of $\theta_{23}$ and $\delta$, and in the medium
term, new reactor and long-baseline experiments such as JUNO \cite{Li:2014qca},
RENO-50 \cite{Kim:2014rfa}, DUNE\footnote{The new name for the LBNF/ELBNF
project.}, T2HK \cite{Abe:2014oxa} and possibly ESS$\nu$B
\cite{Baussan:2012cw,*Baussan:2013zcy} should bring us increased precision on
$\theta_{12}$, $\theta_{23}$ and $\delta$.  Finally, for the most stringent
tests of the models in question, the option remains to construct a more
ambitious facility such as the Neutrino Factory
\cite{Geer:1997iz,*DeRujula:1998hd,*Bandyopadhyay:2007kx}. In a complementary
direction, neutrinoless double beta decay experiments will further sensitivity
to this decay, providing evidence on the Majorana nature of neutrinos and, at
least in principle, the first measurements of the values of Majorana phases.
Many of these observations are largely independent and we can expect
significant evidence either in favour of, or ruling out, the patterns
identified in this paper.
 
\subsection{Precision measurements of $\theta_{12}$}

\begin{figure}
\includegraphics[width=0.99\linewidth, clip, trim=50 80 40 65]{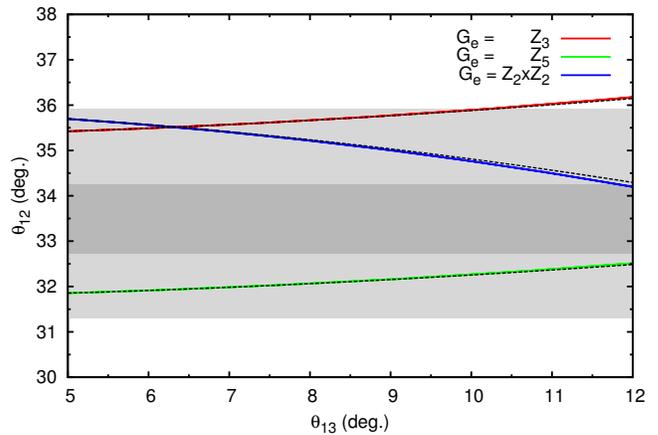}
\caption{\label{fig:th12_precision} Predictions for $\theta_{12}$ as a function
of $\theta_{13}$. All of the patterns of mixing parameters associated with a
given charged-lepton residual symmetry have the same prediction (solid lines).
The dashed line close to each prediction shows the linearised predictions in
\equaref{eq:th12_approx}, \equaref{eq:th12_approx2} and
\equaref{eq:th12_approx3}. The grey regions show the $1$ and $3\sigma$ allowed
regions for $\theta_{12}$ from current global data
\cite{Gonzalez-Garcia:2014bfa}.}
\end{figure}

The viable sets of mixing parameters which we have found above
predict correlations in $\theta_{12}$ and $\theta_{13}$, and therefore very
precise measurements of these angles have the potential to discriminate between
flavour symmetric patterns, or to rule them out entirely~\cite{Ballett:2014uia}.  

The upcoming medium-baseline reactor (MR) neutrino oscillation experiments,
such as JUNO \cite{Li:2014qca} and RENO-50 \cite{Kim:2014rfa}, expect to make
very precise, sub-percent measurements of the oscillation parameter
$\theta_{12}$. 
The precision on $\theta_{13}$, currently dominated by measurements from Daya
Bay \cite{An:2012eh,*An:2013uza} and RENO \cite{Ahn:2012nd}, is not expected to
be significantly improved by the next generation of reactor facilities.
Therefore, the first significant test of the predictions of this paper will
come from increased precision on $\theta_{12}$ independently of $\theta_{13}$.
We have identified 3 distinct predictions for $\theta_{12}$, if we fix
$\theta_{13}$ to its current best-fit \cite{Gonzalez-Garcia:2014bfa} these are
\begin{align*} \theta_{12} = 35.71^\circ,\quad \theta_{12} = 32.11^\circ,\quad
\theta_{12} = 35.14^\circ, \end{align*}
for preserved charged-lepton subgroups $\mathbb{Z}_3$, $\mathbb{Z}_5$ and
$\mathbb{Z}_2\times\mathbb{Z}_2$, respectively. Given that the expected
precision of the MR experiments is at the level of $0.1^\circ$ or around
$0.3\%$ for $\theta_{12}$, a strong discriminatory power exists between the
values of the mixing angles predicted by these correlations. The difference
between the predicted values of all models under consideration is always
greater than $0.26^\circ$ over the current $3\sigma$ interval for
$\theta_{13}$, and in many cases significantly greater. Therefore we can expect
these experiments to identify with considerable confidence if any of the
charged-lepton residual symmetries are consistent with observation.  

In the framework discussed in this article, each model predicts a continuous
correlation between the values of $\theta_{13}$ and $\theta_{12}$. If one of the
predictions above appears to agree with data, it would be desirable to test the
correlation between parameters itself. These correlations can be conveniently
expressed as expansions in the dimensionless parameter $r \equiv
\sqrt{2}\sin\theta_{13}$ \cite{King:2007pr}. The current global best-fits give
$\theta_{13}\approx8.50^\circ$ \cite{Gonzalez-Garcia:2014bfa} which translates
to $r\approx0.2$; the second-order corrections are therefore suppressed by a
factor of $1/25$. 
Expressed in this way, the predictions for $\sin\theta_{12}$ associated with
the charged-lepton subgroups $\mathbb{Z}_3$, $\mathbb{Z}_5$ and
$\mathbb{Z}_2\times\mathbb{Z}_2$ (respectively) can be expanded in the
following relations, 
\begin{align*} \sin\theta_{12} &=
\frac{1}{\sqrt{3}}\left(1+\frac{r^2}{4}\right) + \mathcal{O}\left
(r^4\right),\\ \sin\theta_{12} &=
\frac{1}{\sqrt{1+\varphi^2}}\left(1+\frac{r^2}{4}\right) + \mathcal{O}\left(
r^4\right),\\ \sin\theta_{12} &= \frac{\sqrt{2+\phig}}{2} -
\frac{2-\phig}{\sqrt{2+\phig}}\frac{r^2}{8} + \mathcal{O}\left(r^4\right).
\end{align*}
Expressing these in terms of the angles themselves, we find 
\begin{align} \theta_{12} &= 35.27^\circ + 10.13^\circ\, r^2 +
\mathcal{O}\left(r^4 \right), \label{eq:th12_approx}\\ \theta_{12} &=
31.72^\circ + 8.85^\circ\, r^2 + \mathcal{O}\left(r^4 \right),
\label{eq:th12_approx2}\\ \theta_{12} &= 36.00^\circ - 19.72^\circ\, r^2 +
\mathcal{O}\left(r^4 \right).  \label{eq:th12_approx3} \end{align}
The approximations in
\equassref{eq:th12_approx}{eq:th12_approx2}{eq:th12_approx3} have been plotted
against the unapproximated expressions for $\theta_{12}$ in
\figref{fig:th12_precision}. We see that these relations depend only slightly
on $\theta_{13}$, which first appears at the order $\mathcal{O}(r^2)$, leading
to sub-degree level corrections. 

The formulae above show that the predictions for $\theta_{12}$ only vary by
$0.07^\circ$, $0.06^\circ$ and $0.13^\circ$ (for $\mathbb{Z}_3$, $\mathbb{Z}_5$
and $\mathbb{Z}_2\times\mathbb{Z}_2$, respectively) over the current $3\sigma$
region for $\theta_{13}$. This is of the order of the target precision of the
MR experiments, and it is therefore unlikely that the
$\theta_{12}$--$\theta_{13}$ correlations themselves will be tested at a
significant level even if precision on $\theta_{13}$ were to be greatly
improved. There are no currently planned facilities which could further improve
the precision on $\theta_{12}$. 

\subsection{\label{sec:th23_delta}Maximal-maximal predictions for $\theta_{23}$ and $\delta$}

The current and upcoming generation of long-baseline experiments will be able
to place important constraints on the parameters $\theta_{23}$
and $\delta$. 
Measuring $\theta_{23}$ and $\delta$ independently will provide valuable
information on the viability of flavour symmetric models; however, in the
patterns that we have identified the maximality of $\theta_{23}$ is
significantly correlated with the value of $\delta$. 
In four of these patterns (excluding different Majorana phase assignments), two
from $G_e=\mathbb{Z}_3$ and two from $G_e=\mathbb{Z}_5$, predict a maximal
value of $\theta_{23}$ and a maximal amount of CP violation, 
\[ \theta_{23}=\frac{\pi}{4}\qquad\text{and}\qquad\left|\sin\delta\right|=1.
\]
Therefore, the joint determination of these parameters around these maximal
values would be a particularly interesting measurement from the point of view
of GCP model building.

Testing the maximality of these parameters is an attainable goal for current
and future oscillation experiments. After its full period of data taking, T2K
expects to be able to exclude maximal $\theta_{23}$ at the $90\%$ C.L. for
$\left|\sin^2\left(2\theta_{23}\right)-0.5\right| > 0.05\text{--}0.07$ largely
independently of the value of $\delta$ \cite{Abe:2014tzr}.
Measuring $\delta$ itself is significantly harder; however, the maximal CP
violating values considered here are the most accessible. T2K can expect to be
able to exclude $0\lesssim\delta\lesssim\pi$ ($\pi\lesssim\delta\lesssim2\pi$)
at the $90\%$ C.L. for a true value of $\delta=3\pi/2$ ($\delta=\pi/2$)
\cite{Abe:2014tzr}. This would allow T2K to distinguish between $\delta=\pi/2$
and $\delta=3\pi/2$ if one of them is true at at least the $90\%$ C.L.
NO$\nu$A can also be expected to contribute to this measurement
\cite{Patterson:2012talk,NOVA_web} with a similar power for excluding
$\delta=\pi/2$ and $\delta=3\pi/2$. 
Although these exclusions are expected to be individually statistically weak,
they would constitute valuable information on the validity of the models studied
in this paper and would provide strong encouragement for future work by the
next-generation of oscillation experiments.

\begin{figure}
\includegraphics[angle=0, clip, trim=0 0 120 135, width=0.99\linewidth]{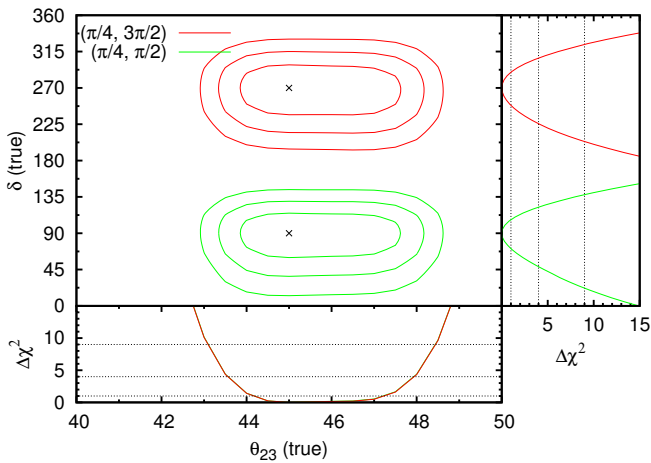}
\caption{\label{fig:LBNF}Red (green) lines show the exclusion regions at $1$,
$2$ and $3\sigma$ for $\theta_{23}=\pi/4$ and $\delta=3\pi/2$ ($\delta=\pi/2$)
expected at LBNF with a $34$~kton LAr detector after $5+5$ years running. In
this regions outside the curves, the two sets of predictions can be
excluded at the given confidence. The side panels show the appropriate
marginalised $\Delta\chi^2$ and the $1$, $2$ and $3\sigma$ confidence levels (1
d.o.f).}
\end{figure}

In the medium term, new long-baseline experiments are expected with
significantly improved sensitivities, in particular to the phase $\delta$,
allowing the maximal-maximal predictions to be further tested.
To estimate the potential for excluding these models with future facilities, we
have run a simulation of LBNE/DUNE using the GLoBES package
\cite{Huber:2004ka,*Huber:2007ji}. Our simulation is based on the detector
responses files and fluxes made available by the LBNE collaboration in
\refref{LBNE_AEDL}. We point out that thanks to its more ambitious design, it
seems likely that the DUNE project can significantly improve the
sensitivity computed here. However, without access to updated experimental
information, making a quantitative assessment of the extent of this improvement
is challenging. We assume a $700$ kW beam operating at $120$~GeV, a detector
based on liquid Argon-TPC technology with a mass of $34$~kton, and overall
systematic errors of $5\%$ for both the signal and background normalizations.
The results of these simulations are shown in \figref{fig:LBNF}, where we
present the regions of true parameter space for which the combinations of
$(\theta_{23}, \delta)=(\pi/4,\pi/2)$ and $(\pi/4,3\pi/2)$ can be excluded
after $5$ years neutrino and $5$ years antineutrino running. We find that these
patterns can be excluded at $3\sigma$ if the true value of $\theta_{23}$
satisfies $\theta_{23}\lesssim43.0^\circ$ or $\theta_{23}\gtrsim48.3^\circ$, or
if the true value of $\delta$ is outside the intervals
${90^\circ}^{+48^\circ}_{-69^\circ}$ or ${270^\circ}^{+53^\circ}_{-67^\circ}$.
Here we see the importance of observing both $\theta_{23}$ and $\delta$ for
excluding our models. A measurement of $\theta_{23}$ alone will not be able to
distinguish between the models which predict $\theta_{23}$-maximality and the
model from $\mathbb{Z}_2\times\mathbb{Z}_2$ which predicts values of
$\theta_{23}$ which differ from maximality by only around $2^\circ$. However,
these models have maximally distinct predictions for $\delta$, and as we have
shown, the measurement of $\delta$ alone would be able to separate these cases
at $3\sigma$.

\subsection{Dirac CP conserving patterns and precision measurements of $\theta_{23}$}

Those patterns which instead make non-maximal predictions of $\theta_{23}$ also
predict CP conserving values of $\delta$, such that
$\left|\cos\delta\right|=1$, and we can expect constraints to be placed on
these models by the attempts to discover leptonic CP violation --- a standard
search for the next-generation of CP-sensitive oscillation experiments
\cite{Agarwalla:2013kaa, Agarwalla:2014tca, Agarwalla:2014ura, Adams:2013qkq,
Abe:2014oxa,Baussan:2012cw,Baussan:2013zcy}. It has been shown that LBNO
running with a beam derived from the SPS accelerator at CERN could rule out
leptonic CP conservation at $3\sigma$ for around $45\%$ ($65\%$) of the
parameter space for a detector mass of $20$~kton ($70$~kton). This could be
increased to $70\%$ ($80\%$) with an upgraded beam power
\cite{Agarwalla:2014tca}.  
LBNE has predicted a similar sensitivity to CP violation \cite{Adams:2013qkq},
with the ultimate reach also depending crucially on the planned series of
upgrades to detector mass and beam power. With a $10$~kton detector and 6 years
of data using a 1.2 MW beam, the measurement could be made for $33\%$ of the
parameter space at $3\sigma$. This rises to $40\%$ of the parameter space at
$5\sigma$ once the detector mass has been increased to $34$~kton and 6 years
more data has been collected. Finally, a beam power upgrade to $2.3$ MW could
increase this to $60\%$ of the parameter space at $5\sigma$
\cite{Adams:2013qkq}.
The T2HK and ESS$\nu$B proposals also show strong sensitivity to CP violation,
both using a megaton-scale water \v{C}erenkov detector and MW power beams. T2HK
has shown that it can expect a discovery of CP violation over $76\%$ ($58\%$)
of the parameter space at $3\sigma$ ($5\sigma$) \cite{Abe:2015qaa}. A similar reach is possible
with ESS$\nu$B, which expects a $3\sigma$ ($5\sigma$) discovery of CP violation
after 10 years of data-taking over $74\%$ ($50\%$) of the parameter space
\cite{Baussan:2013zcy}.

\begin{figure}
\includegraphics[width=0.99\linewidth, clip, trim=50 80 40 65]{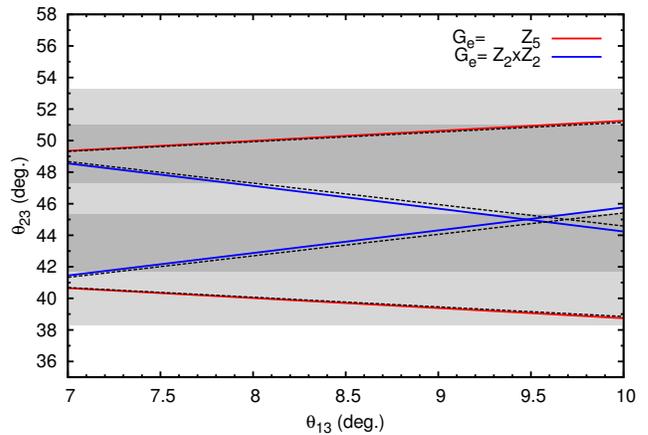}
\caption{\label{fig:th23_precision} $\theta_{23}$ as a function of
$\theta_{13}$ for the patterns which predict CP conservation (solid lines). The
dashed line close to each solid line shows the linearised expression in
Eqs.(\ref{eq:th23_approx})~and~(\ref{eq:th23_approx2}). The grey regions show
the $1$ and $3\sigma$ allowed regions from current global data \cite{Gonzalez-Garcia:2014bfa}.}
\end{figure}

If the current experimental programme fails to discover CP violation in the
leptonic sector and $\sin\delta$ is discovered to be small, there are four
distinct patterns from our model which would remain in agreement with the data.
These models can be tested by the increased precision on measurements of
$\theta_{23}$ expected from next generation long-baseline facilities.
In all cases of this kind, we find predictions coming in pairs. The model
associated with a $\mathbb{Z}_5$ residual symmetry predicts
\begin{equation}\theta_{23} = 45^\circ \pm 25.04^\circ\,r + \mathcal{O}(r^2),
\label{eq:th23_approx}\end{equation}
while the predictions for a residual $\mathbb{Z}_2\times\mathbb{Z}_2$ symmetry
are given by
\begin{equation} \begin{matrix}\theta_{23} = 31.72^\circ + 55.76^\circ\,r +
\mathcal{O}(r^2),\\ \theta_{23} = 58.28^\circ - 55.76^\circ\,r +
\mathcal{O}(r^2).\end{matrix} \label{eq:th23_approx2} \end{equation}
The pairs of sum rules given by \equaref{eq:th23_approx} or
\equaref{eq:th23_approx2} are related by the octant degeneracy, $\theta_{23}
\to \frac{\pi}{2} - \theta_{23}$, as can be seen clearly in
\figref{fig:th23_precision} where the approximations above are shown against
the full predictions.  We note that in contrast to those for
$\theta_{12}$, these relations depend on $r$ at linear order. Therefore, they
are far more sensitive to the precise correlation between parameters, and the
measurement of the correlation itself becomes more accessible.

The first discriminating factor between these solutions will come from improved
precision on $\theta_{23}$ and the resolution of the octant degeneracy.  In
these models we predict non-maximal mixing, and a successful determination of
the octant would provide early evidence in their favour. This would be most
challenging for the model based on $\mathbb{Z}_2\times\mathbb{Z}_2$ for which
$\theta_{23}$ differs from $45^\circ$ by between
$2.5^\circ$ and $0.8^\circ$ over the current $3\sigma$ range of $\theta_{13}$.
The model based on $\mathbb{Z}_5$ instead predicts greater deviations from
maximal atmospheric mixing, ranging between $4.8^\circ$ and $5.6^\circ$ over
the same interval.  
Studies of the potential for the current generation of oscillation experiments,
of which T2K and NO$\nu$A play the most important role, suggest that the octant
can be established at $3\sigma$ ($2\sigma$) for deviations from maximality
greater than around $6^\circ$ ($4^\circ$)
\cite{Chatterjee:2013qus,Agarwalla:2013ju}. This precludes the current
generation from separating between the two predictions of
$\mathbb{Z}_2\times\mathbb{Z}_2$, but would allow for $2\sigma$ evidence for
those predictions coming from our model based on $\mathbb{Z}_5$.
This discovery potential will be improved by the next generation of oscillation
experiments. In \refref{Adams:2013qkq} it is shown that with an exposure of
$60$~kton-years, LBNE could determine the octant at $3\sigma$ if the true value
of $\theta_{23}$ deviates from maximality by more than $4^\circ$--$5^\circ$.
However, the best bounds could come from T2HK by studying atmospheric neutrino
data. A $3\sigma$ determination of the octant is expected to be possible after
10 years of data-taking for true values $\left|\sin^2\theta_{23}-0.5\right| >
0.04\text{--}0.06$ corresponding to deviations between $2^\circ$--$3^\circ$
\cite{Abe:2014oxa}. Although exclusion of the $\mathbb{Z}_2\times\mathbb{Z}_2$
pattern would be unlikely, the two predictions from $\mathbb{Z}_5$ would be
distinguishable.

To go beyond the octant measurement, higher precision will be necessary to
separate between the $\mathbb{Z}_5$ and $\mathbb{Z}_2\times\mathbb{Z}_2$
predictions, or indeed to test their specific correlations with $\theta_{13}$.
The difference between these two predictions varies from between $2.4^\circ$ to
$4.8^\circ$ over the current allowed regions. Therefore degree-level precision
will be required to distinguish between them, even in the presence of greatly
improved knowledge of $\theta_{13}$. 
In \refref{Adams:2013qkq} it is shown that the minimal 10~kton LBNE
configuration running for 6 years would have a precision of around $1^\circ$ at
$1\sigma$ for true values of $\theta_{23}$ around $51^\circ$, which increases
as we approach $\theta_{23}$ maximality to a $1\sigma$ width of around
$2.5^\circ$. Similarly, T2HK shows that around the point expected to give the
worst sensitivity to $\theta_{23}$, the $90\%$ C.L. width is around
$2^\circ$--$3^\circ$ \cite{Abe:2014oxa}.
These results suggest that a significant discrimination between these models
would be challenging with these set-ups; however, evidence in favour of these
models would be possible at low significance and if observed in conjunction
with an absence of observable CP violation, this would present a concrete
hypothesis for future work.  

\subsection{Long-term prospects}

We have seen in the previous sections that although the next generation of
superbeam and reactor experiments will be able to test the consistency of
the patterns that we have identified, much of their discriminatory power relies
on excluding maximal angles and phases. Testing the continuous correlations
predicted in our models, for example between $\theta_{12}$ and $\theta_{13}$ or
between $\theta_{23}$ and $\theta_{13}$, would require higher precision. 

The only proposed experiment capable of pushing the precision frontier beyond
the results of the next-generation superbeams is the Neutrino Factory (NF)
\cite{Huber:2014nga}, which produces a beam with low systematic uncertainties
from the decay of stored muons
\cite{Geer:1997iz,*DeRujula:1998hd,*Bandyopadhyay:2007kx}. 
A NF would be able to improve our knowledge of the mixing parameters in a
number of ways, but for the present purposes it serves two main roles. Firstly,
such a facility would greatly increase the precision on $\delta$, with an
ultimate $1\sigma$ precision estimated at around $5^\circ$ \cite{Coloma:2012wq,
Coloma:2012ji}. This could allow many of our models to be excluded
independently of their other parameter correlations. 
\begin{figure}[t] \includegraphics[width=0.99\linewidth, clip, trim=50 80 40
65]{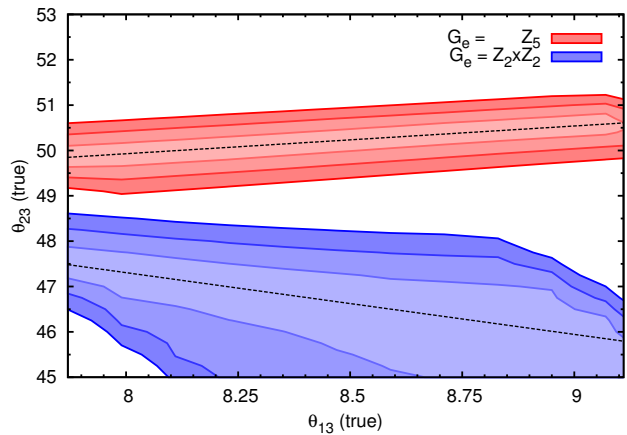} \caption{\label{fig:NF} The regions of true parameter
space for which the upper-octant relations in
\equasref{eq:th23_approx}{eq:th23_approx2} can be excluded at $1$, $2$ and
$3\sigma$ by a $2000$~km, $10$~GeV LENF using a MIND detector. The two regions
start to overlap at low values of $\theta_{13}$ at the $5\sigma$ confidence
level.} \end{figure}
Secondly, a NF would provide a high-precision determination of $\theta_{23}$
\cite{Coloma:2012wq}, allowing it to perform a very stringent test of the
$\theta_{23}$--$\theta_{13}$ correlations discussed in the previous subsection.
To quantify this possibility, we have performed a simulation of a
representative Low-Energy Neutrino Factory
\cite{Bross:2007ts,FernandezMartinez:2010zza} to compute the regions of true
parameter space which would allow two candidate models to be (individually and
collectively) excluded. Our simulations assume a $2000$~km baseline and a
stored-muon energy of $10$~GeV which is close to the optimal configuration for
CP violation discovery
\cite{Huber:2006wb,Agarwalla:2010hk,Choubey:2011zzq,Ballett:2012rz}. For our
detector, we take a Magnetized Iron Neutrino Detector (MIND)
\cite{Michael:2008bc,Bayes:2012ex}, assumed to be in a toroidal magnetic field
allowing for muon charge identification, with a fiducial mass of 100~kton. The
detector response is described by a set of migration matrices provided by
\refref{Bayes:PrivateComm}. Backgrounds to the appearance channel signal come
from both charge- and flavour-misidentified events, as well as the secondary
decay products arising from $\tau^\pm$ decays in the detector. An overall $1\%$
($10\%$) uncertainty is taken on the signal (background).
Our results are shown in \figref{fig:NF}, where the blue (red) shaded regions
show the area of true parameter space for which the
$\mathbb{Z}_2\times\mathbb{Z}_2$ ($\mathbb{Z}_5$) correlation would not be able
to be excluded. We see that the two coloured regions do not overlap, and
therefore all points in this parameter space allow for the exclusion of at
least one of the correlations at $3\sigma$ or higher significance.

\subsection{Neutrinoless double beta decay}

Over the next decade, the new generation of neutrinoless double beta 
($0\nu\beta\beta$) decay experiments will significantly increase the sensitivity to
this rare process. For the first time these experiments will probe the region of
parameter space associated with the inverse hierarchical spectrum.  These
experiments aim to establish that neutrinos are Majorana in nature, but can
also provide valuable information on the neutrino mass spectrum and in
principle, measure the Majorana phases themselves. 
 
\begin{figure*}[t]
\centering
\includegraphics[width=0.47\textwidth]{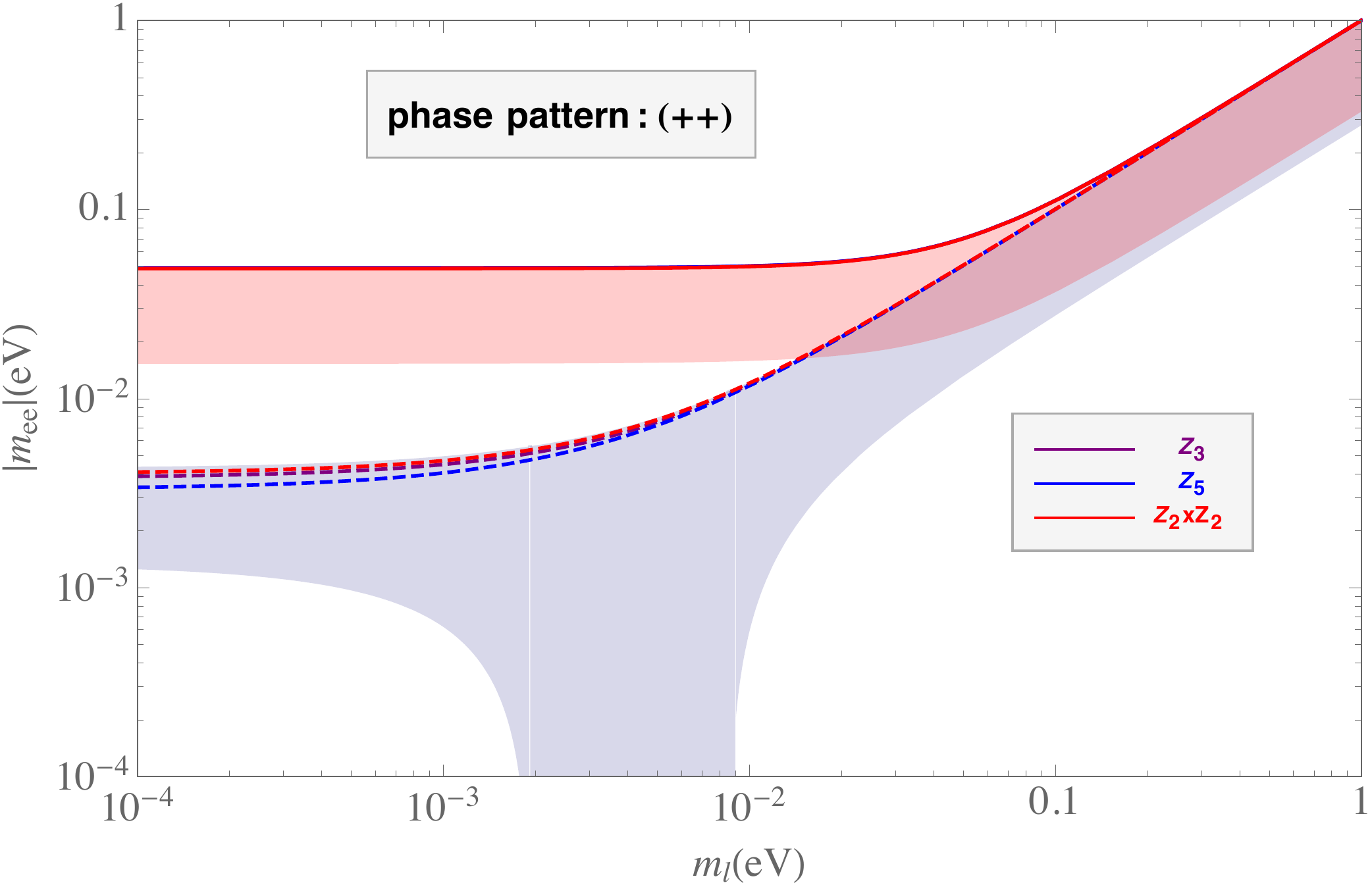}\hspace{0.2cm}% 
\includegraphics[width=0.47\textwidth]{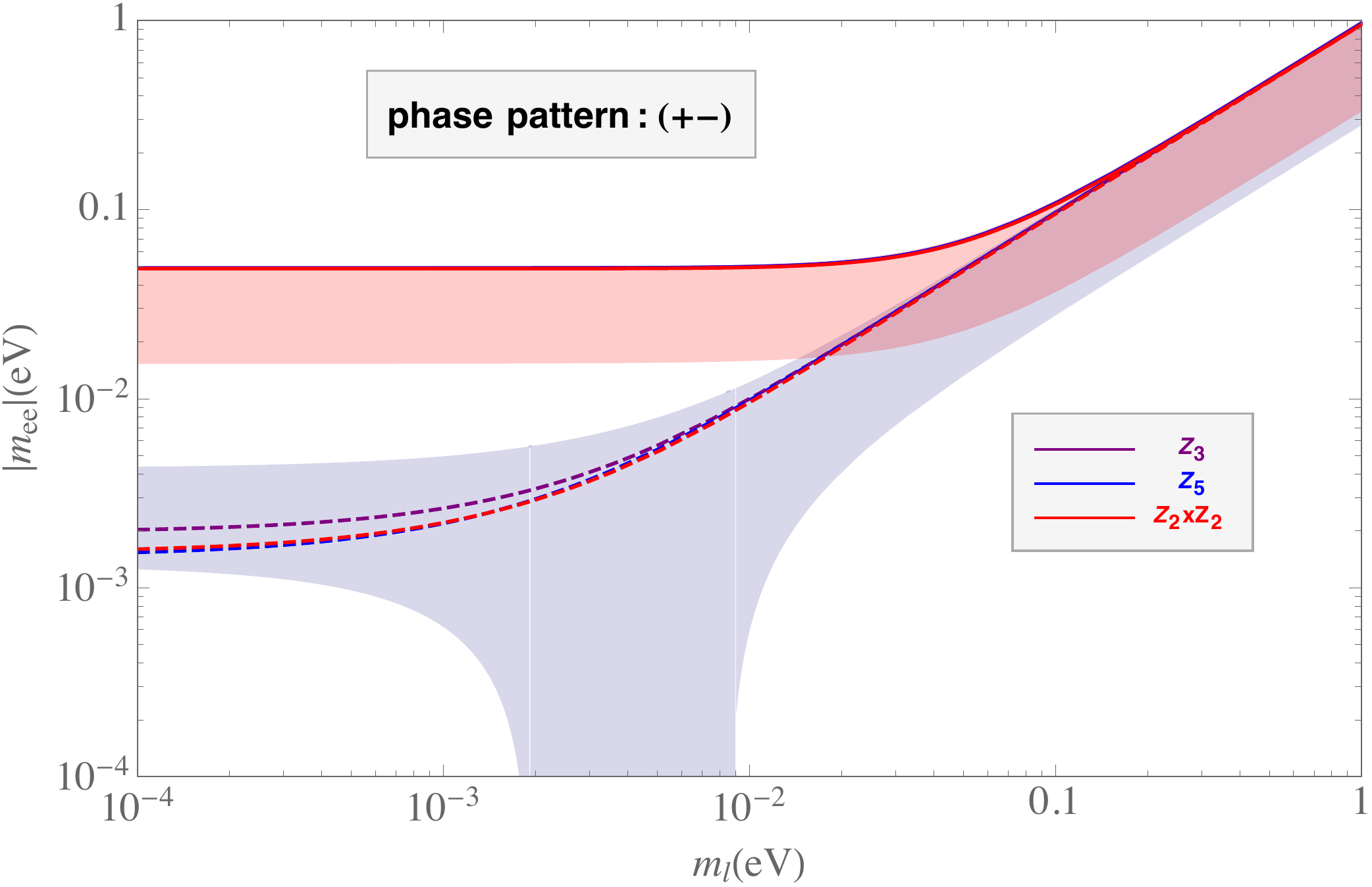}\\
\includegraphics[width=0.47\textwidth]{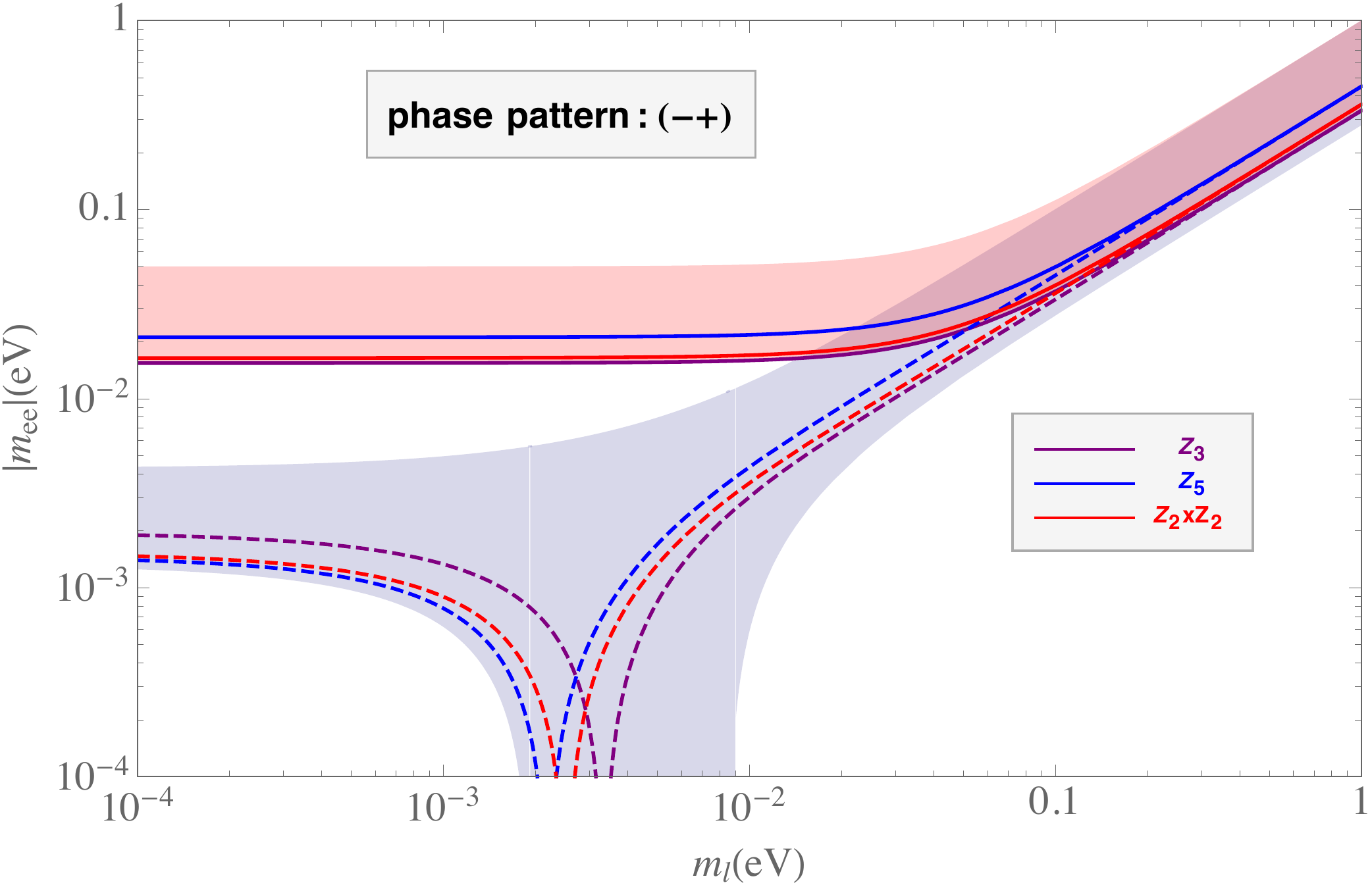}\hspace{0.2cm}% 
\includegraphics[width=0.47\textwidth]{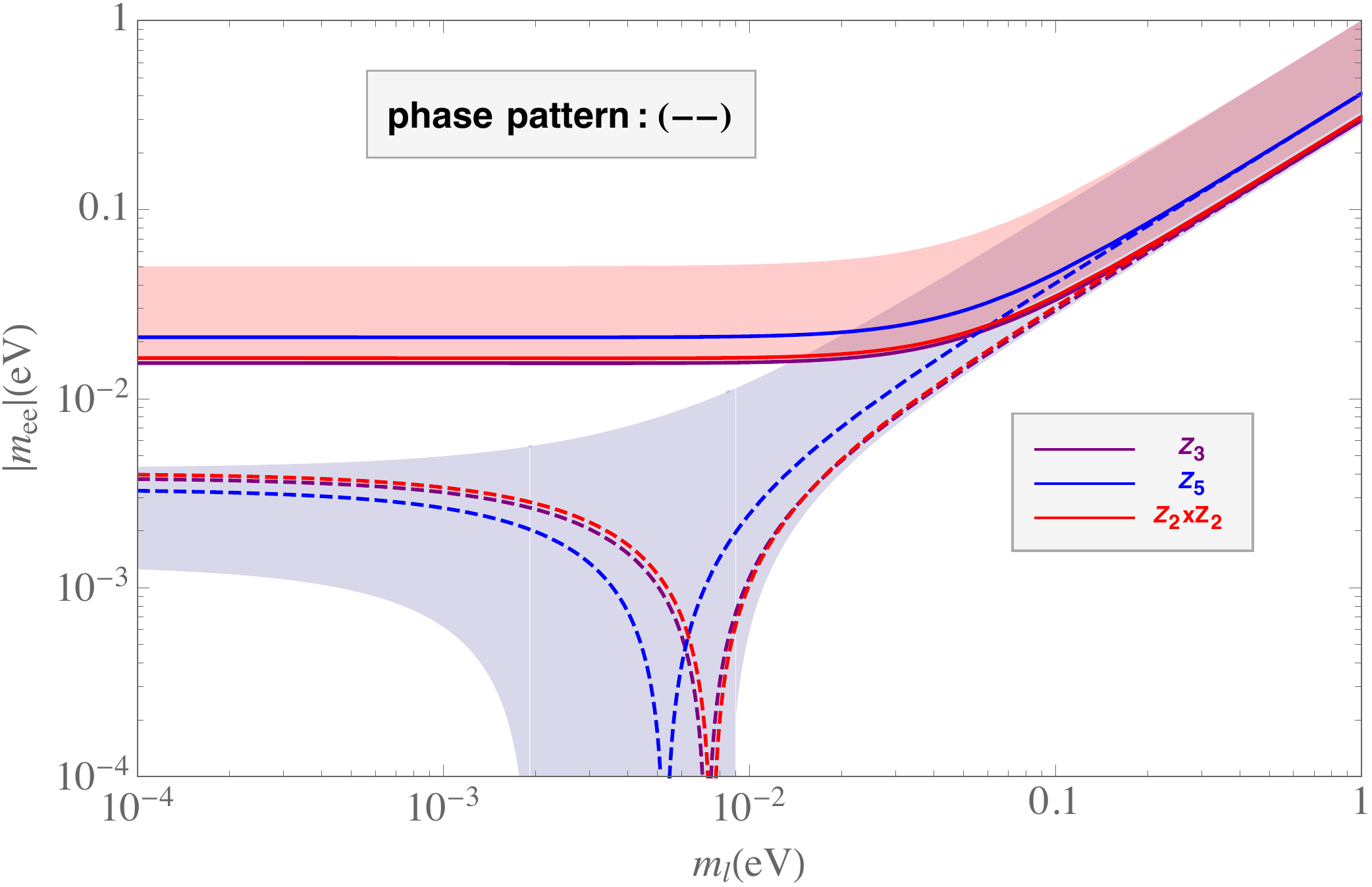}

\caption{\label{fig:NDBD} $\lvert m_{ee}\rvert$ versus the lightest neutrino
mass for the IO (NO) for the solid (dashed) lines. The predictions in a given
panel all have the same phase assignment, shown in the top left of the plot.
The red (blue) shaded region shows the most general predictions for $\left|m_{ee}\right|$
with IO (NO) obtained by varying the oscillation parameters over their current
$3\sigma$ global ranges \cite{Gonzalez-Garcia:2014bfa}. }

\end{figure*}

The $0\nu\beta\beta$ decay rate is proportional to the effective Majorana mass
$\left|m_{ee}\right|$ (see \eg\
\refref{Petcov:1993rk,*Vissani:1999tu,*Pascoli:2003ke,
*Pascoli:2005zb,*Agashe:2014kda,*Choubey:2005rq,*Simkovic:2010ka,*Dell'Oro:2014yca}),
which is given by
\begin{align}\lvert m_{ee} \vert &=\left\lvert \sum_{k=1}^{3}
U_{ek}^2{m_k}\right\rvert,\nonumber\\ =&\Big |
m_1\cos^2\theta_{12}\cos^2\theta_{13} +
m_2\sin^2\theta_{12}\cos^2\theta_{13}e^{i{\alpha_{21}}}\nonumber\\
&\left.+m_3\sin^2\theta_{13}e^{i\left(\alpha_{31}-2\delta\right)}\right\rvert,\label{eq:mee}\end{align}
where $\alpha_{21}$ and  $\alpha_{31}$ are Majorana phases and $\delta$ is the
Dirac  phase. 
The predicted values of $\left|m_{ee}\right|$ depend crucially on the neutrino
masses. The latter can be ordered in two ways: normal ordering (NO;
$m_1<m_2<m_3$) or inverted ordering (IO; $m_3<m_1<m_2$).
As the parameters $\Delta m^2_{21}$ and $\left|\Delta m^2_{31}\right|$ are
known from oscillation physics, there is a single degree of freedom remaining
amongst the masses. This is typically taken to be the lightest neutrino mass,
$m_1$ ($m_3$) for NO (IO), which we will denote in both cases by $m_l$
\cite{Petcov:1993rk}.
The parameter space available to $\left|m_{ee}\right|$ can be further divided
into three particularly interesting regions based on the true value of $m_l$.
The first is for \emph{quasi-degenerate} masses (QD) where $m_l\gtrsim 0.1$ eV,
in which the splitting between masses is a small correction to approximately
degenerate values. For smaller values of $m_{l}$ there are two parameter
regions: one for \emph{normal hierarchical} masses (NH; $m_{1}<m_{2}\ll m_{3}$)
and the other for \emph{inverted hierarchical} masses (IH; $m_{3}\ll
m_{1}<m_{2}$).  
In order to better understand the predictions of our models, we have first
computed the predicted values of $\left|m_{ee}\right|$ in the generic case,
assuming only that the mixing parameters lie in their current $3\sigma$ allowed
ranges \cite{Gonzalez-Garcia:2014bfa}. These predictions for NO (IO) are shown
as the blue (red) region in \figref{fig:NDBD}. For quasi-degenerate and IH
spectra, there exist lower bounds on $\left|m_{ee}\right|$
\cite{Bilenky:2001rz}: 
for IH $\left|m_{ee}\right|\gtrsim0.015$ eV, and for QD
$\left|m_{ee}\right|\gtrsim 0.03$--$0.04$ eV.  For NO there is no non-zero
lower bound as $\left|m_{ee}\right|$ can vanish due to a cancellation between
terms in \equaref{eq:mee}.

There are many experiments that are searching for $0\nu\beta\beta$ decay or are
in various stages of planning and construction. The most recent experiments
that have set upper limits on the effective Majorana mass are CUORICINO, GERDA,
EXO--$200$ and KamLAND-Zen. CUORICINO, based in Gran Sasso National
Laboratories in Italy, was an experiment used to test the feasibility of its
successor, CUORE. Using data taken from $2003$--$2004$ they achieved a bound on
$\left|m_{ee}\right|$ of $200$--$1100$ meV \cite{Capelli:2005jf}, where as with
all $0\nu\beta\beta$ experiments a key uncertainty on their limits comes from
the nuclear matrix element. GERDA (The Germanium Detection Array) is also
located in the Gran Sasso Laboratory LNGS in Italy \cite{BRUGNERA:2014ava}.
During phase I of their data taking period they acquired sufficient data to
attain an upper limit of $\left|m_{ee}\right| \lesssim 200$--$400$ meV.  They
intend to increase their sensitivity by a factor of approximately ten during
GERDA Phase II. EXO (Enriched Xenon Observatory)--$200$, located in Carlsbad,
New Mexico, has placed  upper bounds for $\left|m_{ee}\right|$ of $69$--$163$
meV \cite{Albert:2014awa}.  KamLAND-Zen (Kamioka Liquid Scintillator
Anti-Neutrino Detector-Zero neutrino double beta decay search) has had two
phases of data acquisition and using the combined data they found an upper
limit of $\left|m_{ee}\right| \lesssim 140$--$280$ meV
\cite{TheKamLAND-Zen:2014lma}.  Both EXO--$200$ and KamLAND-Zen have ambitious
long term plans to upgrade their experiments in order  to explore  the inverted
hierarchical region of parameter space.  EXO--$200$ intends to upgrade to nEXO
(next Enriched Xenon Observatory) which, with ten years of data taking, is
expected to cover the full IO region \cite{Albert:2014afa}.  In its next phase,
KamLAND-Zen aims to increase its sensitivity to around $50$ meV after
approximately two years running time \cite{TheKamLAND-Zen:2014lma}. As this
limit will only begin to probe the IH region, KamLAND-Zen has proposed
KamLAND2-Zen.  This upgraded detector (with a running time of 5 years) has a
target sensitivity of $\left|m_{ee}\right| \simeq$ $20$ meV and will allow the
exploration of the majority of the IO region and all of the QD parameter space.
In addition to EXO--$200$ and KamLAND-Zen, other future $0\nu\beta\beta$ decay
experiments are CUORE, SNO+ and NEXT. CUORE plans to start taking data in 2015.
Over the course of five years, they hope to reach a sensitivity of $50$--$120$
meV \cite{Aguirre:2014lua}.  SNO$+$, a multi-purpose experiment located in
Sudbury, Canada, aim to achieve a similar upper bound on $\left|m_{ee}\right|$.
After two years of data taking they expect to be able to set the upper bound
$\left|m_{ee}\right|<100$ meV \cite{Sibley:2014nda}. The NEXT (Neutrino
Experiment with Xenon TPC) experiment, based at CanFranc Underground laboratory
(LSC), will commence data taking in 2018 using the NEXT-100 detector. Despite
their late start compared with that of EXO--$200$ and KamLAND-Zen, they intend
to achieve a $\left|m_{ee}\right|$ sensitivity of approximately $100$ meV by
2020 \cite{DavidLorcafortheNEXT:2014fga}. The next stage will be the
development of BEXT which proposes to fully cover the predicted values of
$\left|m_{ee}\right|$ for IO \cite{Gomez-Cadenas:2014dxa}.  Although they are
not discussed here, there are other experiments which aim to improve the
current bounds on $\left|m_{ee}\right|$; for example, COBRA
\cite{Fritts:2013mwa}, the Majorana Demonstrator \cite{Xu:2015dfa}, SuperNEMO
\cite{Nova:2013ata} and the DCBA experiment \cite{Ishihara:2012pwa}, amongst
others.

The predicted values of $\lvert m_{ee}\rvert$ for the case of A$_5$ with GCP
can be calculated from the leptonic mixing matrices of \equaref{eq:Z3_matrix},
\equaref{eq:Z5_matrix_1} and \equaref{eq:Klein_matrix_1} for both IO and NO. 
The complex phases only influence $\left|m_{ee}\right|$ through the
combinations $e^{\mathrm{i}\alpha_{21}}$ and $e^{\mathrm{i}\left(\alpha_{31} -
2\delta\right)}$, and we will denote the phases of our predictions by an
ordered pair of $\pm$ signs \eg\ $(+-)$ when $\alpha_{21}=0$ and
$\alpha_{21}-2\delta = \pi$.
As $\lvert m_{ee}\rvert$ does not depend on $\theta_{23}$, 
patterns which only differ by this angle will be degenerate and each preserved
charged-lepton subgroup leads to a single prediction for each mass ordering and
phase assignment.
\figref{fig:NDBD} shows the predicted values from the mixing patterns in this
paper for each charged-lepton residual symmetry $G_e$. In these plots, we have
neglected a small width to each line which comes from varying $\theta_{13}$ and
the neutrino mass-squared splittings over their allowed ranges, instead fixing
these at their best-fit values from \refref{Gonzalez-Garcia:2014bfa}. 

We focus first on the IO spectra. The phase assignments $\left(++\right)$ and
$\left(+-\right)$, shown on the top row of \figref{fig:NDBD}, predict large
values of $\lvert m_{ee}\rvert$, close to the upper boundary of the IO region
obtained using the 3$\sigma$ global data. These predictions are very similar
for all models. This can be understood as the term in $\left|m_{ee}\right|$
proportional to $m_3$ only has a subdominant effect: it is not only multiplied
by the small number $\sin^2\theta_{13}$, but is further suppressed for IH by
the small value of $m_3$ itself. If we neglect this term, the resulting
approximation at leading-order is independent of $\theta_{12}$ up to
corrections of the order $\mathcal{O}\left(\Delta m^2_{21}/\Delta
m^2_{31}\right)$.
It is feasible that experiments such as CUORE and KamLAND-Zen, and to a much
greater extent BEXT and nEXO will be able to explore this topmost region of the
IH parameter space and test these predictions. Further distinguishing between
them will be beyond their scope due to the small predicted differences and
substantial experimental and theoretical uncertainties on
$\left|m_{ee}\right|$.
For the phase assignments $\left(-+\right)$ and $\left(--\right)$, shown on the
bottom row of \figref{fig:NDBD}, we see values of $\left|m_{ee}\right|$ that
are further suppressed and which exhibit more model dependence. Once again, the
suppression of the $m_3$ term explains the similarity between the two phase
assignments. The lower values compared to the $\left(++\right)$ case arise from
the relative phase difference between the $m_1$ and $m_2$ terms: at leading
order $\left|m_{ee}\right| = \sqrt{\left|\Delta
m^2_{31}\right|}\cos^2\theta_{13}\cos^2\left(2\theta_{12}\right)$ \cite{Bilenky:1996cb,Bilenky:2001rz}. 
This effect is evident if we compare $\mathbb{Z}_5$ and
$\mathbb{Z}_3$ cases: the larger $\theta_{12}$ value of $\mathbb{Z}_3$ accounts
for the more pronounced cancellation and therefore lower $\lvert m_{ee}\rvert$
than that of $\mathbb{Z}_5$. These predictions are beyond the reach of many of
the facilities discussed so far; although KamLAND2-Zen, aims to set limits near
the predictions for $\mathbb{Z}_5$ and, if capable of testing the full IO
region, should be accessible to nEXO and BEXT.
Although lying in a region of parameter space that is harder to explore, the
greater model dependence for these phase assignments would make it easier to
distinguish between models than with the $\left(++\right)$ and
$\left(+-\right)$ cases. There exists a separation of around $5$~meV between
the predictions for $\mathbb{Z}_5$ and the other subgroups; however, it is
unlikely such a resolution on $\lvert m_{ee}\rvert$ would be attainable in the
foreseeable future. 

For NO, we see quite different behaviour. In the quasi-degenerate region, the
mass-squared splittings are negligible and the predictions for the IO and NO
cases effectively coincide. However, in the limit of vanishing $m_l$ the
situation is very different. In this limit it is the relative phase between the
$m_2$ and $m_3$ terms which dominates the magnitude of $\left|m_{ee}\right|$,
which leads to larger predictions for the phase assignments $\left(++\right)$
and $\left(--\right)$, while suppressing the predictions of $\left(-+\right)$
and $\left(+-\right)$. 
Although exploring the NH region experimentally is beyond the scope of any
planned experiment, if $0\nu\beta\beta$ decays are not observed and oscillation
physics establishes that the neutrino masses are NO, it would be of paramount
importance to try and test $\left|m_{ee}\right|$ values in the NH region. 
Due to the rich interplay between relative phases, these models make quite
different predictions across this parameter space. In fact, all mixing angle
patterns discussed in this paper could accommodate a value of
$\left|m_{ee}\right|$ near the top of the current NH region allowed by global
data. Although such an observation would add further support
to any prediction of this paper which was still consistent with experimental
data, to further discriminate between these models it would be necessary to
provide complementary information on the absolute mass scale.

\begin{table}[t]

\begin{tabular}{ c | c | c | c |c   }
$G_e$ & $\theta_{12} $ & $\theta_{23}$ & $\sin\alpha_{ji}$& $\delta$   \\
\hline\hline

\multirow{2}{*}{$\mathbb{Z}_3$} & \multirow{2}{*}{$35.27^\circ + 10.13^\circ\, r^2$} & \multirow{2}{*}{$45^\circ$} &  \multirow{2}{*}{$0$}& $90^\circ$ \\
\cline{5-5}
 &  &  & & $270^\circ$ \\
\hline
\multirow{4}{*}{$\mathbb{Z}_5$} & \multirow{4}{*}{$31.72^\circ + 8.85^\circ\,r^2$} & \multirow{2}{*}{$45^\circ \pm 25.04^\circ\,r$} & \multirow{2}{*}{$0$} &$0^\circ$ \\
\cline{5-5}
 &  &  & & $180^\circ$ \\
\cline{3-5}
&  & \multirow{2}{*}{$45^\circ$} & \multirow{2}{*}{$0$} & $90^\circ$ \\
\cline{5-5}
 &  &  & & $270^\circ$  \\
\hline
\multirow{4}{*}{$\mathbb{Z}_2\times\mathbb{Z}_2$} & \multirow{4}{*}{$36.00^\circ - 34.78^\circ\,r^2$} & \multirow{2}{*}{$31.72^\circ + 55.76^\circ\,r$} & \multirow{2}{*}{$0$}&$0^\circ$ \\
\cline{5-5}
 &  &  & & $180^\circ$  \\
\cline{3-5}
&  & \multirow{2}{*}{$58.28^\circ - 55.76^\circ\,r$} & \multirow{2}{*}{$0$} & $0^\circ$ \\
\cline{5-5}
 &  &  & & $180^\circ$ 
\end{tabular}
\caption{\label{tab:predictions}Numerical predictions for the correlations
found in this paper. The dimensionless parameter
$r\equiv\sqrt{2}\sin\theta_{13}$ is constrained by global data to lie in the
interval $0.19 \lesssim r \lesssim 0.22 $ at $3\sigma$. The predictions for
$\theta_{12}$ and $\theta_{23}$ shown here neglect terms of order
$\mathcal{O}\!\left(r^4\right)$ and $\mathcal{O}\!\left(r^2\right)$, respectively.
Following the method of this paper, the Majorana phases can only be predicted 
modulo $\pi$ and the values in the fourth column hold for all phases.
}
\end{table}

\section{\label{sec:conclusions}Conclusions}

Assessing the viability of flavour symmetric models of the leptonic sector is
an accessible target for precision measurements from present and future
neutrino oscillation experiments. In this article, we have presented a detailed
analysis of a particular theoretical scenario: the flavour symmetry A$_5$ with
a generalised CP symmetry breaking into residual subgroups at low energies. 
We have identified the most general form of the generalised CP transformation,
and studied the full group for consistent residual symmetries. Our analysis
results in $6$ distinct sets of mixing angle predictions each with an
additional $8$ possible combinations of phases which are shown in
\tabref{tab:predictions}. These depend at most on a single real parameter, and
predict testable correlations between certain parameters.  
In addition, the Majorana phases for all of our predictions are CP conserving.
These patterns can be classified by the residual symmetry in the charged-lepton
mass terms: $\mathbb{Z}_3$, $\mathbb{Z}_5$ and
$\mathbb{Z}_2\times\mathbb{Z}_2$.  A symmetry of $\mathbb{Z}_3$ predicts
maximal $\theta_{23}$, maximal CP violation from $\delta$ and a value of
$\theta_{12}$ that lies close to the upper boundary of the 3$\sigma$ global fit
data. There are two distinct patterns which arise from a preserved
$\mathbb{Z}_5$ residual symmetry. These share a common $\theta_{12}$ prediction
which lies close to the lower boundary of the 3$\sigma$ global fit data;
however, one prediction has maximal $\theta_{23}$ and a maximally CP violating
value of $\delta$ whilst the other has non-maximal $\theta_{23}$ and CP
conserving values of $\delta$.  The patterns arising from a preserved subgroup
$\mathbb{Z}_2\times\mathbb{Z}_2$ also share a common $\theta_{12}$ which lies
above the current $1\sigma$ region. In this case both $\theta_{23}$ predictions
are non-maximal and the value of $\delta$ is CP-conserving. 

We have then discussed the phenomenology of our predictions, focusing on the
role which current and future reactor, superbeam and neutrinoless double beta
decay experiments can play.
The predictions for $\theta_{12}$ should be testable at high significance by
the next generation of reactor neutrino experiments, such a JUNO and RENO-50.
These experiments can be expected to distinguish between the different models;
however, testing the precise correlations between $\theta_{12}$ and
$\theta_{13}$ will most probably remain beyond the reach of any foreseen
experiment.
A particularly interesting feature of the patterns found in this paper is the
correlated maximality of $\theta_{23}$ and $\delta$, and also non-maximal
$\theta_{23}$ and CP conserving values of $\delta$. Testing these correlations
is a feasible goal for current and future superbeam experiments. T2K and
NO$\nu$A can be expected to collect early evidence if such a pattern obtains,
and we have shown that DUNE will be able to identify such a pattern over a
significant part of the parameter space. For the CP conserving patterns, the
deviations from $\theta_{23}=\pi/4$ are expected to be measureable at $3\sigma$
by the next generation of superbeams for the preserved subgroup $\mathbb{Z}_5$,
but not for $\mathbb{Z}_2\times\mathbb{Z}_2$. Ultimately separating between
these models at $3\sigma$ significance across the whole parameter space could
be done using a Neutrino Factory after 10 years of data taking.
An attractive feature of the theoretical scenario in this work is its ability
to predict Majorana phases, and therefore, observables for neutrinoless double
beta decay experiments. We have seen that in the case of inverted mass
ordering, two of the possible Majorana phase combinations predict the discovery
of neutrinoless double beta decay at upcoming experiments. In the longer term,
the exploration of the full parameter space for inverted hierarchical mass
spectra could allow all of our patterns with this mass spectrum to be confirmed 
independently of oscillation physics.  

In conclusion, we find that the combination of the flavour symmetry A$_5$ with
a generalised CP symmetry allows for a number of viable predictions to be made
for the mixing angles and phases. These predictions specify parameter
correlations which present good targets for each stage of the next decade of
the experimental programme.

%%%%

\vspace{0.1cm} \textbf{Note added:}~During the final preparations of this
article, preprints of two similar works were made available
\cite{Li:2015jxa,DiIura:2015kfa}. These works also study the mixing patterns
arising from the residual symmetries of A$_5$ with GCP. The patterns derived in
the first part of our paper confirm those found in
\refsref{Li:2015jxa}{DiIura:2015kfa}. Although, case II in \refref{Li:2015jxa}
is omitted in our analysis as its predictions fall outside the $3\sigma$ global
intervals used in this work. Our phenomenological work, however, significantly
extends the analysis in these papers.

%%%%

\vspace{0.05cm}

\begin{acknowledgments}
We would like to thank Dr Pierre-Philippe Dechant for enlightening discussions
about group theory and for his comments on various stages of this work. 

This work has been supported by the European Research Council under ERC Grant
``NuMass'' (FP7-IDEAS-ERC ERC-CG 617143), and by the European Union FP7
ITN-INVISIBLES (Marie Curie Actions, PITN-GA-2011-289442).

\end{acknowledgments}

%%%%%%%%%%%%%%%%%%%%%%%%%%%%%%%%
%%%%%%%%%%%%%%%%%%%%%%%%%%%%%%%%
%%%%%%%%%%%%%%%%%%%%%%%%%%%%%%%%

\bibliographystyle{apsrev4-1}
\bibliography{CPlib}{}

%Merlin.mbs v4.21 2009-07-09.
\begin{thebibliography}{100}%
\makeatletter
\providecommand \@ifxundefined [1]{%
 \ifx #1\undefined \expandafter \@firstoftwo
 \else \expandafter \@secondoftwo
\fi
}%
\providecommand \@ifnum [1]{%
 \ifnum #1\expandafter \@firstoftwo
 \else \expandafter \@secondoftwo
\fi
}%
\providecommand \enquote [1]{``#1''}%
\providecommand \bibnamefont  [1]{#1}%
\providecommand \bibfnamefont [1]{#1}%
\providecommand \citenamefont [1]{#1}%
\providecommand\href[0]{\@sanitize\@href}%
\providecommand\@href[1]{\endgroup\@@startlink{#1}\endgroup\@@href}%
\providecommand\@@href[1]{#1\@@endlink}%
\providecommand \@sanitize [0]{\begingroup\catcode`\&12\catcode`\#12\relax}%
\@ifxundefined \pdfoutput {\@firstoftwo}{%
 \@ifnum{\z@=\pdfoutput}{\@firstoftwo}{\@secondoftwo}%
}{%
 \providecommand\@@startlink[1]{\leavevmode\special{html:<a href="#1">}}%
 \providecommand\@@endlink[0]{\special{html:</a>}}%
}{%
 \providecommand\@@startlink[1]{%
  \leavevmode
  \pdfstartlink
   attr{/Border[0 0 1 ]/H/I/C[0 1 1]}%
   user{/Subtype/Link/A<</Type/Action/S/URI/URI(#1)>>}%
  \relax
 }%
 \providecommand\@@endlink[0]{\pdfendlink}%
}%
\providecommand \url  [0]{\begingroup\@sanitize \@url }%
\providecommand \@url [1]{\endgroup\@href {#1}{\urlprefix}}%
\providecommand \urlprefix [0]{URL }%
\providecommand \Eprint[0]{\href }%
\@ifxundefined \urlstyle {%
  \providecommand \doi [1]{doi:\discretionary{}{}{}#1}%
}{%
  \providecommand \doi [0]{doi:\discretionary{}{}{}\begingroup
  \urlstyle{rm}\Url }%
}%
\providecommand \doibase [0]{http://dx.doi.org/}%
\providecommand \Doi[1]{\href{\doibase#1}}%
\providecommand \bibAnnote [3]{%
  \BibitemShut{#1}%
  \begin{quotation}\noindent
    \textsc{Key:}\ #2\\\textsc{Annotation:}\ #3%
  \end{quotation}%
}%
\providecommand \bibAnnoteFile [2]{%
  \IfFileExists{#2}{\bibAnnote {#1} {#2} {\input{#2}}}{}%
}%
\providecommand \typeout [0]{\immediate \write \m@ne }%
\providecommand \selectlanguage [0]{\@gobble}%
\providecommand \bibinfo [0]{\@secondoftwo}%
\providecommand \bibfield [0]{\@secondoftwo}%
\providecommand \translation [1]{[#1]}%
\providecommand \BibitemOpen[0]{}%
\providecommand \bibitemStop [0]{}%
\providecommand \bibitemNoStop [0]{.\EOS\space}%
\providecommand \EOS [0]{\spacefactor3000\relax}%
\providecommand \BibitemShut [1]{\csname bibitem#1\endcsname}%
%</preamble>
\bibitem{King:2013eh}%
  \BibitemOpen
  \bibfield{author}{%
  \bibinfo {author} {\bibfnamefont{S.~F.}\ \bibnamefont{King}}\ and\ \bibinfo
  {author} {\bibfnamefont{C.}~\bibnamefont{Luhn}},\ }%
  \bibfield{journal}{%
  \Doi{10.1088/0034-4885/76/5/056201}{\bibinfo {journal} {Rept.Prog.Phys.}}\ }%
  \textbf{\bibinfo {volume} {76}},\ \bibinfo {pages} {056201} (\bibinfo {year}
  {2013}),\ \Eprint{http://arxiv.org/abs/1301.1340}{arXiv:1301.1340 [hep-ph]}%
  \bibAnnoteFile{NoStop}{King:2013eh}%
%%CITATION = ARXIV:1301.1340;%%
\bibitem{An:2012eh}%
  \BibitemOpen
  \bibfield{author}{%
  \bibinfo {author} {\bibfnamefont{F.}~\bibnamefont{An}} \emph{et~al.}
  (\bibinfo {collaboration} {DAYA-BAY Collaboration}),\ }%
  \bibfield{journal}{%
  \Doi{10.1103/PhysRevLett.108.171803}{\bibinfo {journal} {Phys.Rev.Lett.}}\ }%
  \textbf{\bibinfo {volume} {108}},\ \bibinfo {pages} {171803} (\bibinfo {year}
  {2012}),\ \Eprint{http://arxiv.org/abs/1203.1669}{arXiv:1203.1669 [hep-ex]}%
  \bibAnnoteFile{NoStop}{An:2012eh}%
%%CITATION = ARXIV:1203.1669;%%
\bibitem{An:2013uza}%
  \BibitemOpen
  \bibfield{author}{%
  \bibinfo {author} {\bibfnamefont{F.}~\bibnamefont{An}} \emph{et~al.}
  (\bibinfo {collaboration} {Daya Bay}),\ }%
  \bibfield{journal}{%
  \Doi{10.1088/1674-1137/37/1/011001}{\bibinfo {journal} {Chin.Phys.}}\ }%
  \textbf{\bibinfo {volume} {C37}},\ \bibinfo {pages} {011001} (\bibinfo {year}
  {2013}),\ \Eprint{http://arxiv.org/abs/1210.6327}{arXiv:1210.6327 [hep-ex]}%
  \bibAnnoteFile{NoStop}{An:2013uza}%
%%CITATION = ARXIV:1210.6327;%%
\bibitem{Ahn:2012nd}%
  \BibitemOpen
  \bibfield{author}{%
  \bibinfo {author} {\bibfnamefont{J.}~\bibnamefont{Ahn}} \emph{et~al.}
  (\bibinfo {collaboration} {RENO collaboration}),\ }%
  \bibfield{journal}{%
  \Doi{10.1103/PhysRevLett.108.191802}{\bibinfo {journal} {Phys.Rev.Lett.}}\ }%
  \textbf{\bibinfo {volume} {108}},\ \bibinfo {pages} {191802} (\bibinfo {year}
  {2012}),\ \Eprint{http://arxiv.org/abs/1204.0626}{arXiv:1204.0626 [hep-ex]}%
  \bibAnnoteFile{NoStop}{Ahn:2012nd}%
%%CITATION = ARXIV:1204.0626;%%
\bibitem{Abe:2011sj}%
  \BibitemOpen
  \bibfield{author}{%
  \bibinfo {author} {\bibfnamefont{K.}~\bibnamefont{Abe}} \emph{et~al.}
  (\bibinfo {collaboration} {T2K Collaboration}),\ }%
  \bibfield{journal}{%
  \Doi{10.1103/PhysRevLett.107.041801}{\bibinfo {journal} {Phys.Rev.Lett.}}\ }%
  \textbf{\bibinfo {volume} {107}},\ \bibinfo {pages} {041801} (\bibinfo {year}
  {2011}),\ \Eprint{http://arxiv.org/abs/1106.2822}{arXiv:1106.2822 [hep-ex]}%
  \bibAnnoteFile{NoStop}{Abe:2011sj}%
%%CITATION = ARXIV:1106.2822;%%
\bibitem{Abe:2011fz}%
  \BibitemOpen
  \bibfield{author}{%
  \bibinfo {author} {\bibfnamefont{Y.}~\bibnamefont{Abe}} \emph{et~al.}
  (\bibinfo {collaboration} {DOUBLE-CHOOZ Collaboration}),\ }%
  \bibfield{journal}{%
  \Doi{10.1103/PhysRevLett.108.131801}{\bibinfo {journal} {Phys.Rev.Lett.}}\ }%
  \textbf{\bibinfo {volume} {108}},\ \bibinfo {pages} {131801} (\bibinfo {year}
  {2012}),\ \Eprint{http://arxiv.org/abs/1112.6353}{arXiv:1112.6353 [hep-ex]}%
  \bibAnnoteFile{NoStop}{Abe:2011fz}%
%%CITATION = ARXIV:1112.6353;%%
\bibitem{Toorop:2011jn}%
  \BibitemOpen
  \bibfield{author}{%
  \bibinfo {author} {\bibfnamefont{R.~d.~A.}\ \bibnamefont{Toorop}}, \bibinfo
  {author} {\bibfnamefont{F.}~\bibnamefont{Feruglio}},\ and\ \bibinfo {author}
  {\bibfnamefont{C.}~\bibnamefont{Hagedorn}},\ }%
  \bibfield{journal}{%
  \Doi{10.1016/j.physletb.2011.08.013}{\bibinfo {journal} {Phys.Lett.}}\ }%
  \textbf{\bibinfo {volume} {B703}},\ \bibinfo {pages} {447} (\bibinfo {year}
  {2011}),\ \Eprint{http://arxiv.org/abs/1107.3486}{arXiv:1107.3486 [hep-ph]}%
  \bibAnnoteFile{NoStop}{Toorop:2011jn}%
%%CITATION = ARXIV:1107.3486;%%
\bibitem{Ding:2012xx}%
  \BibitemOpen
  \bibfield{author}{%
  \bibinfo {author} {\bibfnamefont{G.-J.}\ \bibnamefont{Ding}},\ }%
  \bibfield{journal}{%
  \Doi{10.1016/j.nuclphysb.2012.04.002}{\bibinfo {journal} {Nucl.Phys.}}\ }%
  \textbf{\bibinfo {volume} {B862}},\ \bibinfo {pages} {1} (\bibinfo {year}
  {2012}),\ \Eprint{http://arxiv.org/abs/1201.3279}{arXiv:1201.3279 [hep-ph]}%
  \bibAnnoteFile{NoStop}{Ding:2012xx}%
%%CITATION = ARXIV:1201.3279;%%
\bibitem{Lam:2012ga}%
  \BibitemOpen
  \bibfield{author}{%
  \bibinfo {author} {\bibfnamefont{C.}~\bibnamefont{Lam}},\ }%
  \bibfield{journal}{%
  \Doi{10.1103/PhysRevD.87.013001}{\bibinfo {journal} {Phys.Rev.}}\ }%
  \textbf{\bibinfo {volume} {D87}},\ \bibinfo {pages} {013001} (\bibinfo {year}
  {2013}),\ \Eprint{http://arxiv.org/abs/1208.5527}{arXiv:1208.5527 [hep-ph]}%
  \bibAnnoteFile{NoStop}{Lam:2012ga}%
%%CITATION = ARXIV:1208.5527;%%
\bibitem{Lam:2013ng}%
  \BibitemOpen
  \bibfield{author}{%
  \bibinfo {author} {\bibfnamefont{C.}~\bibnamefont{Lam}},\ }%
  \bibfield{journal}{%
  \Doi{10.1103/PhysRevD.87.053012}{\bibinfo {journal} {Phys.Rev.}}\ }%
  \textbf{\bibinfo {volume} {D87}},\ \bibinfo {pages} {053012} (\bibinfo {year}
  {2013}),\ \Eprint{http://arxiv.org/abs/1301.1736}{arXiv:1301.1736 [hep-ph]}%
  \bibAnnoteFile{NoStop}{Lam:2013ng}%
%%CITATION = ARXIV:1301.1736;%%
\bibitem{Holthausen:2012wt}%
  \BibitemOpen
  \bibfield{author}{%
  \bibinfo {author} {\bibfnamefont{M.}~\bibnamefont{Holthausen}}, \bibinfo
  {author} {\bibfnamefont{K.~S.}\ \bibnamefont{Lim}},\ and\ \bibinfo {author}
  {\bibfnamefont{M.}~\bibnamefont{Lindner}},\ }%
  \bibfield{journal}{%
  \Doi{10.1016/j.physletb.2013.02.047}{\bibinfo {journal} {Phys.Lett.}}\ }%
  \textbf{\bibinfo {volume} {B721}},\ \bibinfo {pages} {61} (\bibinfo {year}
  {2013}),\ \Eprint{http://arxiv.org/abs/1212.2411}{arXiv:1212.2411 [hep-ph]}%
  \bibAnnoteFile{NoStop}{Holthausen:2012wt}%
%%CITATION = ARXIV:1212.2411;%%
\bibitem{Fonseca:2014koa}%
  \BibitemOpen
  \bibfield{author}{%
  \bibinfo {author} {\bibfnamefont{R.~M.}\ \bibnamefont{Fonseca}}\ and\
  \bibinfo {author} {\bibfnamefont{W.}~\bibnamefont{Grimus}},\ }%
  \bibfield{journal}{%
  \Doi{10.1007/JHEP09(2014)033}{\bibinfo {journal} {JHEP}}\ }%
  \textbf{\bibinfo {volume} {1409}},\ \bibinfo {pages} {033} (\bibinfo {year}
  {2014}),\ \Eprint{http://arxiv.org/abs/1405.3678}{arXiv:1405.3678 [hep-ph]}%
  \bibAnnoteFile{NoStop}{Fonseca:2014koa}%
%%CITATION = ARXIV:1405.3678;%%
\bibitem{King:2007pr}%
  \BibitemOpen
  \bibfield{author}{%
  \bibinfo {author} {\bibfnamefont{S.}~\bibnamefont{King}},\ }%
  \bibfield{journal}{%
  \Doi{10.1016/j.physletb.2007.10.078}{\bibinfo {journal} {Phys.Lett.}}\ }%
  \textbf{\bibinfo {volume} {B659}},\ \bibinfo {pages} {244} (\bibinfo {year}
  {2008}),\ \Eprint{http://arxiv.org/abs/0710.0530}{arXiv:0710.0530 [hep-ph]}%
  \bibAnnoteFile{NoStop}{King:2007pr}%
%%CITATION = ARXIV:0710.0530;%%
\bibitem{King:2005bj}%
  \BibitemOpen
  \bibfield{author}{%
  \bibinfo {author} {\bibfnamefont{S.}~\bibnamefont{King}},\ }%
  \bibfield{journal}{%
  \Doi{10.1088/1126-6708/2005/08/105}{\bibinfo {journal} {JHEP}}\ }%
  \textbf{\bibinfo {volume} {0508}},\ \bibinfo {pages} {105} (\bibinfo {year}
  {2005}),\ \Eprint{http://arxiv.org/abs/hep-ph/0506297}{arXiv:hep-ph/0506297
  [hep-ph]}%
  \bibAnnoteFile{NoStop}{King:2005bj}%
%%CITATION = HEP-PH/0506297;%%
\bibitem{Masina:2005hf}%
  \BibitemOpen
  \bibfield{author}{%
  \bibinfo {author} {\bibfnamefont{I.}~\bibnamefont{Masina}},\ }%
  \bibfield{journal}{%
  \Doi{10.1016/j.physletb.2005.10.097}{\bibinfo {journal} {Phys.Lett.}}\ }%
  \textbf{\bibinfo {volume} {B633}},\ \bibinfo {pages} {134} (\bibinfo {year}
  {2006}),\ \Eprint{http://arxiv.org/abs/hep-ph/0508031}{arXiv:hep-ph/0508031
  [hep-ph]}%
  \bibAnnoteFile{NoStop}{Masina:2005hf}%
%%CITATION = HEP-PH/0508031;%%
\bibitem{Antusch:2005kw}%
  \BibitemOpen
  \bibfield{author}{%
  \bibinfo {author} {\bibfnamefont{S.}~\bibnamefont{Antusch}}\ and\ \bibinfo
  {author} {\bibfnamefont{S.~F.}\ \bibnamefont{King}},\ }%
  \bibfield{journal}{%
  \Doi{10.1016/j.physletb.2005.09.075}{\bibinfo {journal} {Phys.Lett.}}\ }%
  \textbf{\bibinfo {volume} {B631}},\ \bibinfo {pages} {42} (\bibinfo {year}
  {2005}),\ \Eprint{http://arxiv.org/abs/hep-ph/0508044}{arXiv:hep-ph/0508044
  [hep-ph]}%
  \bibAnnoteFile{NoStop}{Antusch:2005kw}%
%%CITATION = HEP-PH/0508044;%%
\bibitem{Hernandez:2012ra}%
  \BibitemOpen
  \bibfield{author}{%
  \bibinfo {author} {\bibfnamefont{D.}~\bibnamefont{Hernandez}}\ and\ \bibinfo
  {author} {\bibfnamefont{A.~Y.}\ \bibnamefont{Smirnov}},\ }%
  \bibfield{journal}{%
  \Doi{10.1103/PhysRevD.86.053014}{\bibinfo {journal} {Phys.Rev.}}\ }%
  \textbf{\bibinfo {volume} {D86}},\ \bibinfo {pages} {053014} (\bibinfo {year}
  {2012}),\ \Eprint{http://arxiv.org/abs/1204.0445}{arXiv:1204.0445 [hep-ph]}%
  \bibAnnoteFile{NoStop}{Hernandez:2012ra}%
%%CITATION = ARXIV:1204.0445;%%
\bibitem{Hernandez:2012sk}%
  \BibitemOpen
  \bibfield{author}{%
  \bibinfo {author} {\bibfnamefont{D.}~\bibnamefont{Hernandez}}\ and\ \bibinfo
  {author} {\bibfnamefont{A.~Y.}\ \bibnamefont{Smirnov}},\ }%
  \bibfield{journal}{%
  \Doi{10.1103/PhysRevD.87.053005}{\bibinfo {journal} {Phys.Rev.}}\ }%
  \textbf{\bibinfo {volume} {D87}},\ \bibinfo {pages} {053005} (\bibinfo {year}
  {2013}),\ \Eprint{http://arxiv.org/abs/1212.2149}{arXiv:1212.2149 [hep-ph]}%
  \bibAnnoteFile{NoStop}{Hernandez:2012sk}%
%%CITATION = ARXIV:1212.2149;%%
\bibitem{Ballett:2013wya}%
  \BibitemOpen
  \bibfield{author}{%
  \bibinfo {author} {\bibfnamefont{P.}~\bibnamefont{Ballett}}, \bibinfo
  {author} {\bibfnamefont{S.~F.}\ \bibnamefont{King}}, \bibinfo {author}
  {\bibfnamefont{C.}~\bibnamefont{Luhn}}, \bibinfo {author}
  {\bibfnamefont{S.}~\bibnamefont{Pascoli}},\ and\ \bibinfo {author}
  {\bibfnamefont{M.~A.}\ \bibnamefont{Schmidt}},\ }%
  \bibfield{journal}{%
  \Doi{10.1103/PhysRevD.89.016016}{\bibinfo {journal} {Phys.Rev.}}\ }%
  \textbf{\bibinfo {volume} {D89}},\ \bibinfo {pages} {016016} (\bibinfo {year}
  {2014}),\ \Eprint{http://arxiv.org/abs/1308.4314}{arXiv:1308.4314 [hep-ph]}%
  \bibAnnoteFile{NoStop}{Ballett:2013wya}%
%%CITATION = ARXIV:1308.4314;%%
\bibitem{Meloni:2013qda}%
  \BibitemOpen
  \bibfield{author}{%
  \bibinfo {author} {\bibfnamefont{D.}~\bibnamefont{Meloni}},\ }%
  \bibfield{journal}{%
  \Doi{10.1016/j.physletb.2013.11.033}{\bibinfo {journal} {Phys.Lett.}}\ }%
  \textbf{\bibinfo {volume} {B728}},\ \bibinfo {pages} {118} (\bibinfo {year}
  {2014}),\ \Eprint{http://arxiv.org/abs/1308.4578}{arXiv:1308.4578 [hep-ph]}%
  \bibAnnoteFile{NoStop}{Meloni:2013qda}%
%%CITATION = ARXIV:1308.4578;%%
\bibitem{Hanlon:2013ska}%
  \BibitemOpen
  \bibfield{author}{%
  \bibinfo {author} {\bibfnamefont{A.~D.}\ \bibnamefont{Hanlon}}, \bibinfo
  {author} {\bibfnamefont{S.-F.}\ \bibnamefont{Ge}},\ and\ \bibinfo {author}
  {\bibfnamefont{W.~W.}\ \bibnamefont{Repko}},\ }%
  \bibfield{journal}{%
  \Doi{10.1016/j.physletb.2013.12.063}{\bibinfo {journal} {Phys.Lett.}}\ }%
  \textbf{\bibinfo {volume} {B729}},\ \bibinfo {pages} {185} (\bibinfo {year}
  {2014}),\ \Eprint{http://arxiv.org/abs/1308.6522}{arXiv:1308.6522 [hep-ph]}%
  \bibAnnoteFile{NoStop}{Hanlon:2013ska}%
%%CITATION = ARXIV:1308.6522;%%
\bibitem{Petcov:2014laa}%
  \BibitemOpen
  \bibfield{author}{%
  \bibinfo {author} {\bibfnamefont{S.}~\bibnamefont{Petcov}},\ }%
  \bibfield{journal}{%
  \Doi{10.1016/j.nuclphysb.2015.01.011}{\bibinfo {journal} {Nucl.Phys.}}\ }%
  \textbf{\bibinfo {volume} {B892}},\ \bibinfo {pages} {400} (\bibinfo {year}
  {2015}),\ \Eprint{http://arxiv.org/abs/1405.6006}{arXiv:1405.6006 [hep-ph]}%
  \bibAnnoteFile{NoStop}{Petcov:2014laa}%
%%CITATION = ARXIV:1405.6006;%%
\bibitem{Ballett:2014dua}%
  \BibitemOpen
  \bibfield{author}{%
  \bibinfo {author} {\bibfnamefont{P.}~\bibnamefont{Ballett}}, \bibinfo
  {author} {\bibfnamefont{S.~F.}\ \bibnamefont{King}}, \bibinfo {author}
  {\bibfnamefont{C.}~\bibnamefont{Luhn}}, \bibinfo {author}
  {\bibfnamefont{S.}~\bibnamefont{Pascoli}},\ and\ \bibinfo {author}
  {\bibfnamefont{M.~A.}\ \bibnamefont{Schmidt}},\ }%
  \bibfield{journal}{%
  \Doi{10.1007/JHEP12(2014)122}{\bibinfo {journal} {JHEP}}\ }%
  \textbf{\bibinfo {volume} {1412}},\ \bibinfo {pages} {122} (\bibinfo {year}
  {2014}),\ \Eprint{http://arxiv.org/abs/1410.7573}{arXiv:1410.7573 [hep-ph]}%
  \bibAnnoteFile{NoStop}{Ballett:2014dua}%
%%CITATION = ARXIV:1410.7573;%%
\bibitem{Antusch:2007rk}%
  \BibitemOpen
  \bibfield{author}{%
  \bibinfo {author} {\bibfnamefont{S.}~\bibnamefont{Antusch}}, \bibinfo
  {author} {\bibfnamefont{P.}~\bibnamefont{Huber}}, \bibinfo {author}
  {\bibfnamefont{S.}~\bibnamefont{King}},\ and\ \bibinfo {author}
  {\bibfnamefont{T.}~\bibnamefont{Schwetz}},\ }%
  \bibfield{journal}{%
  \Doi{10.1088/1126-6708/2007/04/060}{\bibinfo {journal} {JHEP}}\ }%
  \textbf{\bibinfo {volume} {0704}},\ \bibinfo {pages} {060} (\bibinfo {year}
  {2007}),\ \Eprint{http://arxiv.org/abs/hep-ph/0702286}{arXiv:hep-ph/0702286
  [HEP-PH]}%
  \bibAnnoteFile{NoStop}{Antusch:2007rk}%
%%CITATION = HEP-PH/0702286;%%
\bibitem{Ballett:2014uia}%
  \BibitemOpen
  \bibfield{author}{%
  \bibinfo {author} {\bibfnamefont{P.}~\bibnamefont{Ballett}}, \bibinfo
  {author} {\bibfnamefont{S.~F.}\ \bibnamefont{King}}, \bibinfo {author}
  {\bibfnamefont{C.}~\bibnamefont{Luhn}}, \bibinfo {author}
  {\bibfnamefont{S.}~\bibnamefont{Pascoli}},\ and\ \bibinfo {author}
  {\bibfnamefont{M.~A.}\ \bibnamefont{Schmidt}}}%
   (\bibinfo {year} {2014}),\
  \Eprint{http://arxiv.org/abs/1406.0308}{arXiv:1406.0308 [hep-ph]}%
  \bibAnnoteFile{NoStop}{Ballett:2014uia}%
%%CITATION = ARXIV:1406.0308;%%
\bibitem{Girardi:2014faa}%
  \BibitemOpen
  \bibfield{author}{%
  \bibinfo {author} {\bibfnamefont{I.}~\bibnamefont{Girardi}}, \bibinfo
  {author} {\bibfnamefont{S.}~\bibnamefont{Petcov}},\ and\ \bibinfo {author}
  {\bibfnamefont{A.}~\bibnamefont{Titov}}}%
   (\bibinfo {year} {2014}),\
  \Eprint{http://arxiv.org/abs/1410.8056}{arXiv:1410.8056 [hep-ph]}%
  \bibAnnoteFile{NoStop}{Girardi:2014faa}%
%%CITATION = ARXIV:1410.8056;%%
\bibitem{Gonzalez-Garcia:2014bfa}%
  \BibitemOpen
  \bibfield{author}{%
  \bibinfo {author} {\bibfnamefont{M.}~\bibnamefont{Gonzalez-Garcia}}, \bibinfo
  {author} {\bibfnamefont{M.}~\bibnamefont{Maltoni}},\ and\ \bibinfo {author}
  {\bibfnamefont{T.}~\bibnamefont{Schwetz}},\ }%
  \bibfield{journal}{%
  \Doi{10.1007/JHEP11(2014)052}{\bibinfo {journal} {JHEP}}\ }%
  \textbf{\bibinfo {volume} {1411}},\ \bibinfo {pages} {052} (\bibinfo {year}
  {2014}),\ \Eprint{http://arxiv.org/abs/1409.5439}{arXiv:1409.5439 [hep-ph]}%
  \bibAnnoteFile{NoStop}{Gonzalez-Garcia:2014bfa}%
%%CITATION = ARXIV:1409.5439;%%
\bibitem{Capozzi:2013csa}%
  \BibitemOpen
  \bibfield{author}{%
  \bibinfo {author} {\bibfnamefont{F.}~\bibnamefont{Capozzi}}, \bibinfo
  {author} {\bibfnamefont{G.}~\bibnamefont{Fogli}}, \bibinfo {author}
  {\bibfnamefont{E.}~\bibnamefont{Lisi}}, \bibinfo {author}
  {\bibfnamefont{A.}~\bibnamefont{Marrone}}, \bibinfo {author}
  {\bibfnamefont{D.}~\bibnamefont{Montanino}}, \emph{et~al.},\ }%
  \bibfield{journal}{%
  \Doi{10.1103/PhysRevD.89.093018}{\bibinfo {journal} {Phys.Rev.}}\ }%
  \textbf{\bibinfo {volume} {D89}},\ \bibinfo {pages} {093018} (\bibinfo {year}
  {2014}),\ \Eprint{http://arxiv.org/abs/1312.2878}{arXiv:1312.2878 [hep-ph]}%
  \bibAnnoteFile{NoStop}{Capozzi:2013csa}%
%%CITATION = ARXIV:1312.2878;%%
\bibitem{Forero:2014bxa}%
  \BibitemOpen
  \bibfield{author}{%
  \bibinfo {author} {\bibfnamefont{D.}~\bibnamefont{Forero}}, \bibinfo {author}
  {\bibfnamefont{M.}~\bibnamefont{Tortola}},\ and\ \bibinfo {author}
  {\bibfnamefont{J.}~\bibnamefont{Valle}},\ }%
  \bibfield{journal}{%
  \Doi{10.1103/PhysRevD.90.093006}{\bibinfo {journal} {Phys.Rev.}}\ }%
  \textbf{\bibinfo {volume} {D90}},\ \bibinfo {pages} {093006} (\bibinfo {year}
  {2014}),\ \Eprint{http://arxiv.org/abs/1405.7540}{arXiv:1405.7540 [hep-ph]}%
  \bibAnnoteFile{NoStop}{Forero:2014bxa}%
%%CITATION = ARXIV:1405.7540;%%
\bibitem{Feruglio:2012cw}%
  \BibitemOpen
  \bibfield{author}{%
  \bibinfo {author} {\bibfnamefont{F.}~\bibnamefont{Feruglio}}, \bibinfo
  {author} {\bibfnamefont{C.}~\bibnamefont{Hagedorn}},\ and\ \bibinfo {author}
  {\bibfnamefont{R.}~\bibnamefont{Ziegler}},\ }%
  \bibfield{journal}{%
  \Doi{10.1007/JHEP07(2013)027}{\bibinfo {journal} {JHEP}}\ }%
  \textbf{\bibinfo {volume} {1307}},\ \bibinfo {pages} {027} (\bibinfo {year}
  {2013}),\ \Eprint{http://arxiv.org/abs/1211.5560}{arXiv:1211.5560 [hep-ph]}%
  \bibAnnoteFile{NoStop}{Feruglio:2012cw}%
%%CITATION = ARXIV:1211.5560;%%
\bibitem{Ecker:1981wv}%
  \BibitemOpen
  \bibfield{author}{%
  \bibinfo {author} {\bibfnamefont{G.}~\bibnamefont{Ecker}}, \bibinfo {author}
  {\bibfnamefont{W.}~\bibnamefont{Grimus}},\ and\ \bibinfo {author}
  {\bibfnamefont{W.}~\bibnamefont{Konetschny}},\ }%
  \bibfield{journal}{%
  \Doi{10.1016/0550-3213(81)90309-6}{\bibinfo {journal} {Nucl.Phys.}}\ }%
  \textbf{\bibinfo {volume} {B191}},\ \bibinfo {pages} {465} (\bibinfo {year}
  {1981})%
  \bibAnnoteFile{NoStop}{Ecker:1981wv}%
%%CITATION = NUPHA,B191,465;%%
\bibitem{Bernabeu:1986fc}%
  \BibitemOpen
  \bibfield{author}{%
  \bibinfo {author} {\bibfnamefont{J.}~\bibnamefont{Bernabeu}}, \bibinfo
  {author} {\bibfnamefont{G.}~\bibnamefont{Branco}},\ and\ \bibinfo {author}
  {\bibfnamefont{M.}~\bibnamefont{Gronau}},\ }%
  \bibfield{journal}{%
  \Doi{10.1016/0370-2693(86)90659-3}{\bibinfo {journal} {Phys.Lett.}}\ }%
  \textbf{\bibinfo {volume} {B169}},\ \bibinfo {pages} {243} (\bibinfo {year}
  {1986})%
  \bibAnnoteFile{NoStop}{Bernabeu:1986fc}%
%%CITATION = PHLTA,B169,243;%%
\bibitem{Ecker:1989ay}%
  \BibitemOpen
  \bibfield{author}{%
  \bibinfo {author} {\bibfnamefont{G.}~\bibnamefont{Ecker}}, \bibinfo {author}
  {\bibfnamefont{W.}~\bibnamefont{Grimus}},\ and\ \bibinfo {author}
  {\bibfnamefont{H.}~\bibnamefont{Neufeld}},\ }%
  \bibfield{journal}{%
  \Doi{10.1016/0370-2693(89)91567-0}{\bibinfo {journal} {Phys.Lett.}}\ }%
  \textbf{\bibinfo {volume} {B228}},\ \bibinfo {pages} {401} (\bibinfo {year}
  {1989})%
  \bibAnnoteFile{NoStop}{Ecker:1989ay}%
%%CITATION = PHLTA,B228,401;%%
\bibitem{Holthausen:2012dk}%
  \BibitemOpen
  \bibfield{author}{%
  \bibinfo {author} {\bibfnamefont{M.}~\bibnamefont{Holthausen}}, \bibinfo
  {author} {\bibfnamefont{M.}~\bibnamefont{Lindner}},\ and\ \bibinfo {author}
  {\bibfnamefont{M.~A.}\ \bibnamefont{Schmidt}},\ }%
  \bibfield{journal}{%
  \Doi{10.1007/JHEP04(2013)122}{\bibinfo {journal} {JHEP}}\ }%
  \textbf{\bibinfo {volume} {1304}},\ \bibinfo {pages} {122} (\bibinfo {year}
  {2013}),\ \Eprint{http://arxiv.org/abs/1211.6953}{arXiv:1211.6953 [hep-ph]}%
  \bibAnnoteFile{NoStop}{Holthausen:2012dk}%
%%CITATION = ARXIV:1211.6953;%%
\bibitem{Chen:2014tpa}%
  \BibitemOpen
  \bibfield{author}{%
  \bibinfo {author} {\bibfnamefont{M.-C.}\ \bibnamefont{Chen}}, \bibinfo
  {author} {\bibfnamefont{M.}~\bibnamefont{Fallbacher}}, \bibinfo {author}
  {\bibfnamefont{K.}~\bibnamefont{Mahanthappa}}, \bibinfo {author}
  {\bibfnamefont{M.}~\bibnamefont{Ratz}},\ and\ \bibinfo {author}
  {\bibfnamefont{A.}~\bibnamefont{Trautner}},\ }%
  \bibfield{journal}{%
  \Doi{10.1016/j.nuclphysb.2014.03.023}{\bibinfo {journal} {Nucl.Phys.}}\ }%
  \textbf{\bibinfo {volume} {B883}},\ \bibinfo {pages} {267} (\bibinfo {year}
  {2014}),\ \Eprint{http://arxiv.org/abs/1402.0507}{arXiv:1402.0507 [hep-ph]}%
  \bibAnnoteFile{NoStop}{Chen:2014tpa}%
%%CITATION = ARXIV:1402.0507;%%
\bibitem{Ding:2013bpa}%
  \BibitemOpen
  \bibfield{author}{%
  \bibinfo {author} {\bibfnamefont{G.-J.}\ \bibnamefont{Ding}}, \bibinfo
  {author} {\bibfnamefont{S.~F.}\ \bibnamefont{King}},\ and\ \bibinfo {author}
  {\bibfnamefont{A.~J.}\ \bibnamefont{Stuart}},\ }%
  \bibfield{journal}{%
  \Doi{10.1007/JHEP12(2013)006}{\bibinfo {journal} {JHEP}}\ }%
  \textbf{\bibinfo {volume} {1312}},\ \bibinfo {pages} {006} (\bibinfo {year}
  {2013}),\ \Eprint{http://arxiv.org/abs/1307.4212}{arXiv:1307.4212 [hep-ph]}%
  \bibAnnoteFile{NoStop}{Ding:2013bpa}%
%%CITATION = ARXIV:1307.4212;%%
\bibitem{Feruglio:2013hia}%
  \BibitemOpen
  \bibfield{author}{%
  \bibinfo {author} {\bibfnamefont{F.}~\bibnamefont{Feruglio}}, \bibinfo
  {author} {\bibfnamefont{C.}~\bibnamefont{Hagedorn}},\ and\ \bibinfo {author}
  {\bibfnamefont{R.}~\bibnamefont{Ziegler}},\ }%
  \bibfield{journal}{%
  \Doi{10.1140/epjc/s10052-014-2753-2}{\bibinfo {journal} {Eur.Phys.J.}}\ }%
  \textbf{\bibinfo {volume} {C74}},\ \bibinfo {pages} {2753} (\bibinfo {year}
  {2014}),\ \Eprint{http://arxiv.org/abs/1303.7178}{arXiv:1303.7178 [hep-ph]}%
  \bibAnnoteFile{NoStop}{Feruglio:2013hia}%
%%CITATION = ARXIV:1303.7178;%%
\bibitem{Li:2013jya}%
  \BibitemOpen
  \bibfield{author}{%
  \bibinfo {author} {\bibfnamefont{C.-C.}\ \bibnamefont{Li}}\ and\ \bibinfo
  {author} {\bibfnamefont{G.-J.}\ \bibnamefont{Ding}},\ }%
  \bibfield{journal}{%
  \Doi{10.1016/j.nuclphysb.2014.02.002}{\bibinfo {journal} {Nucl.Phys.}}\ }%
  \textbf{\bibinfo {volume} {B881}},\ \bibinfo {pages} {206} (\bibinfo {year}
  {2014}),\ \Eprint{http://arxiv.org/abs/1312.4401}{arXiv:1312.4401 [hep-ph]}%
  \bibAnnoteFile{NoStop}{Li:2013jya}%
%%CITATION = ARXIV:1312.4401;%%
\bibitem{Ding:2013hpa}%
  \BibitemOpen
  \bibfield{author}{%
  \bibinfo {author} {\bibfnamefont{G.-J.}\ \bibnamefont{Ding}}, \bibinfo
  {author} {\bibfnamefont{S.~F.}\ \bibnamefont{King}}, \bibinfo {author}
  {\bibfnamefont{C.}~\bibnamefont{Luhn}},\ and\ \bibinfo {author}
  {\bibfnamefont{A.~J.}\ \bibnamefont{Stuart}},\ }%
  \bibfield{journal}{%
  \Doi{10.1007/JHEP05(2013)084}{\bibinfo {journal} {JHEP}}\ }%
  \textbf{\bibinfo {volume} {1305}},\ \bibinfo {pages} {084} (\bibinfo {year}
  {2013}),\ \Eprint{http://arxiv.org/abs/1303.6180}{arXiv:1303.6180 [hep-ph]}%
  \bibAnnoteFile{NoStop}{Ding:2013hpa}%
%%CITATION = ARXIV:1303.6180;%%
\bibitem{Li:2014eia}%
  \BibitemOpen
  \bibfield{author}{%
  \bibinfo {author} {\bibfnamefont{C.-C.}\ \bibnamefont{Li}}\ and\ \bibinfo
  {author} {\bibfnamefont{G.-J.}\ \bibnamefont{Ding}}}%
   (\bibinfo {year} {2014}),\
  \Eprint{http://arxiv.org/abs/1408.0785}{arXiv:1408.0785 [hep-ph]}%
  \bibAnnoteFile{NoStop}{Li:2014eia}%
%%CITATION = ARXIV:1408.0785;%%
\bibitem{Ding:2013nsa}%
  \BibitemOpen
  \bibfield{author}{%
  \bibinfo {author} {\bibfnamefont{G.-J.}\ \bibnamefont{Ding}}\ and\ \bibinfo
  {author} {\bibfnamefont{Y.-L.}\ \bibnamefont{Zhou}},\ }%
  \bibfield{journal}{%
  \Doi{10.1088/1674-1137/39/2/021001}{\bibinfo {journal} {Chin.Phys.}}\ }%
  \textbf{\bibinfo {volume} {C39}},\ \bibinfo {pages} {021001} (\bibinfo {year}
  {2015}),\ \Eprint{http://arxiv.org/abs/1312.5222}{arXiv:1312.5222 [hep-ph]}%
  \bibAnnoteFile{NoStop}{Ding:2013nsa}%
%%CITATION = ARXIV:1312.5222;%%
\bibitem{Ding:2014hva}%
  \BibitemOpen
  \bibfield{author}{%
  \bibinfo {author} {\bibfnamefont{G.-J.}\ \bibnamefont{Ding}}\ and\ \bibinfo
  {author} {\bibfnamefont{Y.-L.}\ \bibnamefont{Zhou}},\ }%
  \bibfield{journal}{%
  \Doi{10.1007/JHEP06(2014)023}{\bibinfo {journal} {JHEP}}\ }%
  \textbf{\bibinfo {volume} {1406}},\ \bibinfo {pages} {023} (\bibinfo {year}
  {2014}),\ \Eprint{http://arxiv.org/abs/1404.0592}{arXiv:1404.0592 [hep-ph]}%
  \bibAnnoteFile{NoStop}{Ding:2014hva}%
%%CITATION = ARXIV:1404.0592;%%
\bibitem{Ding:2014ssa}%
  \BibitemOpen
  \bibfield{author}{%
  \bibinfo {author} {\bibfnamefont{G.-J.}\ \bibnamefont{Ding}}\ and\ \bibinfo
  {author} {\bibfnamefont{S.~F.}\ \bibnamefont{King}},\ }%
  \bibfield{journal}{%
  \Doi{10.1103/PhysRevD.89.093020}{\bibinfo {journal} {Phys.Rev.}}\ }%
  \textbf{\bibinfo {volume} {D89}},\ \bibinfo {pages} {093020} (\bibinfo {year}
  {2014}),\ \Eprint{http://arxiv.org/abs/1403.5846}{arXiv:1403.5846 [hep-ph]}%
  \bibAnnoteFile{NoStop}{Ding:2014ssa}%
%%CITATION = ARXIV:1403.5846;%%
\bibitem{Hagedorn:2014wha}%
  \BibitemOpen
  \bibfield{author}{%
  \bibinfo {author} {\bibfnamefont{C.}~\bibnamefont{Hagedorn}}, \bibinfo
  {author} {\bibfnamefont{A.}~\bibnamefont{Meroni}},\ and\ \bibinfo {author}
  {\bibfnamefont{E.}~\bibnamefont{Molinaro}}}%
   (\bibinfo {year} {2014}),\
  \Eprint{http://arxiv.org/abs/1408.7118}{arXiv:1408.7118 [hep-ph]}%
  \bibAnnoteFile{NoStop}{Hagedorn:2014wha}%
%%CITATION = ARXIV:1408.7118;%%
\bibitem{Ding:2014ora}%
  \BibitemOpen
  \bibfield{author}{%
  \bibinfo {author} {\bibfnamefont{G.-J.}\ \bibnamefont{Ding}}, \bibinfo
  {author} {\bibfnamefont{S.~F.}\ \bibnamefont{King}},\ and\ \bibinfo {author}
  {\bibfnamefont{T.}~\bibnamefont{Neder}}}%
   (\bibinfo {year} {2014}),\
  \Eprint{http://arxiv.org/abs/1409.8005}{arXiv:1409.8005 [hep-ph]}%
  \bibAnnoteFile{NoStop}{Ding:2014ora}%
%%CITATION = ARXIV:1409.8005;%%
\bibitem{King:2014rwa}%
  \BibitemOpen
  \bibfield{author}{%
  \bibinfo {author} {\bibfnamefont{S.~F.}\ \bibnamefont{King}}\ and\ \bibinfo
  {author} {\bibfnamefont{T.}~\bibnamefont{Neder}},\ }%
  \bibfield{journal}{%
  \Doi{10.1016/j.physletb.2014.07.043}{\bibinfo {journal} {Phys.Lett.}}\ }%
  \textbf{\bibinfo {volume} {B736}},\ \bibinfo {pages} {308} (\bibinfo {year}
  {2014}),\ \Eprint{http://arxiv.org/abs/1403.1758}{arXiv:1403.1758 [hep-ph]}%
  \bibAnnoteFile{NoStop}{King:2014rwa}%
%%CITATION = ARXIV:1403.1758;%%
\bibitem{Everett:2015oka}%
  \BibitemOpen
  \bibfield{author}{%
  \bibinfo {author} {\bibfnamefont{L.~L.}\ \bibnamefont{Everett}}, \bibinfo
  {author} {\bibfnamefont{T.}~\bibnamefont{Garon}},\ and\ \bibinfo {author}
  {\bibfnamefont{A.~J.}\ \bibnamefont{Stuart}}}%
   (\bibinfo {year} {2015}),\
  \Eprint{http://arxiv.org/abs/1501.04336}{arXiv:1501.04336 [hep-ph]}%
  \bibAnnoteFile{NoStop}{Everett:2015oka}%
%%CITATION = ARXIV:1501.04336;%%
\bibitem{Chen:2014wxa}%
  \BibitemOpen
  \bibfield{author}{%
  \bibinfo {author} {\bibfnamefont{P.}~\bibnamefont{Chen}}, \bibinfo {author}
  {\bibfnamefont{C.-C.}\ \bibnamefont{Li}},\ and\ \bibinfo {author}
  {\bibfnamefont{G.-J.}\ \bibnamefont{Ding}},\ }%
  \bibfield{journal}{%
  \Doi{10.1103/PhysRevD.91.033003}{\bibinfo {journal} {Phys.Rev.}}\ }%
  \textbf{\bibinfo {volume} {D91}},\ \bibinfo {pages} {033003} (\bibinfo {year}
  {2015}),\ \Eprint{http://arxiv.org/abs/1412.8352}{arXiv:1412.8352 [hep-ph]}%
  \bibAnnoteFile{NoStop}{Chen:2014wxa}%
%%CITATION = ARXIV:1412.8352;%%
\bibitem{Everett:2008et}%
  \BibitemOpen
  \bibfield{author}{%
  \bibinfo {author} {\bibfnamefont{L.~L.}\ \bibnamefont{Everett}}\ and\
  \bibinfo {author} {\bibfnamefont{A.~J.}\ \bibnamefont{Stuart}},\ }%
  \bibfield{journal}{%
  \Doi{10.1103/PhysRevD.79.085005}{\bibinfo {journal} {Phys.Rev.}}\ }%
  \textbf{\bibinfo {volume} {D79}},\ \bibinfo {pages} {085005} (\bibinfo {year}
  {2009}),\ \Eprint{http://arxiv.org/abs/0812.1057}{arXiv:0812.1057 [hep-ph]}%
  \bibAnnoteFile{NoStop}{Everett:2008et}%
%%CITATION = ARXIV:0812.1057;%%
\bibitem{Kajiyama:2007gx}%
  \BibitemOpen
  \bibfield{author}{%
  \bibinfo {author} {\bibfnamefont{Y.}~\bibnamefont{Kajiyama}}, \bibinfo
  {author} {\bibfnamefont{M.}~\bibnamefont{Raidal}},\ and\ \bibinfo {author}
  {\bibfnamefont{A.}~\bibnamefont{Strumia}},\ }%
  \bibfield{journal}{%
  \Doi{10.1103/PhysRevD.76.117301}{\bibinfo {journal} {Phys.Rev.}}\ }%
  \textbf{\bibinfo {volume} {D76}},\ \bibinfo {pages} {117301} (\bibinfo {year}
  {2007}),\ \Eprint{http://arxiv.org/abs/0705.4559}{arXiv:0705.4559 [hep-ph]}%
  \bibAnnoteFile{NoStop}{Kajiyama:2007gx}%
%%CITATION = ARXIV:0705.4559;%%
\bibitem{Datta:2003qg}%
  \BibitemOpen
  \bibfield{author}{%
  \bibinfo {author} {\bibfnamefont{A.}~\bibnamefont{Datta}}, \bibinfo {author}
  {\bibfnamefont{F.-S.}\ \bibnamefont{Ling}},\ and\ \bibinfo {author}
  {\bibfnamefont{P.}~\bibnamefont{Ramond}},\ }%
  \bibfield{journal}{%
  \Doi{10.1016/j.nuclphysb.2003.08.026}{\bibinfo {journal} {Nucl.Phys.}}\ }%
  \textbf{\bibinfo {volume} {B671}},\ \bibinfo {pages} {383} (\bibinfo {year}
  {2003}),\ \Eprint{http://arxiv.org/abs/hep-ph/0306002}{arXiv:hep-ph/0306002
  [hep-ph]}%
  \bibAnnoteFile{NoStop}{Datta:2003qg}%
%%CITATION = HEP-PH/0306002;%%
\bibitem{Everett:2010rd}%
  \BibitemOpen
  \bibfield{author}{%
  \bibinfo {author} {\bibfnamefont{L.~L.}\ \bibnamefont{Everett}}\ and\
  \bibinfo {author} {\bibfnamefont{A.~J.}\ \bibnamefont{Stuart}},\ }%
  \bibfield{journal}{%
  \Doi{10.1016/j.physletb.2011.02.054}{\bibinfo {journal} {Phys.Lett.}}\ }%
  \textbf{\bibinfo {volume} {B698}},\ \bibinfo {pages} {131} (\bibinfo {year}
  {2011}),\ \Eprint{http://arxiv.org/abs/1011.4928}{arXiv:1011.4928 [hep-ph]}%
  \bibAnnoteFile{NoStop}{Everett:2010rd}%
%%CITATION = ARXIV:1011.4928;%%
\bibitem{Ding:2011cm}%
  \BibitemOpen
  \bibfield{author}{%
  \bibinfo {author} {\bibfnamefont{G.-J.}\ \bibnamefont{Ding}}, \bibinfo
  {author} {\bibfnamefont{L.~L.}\ \bibnamefont{Everett}},\ and\ \bibinfo
  {author} {\bibfnamefont{A.~J.}\ \bibnamefont{Stuart}},\ }%
  \bibfield{journal}{%
  \Doi{10.1016/j.nuclphysb.2011.12.004}{\bibinfo {journal} {Nucl.Phys.}}\ }%
  \textbf{\bibinfo {volume} {B857}},\ \bibinfo {pages} {219} (\bibinfo {year}
  {2012}),\ \Eprint{http://arxiv.org/abs/1110.1688}{arXiv:1110.1688 [hep-ph]}%
  \bibAnnoteFile{NoStop}{Ding:2011cm}%
%%CITATION = ARXIV:1110.1688;%%
\bibitem{Feruglio:2011qq}%
  \BibitemOpen
  \bibfield{author}{%
  \bibinfo {author} {\bibfnamefont{F.}~\bibnamefont{Feruglio}}\ and\ \bibinfo
  {author} {\bibfnamefont{A.}~\bibnamefont{Paris}},\ }%
  \bibfield{journal}{%
  \Doi{10.1007/JHEP03(2011)101}{\bibinfo {journal} {JHEP}}\ }%
  \textbf{\bibinfo {volume} {1103}},\ \bibinfo {pages} {101} (\bibinfo {year}
  {2011}),\ \Eprint{http://arxiv.org/abs/1101.0393}{arXiv:1101.0393 [hep-ph]}%
  \bibAnnoteFile{NoStop}{Feruglio:2011qq}%
%%CITATION = ARXIV:1101.0393;%%
\bibitem{Lam:2011ag}%
  \BibitemOpen
  \bibfield{author}{%
  \bibinfo {author} {\bibfnamefont{C.}~\bibnamefont{Lam}},\ }%
  \bibfield{journal}{%
  \Doi{10.1103/PhysRevD.83.113002}{\bibinfo {journal} {Phys.Rev.}}\ }%
  \textbf{\bibinfo {volume} {D83}},\ \bibinfo {pages} {113002} (\bibinfo {year}
  {2011}),\ \Eprint{http://arxiv.org/abs/1104.0055}{arXiv:1104.0055 [hep-ph]}%
  \bibAnnoteFile{NoStop}{Lam:2011ag}%
%%CITATION = ARXIV:1104.0055;%%
\bibitem{deAdelhartToorop:2011re}%
  \BibitemOpen
  \bibfield{author}{%
  \bibinfo {author} {\bibfnamefont{R.}~\bibnamefont{de~Adelhart~Toorop}},
  \bibinfo {author} {\bibfnamefont{F.}~\bibnamefont{Feruglio}},\ and\ \bibinfo
  {author} {\bibfnamefont{C.}~\bibnamefont{Hagedorn}},\ }%
  \bibfield{journal}{%
  \Doi{10.1016/j.nuclphysb.2012.01.017}{\bibinfo {journal} {Nucl.Phys.}}\ }%
  \textbf{\bibinfo {volume} {B858}},\ \bibinfo {pages} {437} (\bibinfo {year}
  {2012}),\ \Eprint{http://arxiv.org/abs/1112.1340}{arXiv:1112.1340 [hep-ph]}%
  \bibAnnoteFile{NoStop}{deAdelhartToorop:2011re}%
%%CITATION = ARXIV:1112.1340;%%
\bibitem{Agashe:2014kda}%
  \BibitemOpen
  \bibfield{author}{%
  \bibinfo {author} {\bibfnamefont{K.}~\bibnamefont{Olive}} \emph{et~al.}
  (\bibinfo {collaboration} {Particle Data Group}),\ }%
  \bibfield{journal}{%
  \Doi{10.1088/1674-1137/38/9/090001}{\bibinfo {journal} {Chin.Phys.}}\ }%
  \textbf{\bibinfo {volume} {C38}},\ \bibinfo {pages} {090001} (\bibinfo {year}
  {2014})%
  \bibAnnoteFile{NoStop}{Agashe:2014kda}%
%%CITATION = CHPHD,C38,090001;%%
\bibitem{simple}%
  \BibitemOpen
  \bibfield{author}{%
  \bibinfo {author} {\bibfnamefont{R.}~\bibnamefont{Wilson}},\ }%
  \emph{\bibinfo {title} {The Simple Finite Groups}}\ (\bibinfo {publisher}
  {Springer},\ \bibinfo {address} {LONDON},\ \bibinfo {year} {2009})%
  \bibAnnoteFile{NoStop}{simple}%
\bibitem{Burnside:1913aa}%
  \BibitemOpen
  \bibfield{author}{%
  \bibinfo {author} {\bibfnamefont{W.}~\bibnamefont{Burnside}},\ }%
  \bibfield{journal}{%
  \bibinfo {journal} {Proc.London Math.Soc}\ }%
  \textbf{\bibinfo {volume} {11}},\ \bibinfo {pages} {40} (\bibinfo {year}
  {1913})%
  \bibAnnoteFile{NoStop}{Burnside:1913aa}%
\bibitem{Brooksbank:2014aa}%
  \BibitemOpen
  \bibfield{author}{%
  \bibinfo {author} {\bibfnamefont{P.}~\bibnamefont{Brooksbank}}\ and\ \bibinfo
  {author} {\bibfnamefont{M.}~\bibnamefont{Mizuhara}},\ }%
  \bibfield{journal}{%
  \bibinfo {journal} {Involve 7}\ }%
  \textbf{\bibinfo {volume} {2}},\ \bibinfo {pages} {171} (\bibinfo {year}
  {2014})%
  \bibAnnoteFile{NoStop}{Brooksbank:2014aa}%
\bibitem{Harrison:2002er}%
  \BibitemOpen
  \bibfield{author}{%
  \bibinfo {author} {\bibfnamefont{P.}~\bibnamefont{Harrison}}, \bibinfo
  {author} {\bibfnamefont{D.}~\bibnamefont{Perkins}},\ and\ \bibinfo {author}
  {\bibfnamefont{W.}~\bibnamefont{Scott}},\ }%
  \bibfield{journal}{%
  \Doi{10.1016/S0370-2693(02)01336-9}{\bibinfo {journal} {Phys.Lett.}}\ }%
  \textbf{\bibinfo {volume} {B530}},\ \bibinfo {pages} {167} (\bibinfo {year}
  {2002}),\ \Eprint{http://arxiv.org/abs/hep-ph/0202074}{arXiv:hep-ph/0202074
  [hep-ph]}%
  \bibAnnoteFile{NoStop}{Harrison:2002er}%
%%CITATION = HEP-PH/0202074;%%
\bibitem{Haba:2006dz}%
  \BibitemOpen
  \bibfield{author}{%
  \bibinfo {author} {\bibfnamefont{N.}~\bibnamefont{Haba}}, \bibinfo {author}
  {\bibfnamefont{A.}~\bibnamefont{Watanabe}},\ and\ \bibinfo {author}
  {\bibfnamefont{K.}~\bibnamefont{Yoshioka}},\ }%
  \bibfield{journal}{%
  \Doi{10.1103/PhysRevLett.97.041601}{\bibinfo {journal} {Phys.Rev.Lett.}}\ }%
  \textbf{\bibinfo {volume} {97}},\ \bibinfo {pages} {041601} (\bibinfo {year}
  {2006}),\ \Eprint{http://arxiv.org/abs/hep-ph/0603116}{arXiv:hep-ph/0603116
  [hep-ph]}%
  \bibAnnoteFile{NoStop}{Haba:2006dz}%
%%CITATION = HEP-PH/0603116;%%
\bibitem{He:2006qd}%
  \BibitemOpen
  \bibfield{author}{%
  \bibinfo {author} {\bibfnamefont{X.-G.}\ \bibnamefont{He}}\ and\ \bibinfo
  {author} {\bibfnamefont{A.}~\bibnamefont{Zee}},\ }%
  \bibfield{journal}{%
  \Doi{10.1016/j.physletb.2006.11.055}{\bibinfo {journal} {Phys.Lett.}}\ }%
  \textbf{\bibinfo {volume} {B645}},\ \bibinfo {pages} {427} (\bibinfo {year}
  {2007}),\ \Eprint{http://arxiv.org/abs/hep-ph/0607163}{arXiv:hep-ph/0607163
  [hep-ph]}%
  \bibAnnoteFile{NoStop}{He:2006qd}%
%%CITATION = HEP-PH/0607163;%%
\bibitem{Grimus:2008tt}%
  \BibitemOpen
  \bibfield{author}{%
  \bibinfo {author} {\bibfnamefont{W.}~\bibnamefont{Grimus}}\ and\ \bibinfo
  {author} {\bibfnamefont{L.}~\bibnamefont{Lavoura}},\ }%
  \bibfield{journal}{%
  \Doi{10.1088/1126-6708/2008/09/106}{\bibinfo {journal} {JHEP}}\ }%
  \textbf{\bibinfo {volume} {0809}},\ \bibinfo {pages} {106} (\bibinfo {year}
  {2008}),\ \Eprint{http://arxiv.org/abs/0809.0226}{arXiv:0809.0226 [hep-ph]}%
  \bibAnnoteFile{NoStop}{Grimus:2008tt}%
%%CITATION = ARXIV:0809.0226;%%
\bibitem{Abe:2014tzr}%
  \BibitemOpen
  \bibfield{author}{%
  \bibinfo {author} {\bibfnamefont{K.}~\bibnamefont{Abe}} \emph{et~al.}
  (\bibinfo {collaboration} {T2K Collaboration})}%
   (\bibinfo {year} {2014}),\
  \Eprint{http://arxiv.org/abs/1409.7469}{arXiv:1409.7469 [hep-ex]}%
  \bibAnnoteFile{NoStop}{Abe:2014tzr}%
%%CITATION = ARXIV:1409.7469;%%
\bibitem{Patterson:2012talk}%
  \BibitemOpen
  \bibfield{author}{%
  \bibinfo {author} {\bibfnamefont{R.}~\bibnamefont{Patterson}}\ }%
  (\bibinfo {publisher} {Presented at XXV International Conference on Neutrino
  Physics and Astrophysics, Jun. 2012},\ \bibinfo {year} {2012})%
  \bibAnnoteFile{NoStop}{Patterson:2012talk}%
\bibitem{NOVA_web}%
  \BibitemOpen
  \enquote{\bibinfo {title} {{NO$\nu$A Official Plots and Figures}},}\
  (\bibinfo {year} {accessed 2015}),\
  \url{http://www-nova.fnal.gov/plots_and_figures.html}%
  \bibAnnoteFile{NoStop}{NOVA_web}%
\bibitem{Li:2014qca}%
  \BibitemOpen
  \bibfield{author}{%
  \bibinfo {author} {\bibfnamefont{Y.-F.}\ \bibnamefont{Li}},\ }%
  \bibfield{journal}{%
  \Doi{10.1142/S2010194514603007}{\bibinfo {journal}
  {Int.J.Mod.Phys.Conf.Ser.}}\ }%
  \textbf{\bibinfo {volume} {31}},\ \bibinfo {pages} {1460300} (\bibinfo {year}
  {2014}),\ \Eprint{http://arxiv.org/abs/1402.6143}{arXiv:1402.6143
  [physics.ins-det]}%
  \bibAnnoteFile{NoStop}{Li:2014qca}%
%%CITATION = ARXIV:1402.6143;%%
\bibitem{Kim:2014rfa}%
  \BibitemOpen
  \bibfield{author}{%
  \bibinfo {author} {\bibfnamefont{S.-B.}\ \bibnamefont{Kim}}}%
   (\bibinfo {year} {2014}),\
  \Eprint{http://arxiv.org/abs/1412.2199}{arXiv:1412.2199 [hep-ex]}%
  \bibAnnoteFile{NoStop}{Kim:2014rfa}%
%%CITATION = ARXIV:1412.2199;%%
\bibitem{Abe:2014oxa}%
  \BibitemOpen
  \bibfield{author}{%
  \bibinfo {author} {\bibfnamefont{K.}~\bibnamefont{Abe}} \emph{et~al.}
  (\bibinfo {collaboration} {Hyper-Kamiokande Working Group})}%
   (\bibinfo {year} {2014}),\
  \Eprint{http://arxiv.org/abs/1412.4673}{arXiv:1412.4673 [physics.ins-det]}%
  \bibAnnoteFile{NoStop}{Abe:2014oxa}%
%%CITATION = ARXIV:1412.4673;%%
\bibitem{Baussan:2012cw}%
  \BibitemOpen
  \bibfield{author}{%
  \bibinfo {author} {\bibfnamefont{E.}~\bibnamefont{Baussan}}, \bibinfo
  {author} {\bibfnamefont{M.}~\bibnamefont{Dracos}}, \bibinfo {author}
  {\bibfnamefont{T.}~\bibnamefont{Ekelof}}, \bibinfo {author}
  {\bibfnamefont{E.~F.}\ \bibnamefont{Martinez}}, \bibinfo {author}
  {\bibfnamefont{H.}~\bibnamefont{Ohman}}, \emph{et~al.}}%
   (\bibinfo {year} {2012}),\
  \Eprint{http://arxiv.org/abs/1212.5048}{arXiv:1212.5048 [hep-ex]}%
  \bibAnnoteFile{NoStop}{Baussan:2012cw}%
%%CITATION = ARXIV:1212.5048;%%
\bibitem{Baussan:2013zcy}%
  \BibitemOpen
  \bibfield{author}{%
  \bibinfo {author} {\bibfnamefont{E.}~\bibnamefont{Baussan}} \emph{et~al.}
  (\bibinfo {collaboration} {ESSnuSB Collaboration}),\ }%
  \bibfield{journal}{%
  \Doi{10.1016/j.nuclphysb.2014.05.016}{\bibinfo {journal} {Nucl.Phys.}}\ }%
  \textbf{\bibinfo {volume} {B885}},\ \bibinfo {pages} {127} (\bibinfo {year}
  {2014}),\ \Eprint{http://arxiv.org/abs/1309.7022}{arXiv:1309.7022 [hep-ex]}%
  \bibAnnoteFile{NoStop}{Baussan:2013zcy}%
%%CITATION = ARXIV:1309.7022;%%
\bibitem{Geer:1997iz}%
  \BibitemOpen
  \bibfield{author}{%
  \bibinfo {author} {\bibfnamefont{S.}~\bibnamefont{Geer}},\ }%
  \bibfield{journal}{%
  \Doi{10.1103/PhysRevD.57.6989 10.1103/PhysRevD.59.039903,
  10.1103/PhysRevD.57.6989, 10.1103/PhysRevD.59.039903}{\bibinfo {journal}
  {Phys.Rev.}}\ }%
  \textbf{\bibinfo {volume} {D57}},\ \bibinfo {pages} {6989} (\bibinfo {year}
  {1998}),\ \Eprint{http://arxiv.org/abs/hep-ph/9712290}{arXiv:hep-ph/9712290
  [hep-ph]}%
  \bibAnnoteFile{NoStop}{Geer:1997iz}%
%%CITATION = HEP-PH/9712290;%%
\bibitem{DeRujula:1998hd}%
  \BibitemOpen
  \bibfield{author}{%
  \bibinfo {author} {\bibfnamefont{A.}~\bibnamefont{De~Rujula}}, \bibinfo
  {author} {\bibfnamefont{M.}~\bibnamefont{Gavela}},\ and\ \bibinfo {author}
  {\bibfnamefont{P.}~\bibnamefont{Hernandez}},\ }%
  \bibfield{journal}{%
  \Doi{10.1016/S0550-3213(99)00070-X}{\bibinfo {journal} {Nucl.Phys.}}\ }%
  \textbf{\bibinfo {volume} {B547}},\ \bibinfo {pages} {21} (\bibinfo {year}
  {1999}),\ \Eprint{http://arxiv.org/abs/hep-ph/9811390}{arXiv:hep-ph/9811390
  [hep-ph]}%
  \bibAnnoteFile{NoStop}{DeRujula:1998hd}%
%%CITATION = HEP-PH/9811390;%%
\bibitem{Bandyopadhyay:2007kx}%
  \BibitemOpen
  \bibfield{author}{%
  \bibinfo {author} {\bibfnamefont{A.}~\bibnamefont{Bandyopadhyay}}
  \emph{et~al.} (\bibinfo {collaboration} {ISS Physics Working Group}),\ }%
  \bibfield{journal}{%
  \Doi{10.1088/0034-4885/72/10/106201}{\bibinfo {journal} {Rept.Prog.Phys.}}\
  }%
  \textbf{\bibinfo {volume} {72}},\ \bibinfo {pages} {106201} (\bibinfo {year}
  {2009}),\ \Eprint{http://arxiv.org/abs/0710.4947}{arXiv:0710.4947 [hep-ph]}%
  \bibAnnoteFile{NoStop}{Bandyopadhyay:2007kx}%
%%CITATION = ARXIV:0710.4947;%%
\bibitem{Huber:2004ka}%
  \BibitemOpen
  \bibfield{author}{%
  \bibinfo {author} {\bibfnamefont{P.}~\bibnamefont{Huber}}, \bibinfo {author}
  {\bibfnamefont{M.}~\bibnamefont{Lindner}},\ and\ \bibinfo {author}
  {\bibfnamefont{W.}~\bibnamefont{Winter}},\ }%
  \bibfield{journal}{%
  \Doi{10.1016/j.cpc.2005.01.003}{\bibinfo {journal} {Comput.Phys.Commun.}}\ }%
  \textbf{\bibinfo {volume} {167}},\ \bibinfo {pages} {195} (\bibinfo {year}
  {2005}),\ \Eprint{http://arxiv.org/abs/hep-ph/0407333}{arXiv:hep-ph/0407333
  [hep-ph]}%
  \bibAnnoteFile{NoStop}{Huber:2004ka}%
%%CITATION = HEP-PH/0407333;%%
\bibitem{Huber:2007ji}%
  \BibitemOpen
  \bibfield{author}{%
  \bibinfo {author} {\bibfnamefont{P.}~\bibnamefont{Huber}}, \bibinfo {author}
  {\bibfnamefont{J.}~\bibnamefont{Kopp}}, \bibinfo {author}
  {\bibfnamefont{M.}~\bibnamefont{Lindner}}, \bibinfo {author}
  {\bibfnamefont{M.}~\bibnamefont{Rolinec}},\ and\ \bibinfo {author}
  {\bibfnamefont{W.}~\bibnamefont{Winter}},\ }%
  \bibfield{journal}{%
  \Doi{10.1016/j.cpc.2007.05.004}{\bibinfo {journal} {Comput.Phys.Commun.}}\ }%
  \textbf{\bibinfo {volume} {177}},\ \bibinfo {pages} {432} (\bibinfo {year}
  {2007}),\ \Eprint{http://arxiv.org/abs/hep-ph/0701187}{arXiv:hep-ph/0701187
  [hep-ph]}%
  \bibAnnoteFile{NoStop}{Huber:2007ji}%
%%CITATION = HEP-PH/0701187;%%
\bibitem{LBNE_AEDL}%
  \BibitemOpen
  \bibfield{author}{%
  \bibinfo {author} {\bibfnamefont{S.}~\bibnamefont{Zeller}},\ }%
  \enquote{\bibinfo {title} {{LBNE-doc-5823-v9}},}\  (\bibinfo {year} {2012}),\
  \url{http://lbne2-docdb.fnal.gov/cgi-bin/ShowDocument?docid=5823}%
  \bibAnnoteFile{NoStop}{LBNE_AEDL}%
\bibitem{Agarwalla:2013kaa}%
  \BibitemOpen
  \bibfield{author}{%
  \bibinfo {author} {\bibfnamefont{S.}~\bibnamefont{Agarwalla}} \emph{et~al.}
  (\bibinfo {collaboration} {LAGUNA-LBNO Collaboration}),\ }%
  \bibfield{journal}{%
  \Doi{10.1007/JHEP05(2014)094}{\bibinfo {journal} {JHEP}}\ }%
  \textbf{\bibinfo {volume} {1405}},\ \bibinfo {pages} {094} (\bibinfo {year}
  {2014}),\ \Eprint{http://arxiv.org/abs/1312.6520}{arXiv:1312.6520 [hep-ph]}%
  \bibAnnoteFile{NoStop}{Agarwalla:2013kaa}%
%%CITATION = ARXIV:1312.6520;%%
\bibitem{Agarwalla:2014tca}%
  \BibitemOpen
  \bibfield{author}{%
  \bibinfo {author} {\bibfnamefont{S.}~\bibnamefont{Agarwalla}} \emph{et~al.}
  (\bibinfo {collaboration} {LAGUNA-LBNO Collaboration})}%
   (\bibinfo {year} {2014}),\
  \Eprint{http://arxiv.org/abs/1412.0593}{arXiv:1412.0593 [hep-ph]}%
  \bibAnnoteFile{NoStop}{Agarwalla:2014tca}%
%%CITATION = ARXIV:1412.0593;%%
\bibitem{Agarwalla:2014ura}%
  \BibitemOpen
  \bibfield{author}{%
  \bibinfo {author} {\bibfnamefont{S.}~\bibnamefont{Agarwalla}} \emph{et~al.}
  (\bibinfo {collaboration} {LAGUNA-LBNO Collaboration})}%
   (\bibinfo {year} {2014}),\
  \Eprint{http://arxiv.org/abs/1412.0804}{arXiv:1412.0804 [hep-ph]}%
  \bibAnnoteFile{NoStop}{Agarwalla:2014ura}%
%%CITATION = ARXIV:1412.0804;%%
\bibitem{Adams:2013qkq}%
  \BibitemOpen
  \bibfield{author}{%
  \bibinfo {author} {\bibfnamefont{C.}~\bibnamefont{Adams}} \emph{et~al.}
  (\bibinfo {collaboration} {LBNE Collaboration})}%
   (\bibinfo {year} {2013}),\
  \Eprint{http://arxiv.org/abs/1307.7335}{arXiv:1307.7335 [hep-ex]}%
  \bibAnnoteFile{NoStop}{Adams:2013qkq}%
%%CITATION = ARXIV:1307.7335;%%
\bibitem{Abe:2015qaa}%
  \BibitemOpen
  \bibfield{author}{%
  \bibinfo {author} {\bibfnamefont{K.}~\bibnamefont{Abe}} \emph{et~al.}
  (\bibinfo {collaboration} {Hyper-Kamiokande Proto-Collaboration}),\ }%
  \bibfield{journal}{%
  \bibinfo {journal} {PTEP}}%
   (\bibinfo {year} {2015}),\
  \Eprint{http://arxiv.org/abs/1502.05199}{arXiv:1502.05199 [hep-ex]}%
  \bibAnnoteFile{NoStop}{Abe:2015qaa}%
%%CITATION = ARXIV:1502.05199;%%
\bibitem{Chatterjee:2013qus}%
  \BibitemOpen
  \bibfield{author}{%
  \bibinfo {author} {\bibfnamefont{A.}~\bibnamefont{Chatterjee}}, \bibinfo
  {author} {\bibfnamefont{P.}~\bibnamefont{Ghoshal}}, \bibinfo {author}
  {\bibfnamefont{S.}~\bibnamefont{Goswami}},\ and\ \bibinfo {author}
  {\bibfnamefont{S.~K.}\ \bibnamefont{Raut}},\ }%
  \bibfield{journal}{%
  \Doi{10.1007/JHEP06(2013)010}{\bibinfo {journal} {JHEP}}\ }%
  \textbf{\bibinfo {volume} {1306}},\ \bibinfo {pages} {010} (\bibinfo {year}
  {2013}),\ \Eprint{http://arxiv.org/abs/1302.1370}{arXiv:1302.1370 [hep-ph]}%
  \bibAnnoteFile{NoStop}{Chatterjee:2013qus}%
%%CITATION = ARXIV:1302.1370;%%
\bibitem{Agarwalla:2013ju}%
  \BibitemOpen
  \bibfield{author}{%
  \bibinfo {author} {\bibfnamefont{S.~K.}\ \bibnamefont{Agarwalla}}, \bibinfo
  {author} {\bibfnamefont{S.}~\bibnamefont{Prakash}},\ and\ \bibinfo {author}
  {\bibfnamefont{S.~U.}\ \bibnamefont{Sankar}},\ }%
  \bibfield{journal}{%
  \Doi{10.1007/JHEP07(2013)131}{\bibinfo {journal} {JHEP}}\ }%
  \textbf{\bibinfo {volume} {1307}},\ \bibinfo {pages} {131} (\bibinfo {year}
  {2013}),\ \Eprint{http://arxiv.org/abs/1301.2574}{arXiv:1301.2574 [hep-ph]}%
  \bibAnnoteFile{NoStop}{Agarwalla:2013ju}%
%%CITATION = ARXIV:1301.2574;%%
\bibitem{Huber:2014nga}%
  \BibitemOpen
  \bibfield{author}{%
  \bibinfo {author} {\bibfnamefont{P.}~\bibnamefont{Huber}}, \bibinfo {author}
  {\bibfnamefont{A.}~\bibnamefont{Bross}},\ and\ \bibinfo {author}
  {\bibfnamefont{M.}~\bibnamefont{Palmer}}}%
   (\bibinfo {year} {2014}),\
  \Eprint{http://arxiv.org/abs/1411.0629}{arXiv:1411.0629 [hep-ex]}%
  \bibAnnoteFile{NoStop}{Huber:2014nga}%
%%CITATION = ARXIV:1411.0629;%%
\bibitem{Coloma:2012wq}%
  \BibitemOpen
  \bibfield{author}{%
  \bibinfo {author} {\bibfnamefont{P.}~\bibnamefont{Coloma}}, \bibinfo {author}
  {\bibfnamefont{A.}~\bibnamefont{Donini}}, \bibinfo {author}
  {\bibfnamefont{E.}~\bibnamefont{Fernandez-Martinez}},\ and\ \bibinfo {author}
  {\bibfnamefont{P.}~\bibnamefont{Hernandez}},\ }%
  \bibfield{journal}{%
  \Doi{10.1007/JHEP06(2012)073}{\bibinfo {journal} {JHEP}}\ }%
  \textbf{\bibinfo {volume} {1206}},\ \bibinfo {pages} {073} (\bibinfo {year}
  {2012}),\ \Eprint{http://arxiv.org/abs/1203.5651}{arXiv:1203.5651 [hep-ph]}%
  \bibAnnoteFile{NoStop}{Coloma:2012wq}%
%%CITATION = ARXIV:1203.5651;%%
\bibitem{Coloma:2012ji}%
  \BibitemOpen
  \bibfield{author}{%
  \bibinfo {author} {\bibfnamefont{P.}~\bibnamefont{Coloma}}, \bibinfo {author}
  {\bibfnamefont{P.}~\bibnamefont{Huber}}, \bibinfo {author}
  {\bibfnamefont{J.}~\bibnamefont{Kopp}},\ and\ \bibinfo {author}
  {\bibfnamefont{W.}~\bibnamefont{Winter}},\ }%
  \bibfield{journal}{%
  \Doi{10.1103/PhysRevD.87.033004}{\bibinfo {journal} {Phys.Rev.}}\ }%
  \textbf{\bibinfo {volume} {D87}},\ \bibinfo {pages} {033004} (\bibinfo {year}
  {2013}),\ \Eprint{http://arxiv.org/abs/1209.5973}{arXiv:1209.5973 [hep-ph]}%
  \bibAnnoteFile{NoStop}{Coloma:2012ji}%
%%CITATION = ARXIV:1209.5973;%%
\bibitem{Bross:2007ts}%
  \BibitemOpen
  \bibfield{author}{%
  \bibinfo {author} {\bibfnamefont{A.~D.}\ \bibnamefont{Bross}}, \bibinfo
  {author} {\bibfnamefont{M.}~\bibnamefont{Ellis}}, \bibinfo {author}
  {\bibfnamefont{S.}~\bibnamefont{Geer}}, \bibinfo {author}
  {\bibfnamefont{O.}~\bibnamefont{Mena}},\ and\ \bibinfo {author}
  {\bibfnamefont{S.}~\bibnamefont{Pascoli}},\ }%
  \bibfield{journal}{%
  \Doi{10.1103/PhysRevD.77.093012}{\bibinfo {journal} {Phys.Rev.}}\ }%
  \textbf{\bibinfo {volume} {D77}},\ \bibinfo {pages} {093012} (\bibinfo {year}
  {2008}),\ \Eprint{http://arxiv.org/abs/0709.3889}{arXiv:0709.3889 [hep-ph]}%
  \bibAnnoteFile{NoStop}{Bross:2007ts}%
%%CITATION = ARXIV:0709.3889;%%
\bibitem{FernandezMartinez:2010zza}%
  \BibitemOpen
  \bibfield{author}{%
  \bibinfo {author} {\bibfnamefont{E.}~\bibnamefont{Fernandez~Martinez}},
  \bibinfo {author} {\bibfnamefont{T.}~\bibnamefont{Li}}, \bibinfo {author}
  {\bibfnamefont{S.}~\bibnamefont{Pascoli}},\ and\ \bibinfo {author}
  {\bibfnamefont{O.}~\bibnamefont{Mena}},\ }%
  \bibfield{journal}{%
  \Doi{10.1103/PhysRevD.81.073010}{\bibinfo {journal} {Phys.Rev.}}\ }%
  \textbf{\bibinfo {volume} {D81}},\ \bibinfo {pages} {073010} (\bibinfo {year}
  {2010}),\ \Eprint{http://arxiv.org/abs/0911.3776}{arXiv:0911.3776 [hep-ph]}%
  \bibAnnoteFile{NoStop}{FernandezMartinez:2010zza}%
%%CITATION = ARXIV:0911.3776;%%
\bibitem{Huber:2006wb}%
  \BibitemOpen
  \bibfield{author}{%
  \bibinfo {author} {\bibfnamefont{P.}~\bibnamefont{Huber}}, \bibinfo {author}
  {\bibfnamefont{M.}~\bibnamefont{Lindner}}, \bibinfo {author}
  {\bibfnamefont{M.}~\bibnamefont{Rolinec}},\ and\ \bibinfo {author}
  {\bibfnamefont{W.}~\bibnamefont{Winter}},\ }%
  \bibfield{journal}{%
  \Doi{10.1103/PhysRevD.74.073003}{\bibinfo {journal} {Phys.Rev.}}\ }%
  \textbf{\bibinfo {volume} {D74}},\ \bibinfo {pages} {073003} (\bibinfo {year}
  {2006}),\ \Eprint{http://arxiv.org/abs/hep-ph/0606119}{arXiv:hep-ph/0606119
  [hep-ph]}%
  \bibAnnoteFile{NoStop}{Huber:2006wb}%
%%CITATION = HEP-PH/0606119;%%
\bibitem{Agarwalla:2010hk}%
  \BibitemOpen
  \bibfield{author}{%
  \bibinfo {author} {\bibfnamefont{S.~K.}\ \bibnamefont{Agarwalla}}, \bibinfo
  {author} {\bibfnamefont{P.}~\bibnamefont{Huber}}, \bibinfo {author}
  {\bibfnamefont{J.}~\bibnamefont{Tang}},\ and\ \bibinfo {author}
  {\bibfnamefont{W.}~\bibnamefont{Winter}},\ }%
  \bibfield{journal}{%
  \Doi{10.1007/JHEP01(2011)120}{\bibinfo {journal} {JHEP}}\ }%
  \textbf{\bibinfo {volume} {1101}},\ \bibinfo {pages} {120} (\bibinfo {year}
  {2011}),\ \Eprint{http://arxiv.org/abs/1012.1872}{arXiv:1012.1872 [hep-ph]}%
  \bibAnnoteFile{NoStop}{Agarwalla:2010hk}%
%%CITATION = ARXIV:1012.1872;%%
\bibitem{Choubey:2011zzq}%
  \BibitemOpen
  \bibfield{author}{%
  \bibinfo {author} {\bibfnamefont{S.}~\bibnamefont{Choubey}} \emph{et~al.}
  (\bibinfo {collaboration} {IDS-NF Collaboration})}%
   (\bibinfo {year} {2011}),\
  \Eprint{http://arxiv.org/abs/1112.2853}{arXiv:1112.2853 [hep-ex]}%
  \bibAnnoteFile{NoStop}{Choubey:2011zzq}%
%%CITATION = ARXIV:1112.2853;%%
\bibitem{Ballett:2012rz}%
  \BibitemOpen
  \bibfield{author}{%
  \bibinfo {author} {\bibfnamefont{P.}~\bibnamefont{Ballett}}\ and\ \bibinfo
  {author} {\bibfnamefont{S.}~\bibnamefont{Pascoli}},\ }%
  \bibfield{journal}{%
  \Doi{10.1103/PhysRevD.86.053002}{\bibinfo {journal} {Phys.Rev.}}\ }%
  \textbf{\bibinfo {volume} {D86}},\ \bibinfo {pages} {053002} (\bibinfo {year}
  {2012}),\ \Eprint{http://arxiv.org/abs/1201.6299}{arXiv:1201.6299 [hep-ph]}%
  \bibAnnoteFile{NoStop}{Ballett:2012rz}%
%%CITATION = ARXIV:1201.6299;%%
\bibitem{Michael:2008bc}%
  \BibitemOpen
  \bibfield{author}{%
  \bibinfo {author} {\bibfnamefont{D.}~\bibnamefont{Michael}} \emph{et~al.}
  (\bibinfo {collaboration} {MINOS}),\ }%
  \bibfield{journal}{%
  \Doi{10.1016/j.nima.2008.08.003}{\bibinfo {journal} {Nucl.Instrum.Meth.}}\ }%
  \textbf{\bibinfo {volume} {A596}},\ \bibinfo {pages} {190} (\bibinfo {year}
  {2008}),\ \Eprint{http://arxiv.org/abs/0805.3170}{arXiv:0805.3170
  [physics.ins-det]}%
  \bibAnnoteFile{NoStop}{Michael:2008bc}%
%%CITATION = ARXIV:0805.3170;%%
\bibitem{Bayes:2012ex}%
  \BibitemOpen
  \bibfield{author}{%
  \bibinfo {author} {\bibfnamefont{R.}~\bibnamefont{Bayes}}, \bibinfo {author}
  {\bibfnamefont{A.}~\bibnamefont{Laing}}, \bibinfo {author}
  {\bibfnamefont{F.}~\bibnamefont{Soler}}, \bibinfo {author}
  {\bibfnamefont{A.}~\bibnamefont{Cervera~Villanueva}}, \bibinfo {author}
  {\bibfnamefont{J.}~\bibnamefont{Gomez~Cadenas}}, \emph{et~al.},\ }%
  \bibfield{journal}{%
  \Doi{10.1103/PhysRevD.86.093015}{\bibinfo {journal} {Phys.Rev.}}\ }%
  \textbf{\bibinfo {volume} {D86}},\ \bibinfo {pages} {093015} (\bibinfo {year}
  {2012}),\ \Eprint{http://arxiv.org/abs/1208.2735}{arXiv:1208.2735 [hep-ex]}%
  \bibAnnoteFile{NoStop}{Bayes:2012ex}%
%%CITATION = ARXIV:1208.2735;%%
\bibitem{Bayes:PrivateComm}%
  \BibitemOpen
  \bibfield{author}{%
  \bibinfo {author} {\bibfnamefont{R.}~\bibnamefont{Bayes}},\ }%
  \bibinfo {note} {{private communication.}}%
  \bibAnnoteFile{Stop}{Bayes:PrivateComm}%
\bibitem{Petcov:1993rk}%
  \BibitemOpen
  \bibfield{author}{%
  \bibinfo {author} {\bibfnamefont{S.}~\bibnamefont{Petcov}}\ and\ \bibinfo
  {author} {\bibfnamefont{A.~Y.}\ \bibnamefont{Smirnov}},\ }%
  \bibfield{journal}{%
  \Doi{10.1016/0370-2693(94)90498-7}{\bibinfo {journal} {Phys.Lett.}}\ }%
  \textbf{\bibinfo {volume} {B322}},\ \bibinfo {pages} {109} (\bibinfo {year}
  {1994}),\ \Eprint{http://arxiv.org/abs/hep-ph/9311204}{arXiv:hep-ph/9311204
  [hep-ph]}%
  \bibAnnoteFile{NoStop}{Petcov:1993rk}%
%%CITATION = HEP-PH/9311204;%%
\bibitem{Vissani:1999tu}%
  \BibitemOpen
  \bibfield{author}{%
  \bibinfo {author} {\bibfnamefont{F.}~\bibnamefont{Vissani}},\ }%
  \bibfield{journal}{%
  \Doi{10.1088/1126-6708/1999/06/022}{\bibinfo {journal} {JHEP}}\ }%
  \textbf{\bibinfo {volume} {9906}},\ \bibinfo {pages} {022} (\bibinfo {year}
  {1999}),\ \Eprint{http://arxiv.org/abs/hep-ph/9906525}{arXiv:hep-ph/9906525
  [hep-ph]}%
  \bibAnnoteFile{NoStop}{Vissani:1999tu}%
%%CITATION = HEP-PH/9906525;%%
\bibitem{Pascoli:2003ke}%
  \BibitemOpen
  \bibfield{author}{%
  \bibinfo {author} {\bibfnamefont{S.}~\bibnamefont{Pascoli}}\ and\ \bibinfo
  {author} {\bibfnamefont{S.}~\bibnamefont{Petcov}},\ }%
  \bibfield{journal}{%
  \Doi{10.1016/j.physletb.2003.11.030}{\bibinfo {journal} {Phys.Lett.}}\ }%
  \textbf{\bibinfo {volume} {B580}},\ \bibinfo {pages} {280} (\bibinfo {year}
  {2004}),\ \Eprint{http://arxiv.org/abs/hep-ph/0310003}{arXiv:hep-ph/0310003
  [hep-ph]}%
  \bibAnnoteFile{NoStop}{Pascoli:2003ke}%
%%CITATION = HEP-PH/0310003;%%
\bibitem{Pascoli:2005zb}%
  \BibitemOpen
  \bibfield{author}{%
  \bibinfo {author} {\bibfnamefont{S.}~\bibnamefont{Pascoli}}, \bibinfo
  {author} {\bibfnamefont{S.}~\bibnamefont{Petcov}},\ and\ \bibinfo {author}
  {\bibfnamefont{T.}~\bibnamefont{Schwetz}},\ }%
  \bibfield{journal}{%
  \Doi{10.1016/j.nuclphysb.2005.11.003}{\bibinfo {journal} {Nucl.Phys.}}\ }%
  \textbf{\bibinfo {volume} {B734}},\ \bibinfo {pages} {24} (\bibinfo {year}
  {2006}),\ \Eprint{http://arxiv.org/abs/hep-ph/0505226}{arXiv:hep-ph/0505226
  [hep-ph]}%
  \bibAnnoteFile{NoStop}{Pascoli:2005zb}%
%%CITATION = HEP-PH/0505226;%%
\bibitem{Choubey:2005rq}%
  \BibitemOpen
  \bibfield{author}{%
  \bibinfo {author} {\bibfnamefont{S.}~\bibnamefont{Choubey}}\ and\ \bibinfo
  {author} {\bibfnamefont{W.}~\bibnamefont{Rodejohann}},\ }%
  \bibfield{journal}{%
  \Doi{10.1103/PhysRevD.72.033016}{\bibinfo {journal} {Phys.Rev.}}\ }%
  \textbf{\bibinfo {volume} {D72}},\ \bibinfo {pages} {033016} (\bibinfo {year}
  {2005}),\ \Eprint{http://arxiv.org/abs/hep-ph/0506102}{arXiv:hep-ph/0506102
  [hep-ph]}%
  \bibAnnoteFile{NoStop}{Choubey:2005rq}%
%%CITATION = HEP-PH/0506102;%%
\bibitem{Simkovic:2010ka}%
  \BibitemOpen
  \bibfield{author}{%
  \bibinfo {author} {\bibfnamefont{F.}~\bibnamefont{Simkovic}}, \bibinfo
  {author} {\bibfnamefont{J.}~\bibnamefont{Vergados}},\ and\ \bibinfo {author}
  {\bibfnamefont{A.}~\bibnamefont{Faessler}},\ }%
  \bibfield{journal}{%
  \Doi{10.1103/PhysRevD.82.113015}{\bibinfo {journal} {Phys.Rev.}}\ }%
  \textbf{\bibinfo {volume} {D82}},\ \bibinfo {pages} {113015} (\bibinfo {year}
  {2010}),\ \Eprint{http://arxiv.org/abs/1006.0571}{arXiv:1006.0571 [hep-ph]}%
  \bibAnnoteFile{NoStop}{Simkovic:2010ka}%
%%CITATION = ARXIV:1006.0571;%%
\bibitem{Dell'Oro:2014yca}%
  \BibitemOpen
  \bibfield{author}{%
  \bibinfo {author} {\bibfnamefont{S.}~\bibnamefont{Dell'Oro}}, \bibinfo
  {author} {\bibfnamefont{S.}~\bibnamefont{Marcocci}},\ and\ \bibinfo {author}
  {\bibfnamefont{F.}~\bibnamefont{Vissani}},\ }%
  \bibfield{journal}{%
  \Doi{10.1103/PhysRevD.90.033005}{\bibinfo {journal} {Phys.Rev.}}\ }%
  \textbf{\bibinfo {volume} {D90}},\ \bibinfo {pages} {033005} (\bibinfo {year}
  {2014}),\ \Eprint{http://arxiv.org/abs/1404.2616}{arXiv:1404.2616 [hep-ph]}%
  \bibAnnoteFile{NoStop}{Dell'Oro:2014yca}%
%%CITATION = ARXIV:1404.2616;%%
\bibitem{Bilenky:2001rz}%
  \BibitemOpen
  \bibfield{author}{%
  \bibinfo {author} {\bibfnamefont{S.~M.}\ \bibnamefont{Bilenky}}, \bibinfo
  {author} {\bibfnamefont{S.}~\bibnamefont{Pascoli}},\ and\ \bibinfo {author}
  {\bibfnamefont{S.}~\bibnamefont{Petcov}},\ }%
  \bibfield{journal}{%
  \Doi{10.1103/PhysRevD.64.053010}{\bibinfo {journal} {Phys.Rev.}}\ }%
  \textbf{\bibinfo {volume} {D64}},\ \bibinfo {pages} {053010} (\bibinfo {year}
  {2001}),\ \Eprint{http://arxiv.org/abs/hep-ph/0102265}{arXiv:hep-ph/0102265
  [hep-ph]}%
  \bibAnnoteFile{NoStop}{Bilenky:2001rz}%
%%CITATION = HEP-PH/0102265;%%
\bibitem{Capelli:2005jf}%
  \BibitemOpen
  \bibfield{author}{%
  \bibinfo {author} {\bibfnamefont{S.}~\bibnamefont{Capelli}} \emph{et~al.}
  (\bibinfo {collaboration} {CUORE})}%
   (\bibinfo {year} {2005}),\
  \Eprint{http://arxiv.org/abs/hep-ex/0505045}{arXiv:hep-ex/0505045 [hep-ex]}%
  \bibAnnoteFile{NoStop}{Capelli:2005jf}%
%%CITATION = HEP-EX/0505045;%%
\bibitem{BRUGNERA:2014ava}%
  \BibitemOpen
  \bibfield{author}{%
  \bibinfo {author} {\bibfnamefont{R.}~\bibnamefont{Brugnera}} \emph{et~al.}
  (\bibinfo {collaboration} {GERDA}),\ }%
  \bibfield{journal}{%
  \bibinfo {journal} {PoS}\ }%
  \textbf{\bibinfo {volume} {Neutel2013}},\ \bibinfo {pages} {039} (\bibinfo
  {year} {2013})%
  \bibAnnoteFile{NoStop}{BRUGNERA:2014ava}%
%%CITATION = POSCI,Neutel2013,039;%%
\bibitem{Albert:2014awa}%
  \BibitemOpen
  \bibfield{author}{%
  \bibinfo {author} {\bibfnamefont{J.}~\bibnamefont{Albert}} \emph{et~al.}
  (\bibinfo {collaboration} {EXO-200 Collaboration}),\ }%
  \bibfield{journal}{%
  \Doi{10.1038/nature13432}{\bibinfo {journal} {Nature}}\ }%
  \textbf{\bibinfo {volume} {510}},\ \bibinfo {pages} {229} (\bibinfo {year}
  {2014}),\ \Eprint{http://arxiv.org/abs/1402.6956}{arXiv:1402.6956 [nucl-ex]}%
  \bibAnnoteFile{NoStop}{Albert:2014awa}%
%%CITATION = ARXIV:1402.6956;%%
\bibitem{TheKamLAND-Zen:2014lma}%
  \BibitemOpen
   (\bibinfo {year} {2014}),\
  \Eprint{http://arxiv.org/abs/1409.0077}{arXiv:1409.0077 [physics.ins-det]}%
  \bibAnnoteFile{NoStop}{TheKamLAND-Zen:2014lma}%
%%CITATION = ARXIV:1409.0077;%%
\bibitem{Albert:2014afa}%
  \BibitemOpen
  \bibfield{author}{%
  \bibinfo {author} {\bibfnamefont{J.}~\bibnamefont{Albert}},\ }%
  \bibfield{journal}{%
  \Doi{10.1051/epjconf/20146608001}{\bibinfo {journal} {EPJ Web Conf.}}\ }%
  \textbf{\bibinfo {volume} {66}},\ \bibinfo {pages} {08001} (\bibinfo {year}
  {2014})%
  \bibAnnoteFile{NoStop}{Albert:2014afa}%
%%CITATION = 00776,66,08001;%%
\bibitem{Aguirre:2014lua}%
  \BibitemOpen
  \bibfield{author}{%
  \bibinfo {author} {\bibfnamefont{D.}~\bibnamefont{Artusa}} \emph{et~al.}
  (\bibinfo {collaboration} {CUORE Collaboration}),\ }%
  \bibfield{journal}{%
  \Doi{10.1140/epjc/s10052-014-2956-6}{\bibinfo {journal} {Eur.Phys.J.}}\ }%
  \textbf{\bibinfo {volume} {C74}},\ \bibinfo {pages} {2956} (\bibinfo {year}
  {2014}),\ \Eprint{http://arxiv.org/abs/1402.0922}{arXiv:1402.0922
  [physics.ins-det]}%
  \bibAnnoteFile{NoStop}{Aguirre:2014lua}%
%%CITATION = ARXIV:1402.0922;%%
\bibitem{Sibley:2014nda}%
  \BibitemOpen
  \bibfield{author}{%
  \bibinfo {author} {\bibfnamefont{L.}~\bibnamefont{Sibley}} (\bibinfo
  {collaboration} {SNO+}),\ }%
  \bibfield{journal}{%
  \Doi{10.1063/1.4883464}{\bibinfo {journal} {AIP Conf.Proc.}}\ }%
  \textbf{\bibinfo {volume} {1604}},\ \bibinfo {pages} {449} (\bibinfo {year}
  {2014})%
  \bibAnnoteFile{NoStop}{Sibley:2014nda}%
%%CITATION = APCPC,1604,449;%%
\bibitem{DavidLorcafortheNEXT:2014fga}%
  \BibitemOpen
  \bibfield{author}{%
  \bibinfo {author} {\bibfnamefont{D.}~\bibnamefont{Lorca}} (\bibinfo
  {collaboration} {NEXT Collaboration})}%
   (\bibinfo {year} {2014}),\
  \Eprint{http://arxiv.org/abs/1411.0475}{arXiv:1411.0475 [physics.ins-det]}%
  \bibAnnoteFile{NoStop}{DavidLorcafortheNEXT:2014fga}%
%%CITATION = ARXIV:1411.0475;%%
\bibitem{Gomez-Cadenas:2014dxa}%
  \BibitemOpen
  \bibfield{author}{%
  \bibinfo {author} {\bibfnamefont{J.~J.}\ \bibnamefont{Gomez-Cadenas}}}%
   (\bibinfo {year} {2014}),\
  \Eprint{http://arxiv.org/abs/1411.2433}{arXiv:1411.2433 [physics.ins-det]}%
  \bibAnnoteFile{NoStop}{Gomez-Cadenas:2014dxa}%
%%CITATION = ARXIV:1411.2433;%%
\bibitem{Fritts:2013mwa}%
  \BibitemOpen
  \bibfield{author}{%
  \bibinfo {author} {\bibfnamefont{M.}~\bibnamefont{Fritts}}\ and\ \bibinfo
  {author} {\bibfnamefont{K.}~\bibnamefont{Zuber}} (\bibinfo {collaboration}
  {COBRA}),\ }%
  \bibfield{journal}{%
  \Doi{10.1016/j.nuclphysbps.2013.04.052}{\bibinfo {journal}
  {Nucl.Phys.Proc.Suppl.}}\ }%
  \textbf{\bibinfo {volume} {237-238}},\ \bibinfo {pages} {37} (\bibinfo {year}
  {2013})%
  \bibAnnoteFile{NoStop}{Fritts:2013mwa}%
%%CITATION = NUPHZ,237-238,37;%%
\bibitem{Xu:2015dfa}%
  \BibitemOpen
  \bibfield{author}{%
  \bibinfo {author} {\bibfnamefont{W.}~\bibnamefont{Xu}} \emph{et~al.}
  (\bibinfo {collaboration} {Majorana})}%
   (\bibinfo {year} {2015}),\
  \Eprint{http://arxiv.org/abs/1501.03089}{arXiv:1501.03089 [nucl-ex]}%
  \bibAnnoteFile{NoStop}{Xu:2015dfa}%
%%CITATION = ARXIV:1501.03089;%%
\bibitem{Nova:2013ata}%
  \BibitemOpen
  \bibfield{author}{%
  \bibinfo {author} {\bibfnamefont{F.}~\bibnamefont{Nova}},\ }%
  \bibfield{journal}{%
  \Doi{10.1063/1.4826748}{\bibinfo {journal} {AIP Conf.Proc.}}\ }%
  \textbf{\bibinfo {volume} {1560}},\ \bibinfo {pages} {184} (\bibinfo {year}
  {2013})%
  \bibAnnoteFile{NoStop}{Nova:2013ata}%
%%CITATION = APCPC,1560,184;%%
\bibitem{Ishihara:2012pwa}%
  \BibitemOpen
  \bibfield{author}{%
  \bibinfo {author} {\bibfnamefont{N.}~\bibnamefont{Ishihara}} (\bibinfo
  {collaboration} {DCBA}),\ }%
  \bibfield{journal}{%
  \Doi{10.1016/j.nuclphysbps.2012.09.118}{\bibinfo {journal}
  {Nucl.Phys.Proc.Suppl.}}\ }%
  \textbf{\bibinfo {volume} {229-232}},\ \bibinfo {pages} {481} (\bibinfo
  {year} {2012})%
  \bibAnnoteFile{NoStop}{Ishihara:2012pwa}%
%%CITATION = NUPHZ,229-232,481;%%
\bibitem{Bilenky:1996cb}%
  \BibitemOpen
  \bibfield{author}{%
  \bibinfo {author} {\bibfnamefont{S.~M.}\ \bibnamefont{Bilenky}}, \bibinfo
  {author} {\bibfnamefont{C.}~\bibnamefont{Giunti}}, \bibinfo {author}
  {\bibfnamefont{C.}~\bibnamefont{Kim}},\ and\ \bibinfo {author}
  {\bibfnamefont{S.}~\bibnamefont{Petcov}},\ }%
  \bibfield{journal}{%
  \Doi{10.1103/PhysRevD.54.4432}{\bibinfo {journal} {Phys.Rev.}}\ }%
  \textbf{\bibinfo {volume} {D54}},\ \bibinfo {pages} {4432} (\bibinfo {year}
  {1996}),\ \Eprint{http://arxiv.org/abs/hep-ph/9604364}{arXiv:hep-ph/9604364
  [hep-ph]}%
  \bibAnnoteFile{NoStop}{Bilenky:1996cb}%
%%CITATION = HEP-PH/9604364;%%
\bibitem{Li:2015jxa}%
  \BibitemOpen
  \bibfield{author}{%
  \bibinfo {author} {\bibfnamefont{C.-C.}\ \bibnamefont{Li}}\ and\ \bibinfo
  {author} {\bibfnamefont{G.-J.}\ \bibnamefont{Ding}}}%
   (\bibinfo {year} {2015}),\
  \Eprint{http://arxiv.org/abs/1503.03711}{arXiv:1503.03711 [hep-ph]}%
  \bibAnnoteFile{NoStop}{Li:2015jxa}%
%%CITATION = ARXIV:1503.03711;%%
\bibitem{DiIura:2015kfa}%
  \BibitemOpen
  \bibfield{author}{%
  \bibinfo {author} {\bibfnamefont{A.}~\bibnamefont{Di~Iura}}, \bibinfo
  {author} {\bibfnamefont{C.}~\bibnamefont{Hagedorn}},\ and\ \bibinfo {author}
  {\bibfnamefont{D.}~\bibnamefont{Meloni}}}%
   (\bibinfo {year} {2015}),\
  \Eprint{http://arxiv.org/abs/1503.04140}{arXiv:1503.04140 [hep-ph]}%
  \bibAnnoteFile{NoStop}{DiIura:2015kfa}%
%%CITATION = ARXIV:1503.04140;%%
\end{thebibliography}%

\end{document}